\documentclass[preprint,showpacs,amsmath,amssymb,pra]{revtex4-1}

\usepackage{graphicx}
\usepackage{dcolumn}
\usepackage{bm}
\usepackage{overpic,multirow,subfigure}
\usepackage{longtable}

\newcommand{\avg}[1]{\langle #1 \rangle}

\begin{document}

\preprint{AIP}

\title{Late-time growth rate, mixing and anisotropy in the multimode narrowband Richtmyer--Meshkov Instability: the $\theta$-Group Collaboration}

\author{B. Thornber}
 \email{ben.thornber@sydney.edu.au}
\affiliation{%
School of Aerospace, Mechanical and Mechatronic Engineering,\\ The University of Sydney, Australia
}%
\author{J. Griffond, O. Poujade}
\affiliation{%
CEA, DAM, DIF, F-91297 Arpajon, France
}%
\author{N. Attal, H. Varshochi, P. Bigdelou, P. Ramaprabhu}
\affiliation{%
Mechanical Engineering and Engineering Science, UNC Charlotte, USA
}%
\author{B. Olson, J. Greenough, Y. Zhou, O.Schilling}
\affiliation{%
Lawrence Livermore National Laboratory, Livermore, USA
}%
\author{K. A. Garside, R. J. R. Williams, C. A. Batha}
\affiliation{%
AWE, Aldermaston, UK
}%
\author{P. A. Kuchugov$^{a,b}$, M. E. Ladonkina$^a$, V. F. Tishkin$^a$, N. V. Zmitrenko$^a$, \& V. B. Rozanov$^b$}
\affiliation{%
$^a$ Keldysh Institute of Applied Mathematics, Russian Academy of Sciences, Moscow, Russia, $^b$ P.N. Lebedev Physical Institute, Russian Academy of Sciences, Moscow, Russia
}%
\author{D. L. Youngs}
\affiliation{%
University of Strathclyde, Department of Mechanical and Aerospace Engineering,Glasgow, G1 1XJ, United Kingdom
}%

\date{\today}

\begin{abstract}
Turbulent Richtmyer--Meshkov instability (RMI) is investigated through a series of high resolution three dimensional smulations of two initial conditions with eight independent codes. The simulations are initialised with a narrowband perturbation such that instability growth is due to non-linear coupling/backscatter from the energetic modes, thus generating the lowest expected growth rate from a pure RMI. By independently assessing the results from each algorithm, and computing ensemble averages of multiple algorithms, the results allow a quantification of key flow properties as well as the uncertainty due to differing numerical approaches. A new analytical model predicting the initial layer growth for a multimode narrowband perturbation is presented, along with two models for the linear and non-linear regime combined. Overall, the growth rate exponent is determined as $\theta=0.292 \pm 0.009$, in good agreement with prior studies; however, the exponent is decaying slowly in time. Also, $\theta$ is shown to be relatively insensitive to the choice of mixing layer width measurement. The asymptotic integral molecular mixing measures $\Theta=0.792\pm 0.014$, $\Xi=0.800 \pm 0.014$ and $\Psi=0.782\pm 0.013$ which are lower than some experimental measurements but within the range of prior numerical studies. The flow field is shown to be persistently anisotropic for all algorithms, at the latest time having between 49\% and 66\% higher kinetic energy in the shock parallel direction compared to perpendicular and does not show any return to isotropy. The plane averaged volume fraction profiles at different time instants collapse reasonably well when scaled by the integral width, implying that the layer can be described by a single length scale and thus a single $\theta$. Quantitative data given for both ensemble averages and individual algorithms provide useful benchmark results for future research.
\end{abstract}

\pacs{47.27.Gs,47.27.wj,47.27.ep,47.27.Jv,47.40.Nm,47.20.Ma}
\maketitle

\section{Introduction}

Richtmyer--Meshkov instability (RMI) occurs when a perturbed interface between two materials is accelerated impulsively, usually by an incident shock wave \cite{Richtmyer1960, Meshkov1969}. During the passage of the shock-wave, a layer of vorticity is deposited by baroclinic source terms at the interface of the gas, and in the nearby region as the shock returns to planarity. This deposition causes a net growth of the layer, which during the linear stage for a single mode may be estimated by $\dot {a}\approx k At^+ a_0^+ \Delta u$ \cite{Richtmyer1960} where $\dot {a}$ is the growth rate of the perturbation, $k$ the wavenumber, $At^+$ the post-shock Atwood number, $a_0^+$ the post-shock amplitude and $\Delta u$ the velocity jump imparted by the incident shock. 

The linear growth rate reduces as the amplitude of the pertubation becomes large relative to the wavelength, and characteristic bubble/spike (mushrooms) features develop. At later times, shear layers along the contact surface break down due to Kelvin--Helmholtz instabilities which drive a transfer of kinetic energy from the shock-parallel direction to the perpendicular components. It is reasonably well established that a multimode perturbation will transition to a turbulent mixing layer with a growth rate  $\propto t^\theta$; however, the exact behaviour of the layer is complicated by a strong dependence on initial conditions, and chaotic turbulence physics. For a comprehensive and up to date survey of research on Richtmyer--Meshkov and the closely related Rayleigh--Taylor instability see Zhou \cite{Zhou2017}. 

RMI occurs in man-made events such as inertial confinement implosions (e.g. \cite{Clark2016}), explosions (e.g. \cite{Kuhl2013}), and fuel injection in supersonic flows \cite{yang1993}, or in large astrophysical events such as supernovae (e.g. \cite{Burrows2000}). Such events are by their nature relatively inaccessible to detailed experimental diagnostics, due to short time scales, extremes of temperature and pressure, or very small or very large observational distances. Thus the understanding of RMI relies on a judicious use of theory informed by careful experimental and numerical studies, the latter of which is the focus of this paper.

There have been multiple numerical studies of three dimensional multimode RMI which have improved our understanding of this complex flow \cite{Youngs1994,Youngs2004,Hill2006,Thornber2010,Lombardini2012,tritschler2014richtmyer,Oggian2015,Liu2016,Weber2013,Thornber2016}. Each study has explored the integral properties of the flow field, such as width, mixedness and fluctuating kinetic energy, along with higher order quantities. Even for the lower order quantities, a consensus on the measured physical behaviour has been difficult, given the acknowledged sensitivity of the late time behaviour to the initial conditions, and the challenging flow physics for even state of the art algorithms. 

As summarized by Zhou \cite{Zhou2017}, the growth exponent $\theta$ for the mixing layer width lies between $0.213$ and $2/3$ in theoretical, experimental and numerical studies. Of the ten prior studies, four of them have similar idealised narrowband initial conditions and thus may be compared \cite{Youngs2004,Thornber2010,Oggian2015,Thornber2016}, however the other studies have individually tailored the initial conditions to match specific experiments or for numerical reasons. Thus a direct comparison of integral properties between all prior studies is not possible. 

Furthermore, of all prior studies, only four include computations using more than one algorithm  for the same initial condition \cite{Thornber2010,Oggian2015,tritschler2014richtmyer,Olson2014}, each of them using two independent algorithms. It is not possible to obtain a robust estimate of numerical uncertainty with such a limited sample. As all other studies used differing initial conditions, information about numerical uncertainty cannot be obtained from the existing literature. This is particularly important as all algorithms are challenged by the need to (i) capture shock waves, (ii) capture moving contact surfaces and (iii) represent fine scale turbulent motion.

This paper considers RMI generated from an initial high wavenumber perturbation on a contact surface. Although most realistic problems have a larger range of initial perturbation lengthscales, this perturbation form is of interest as it represents a lower limit on the growth rate due to RMI, since the late time growth at large lengthscales is driven solely through mode coupling. Based on the previous literature, it is expected that the growth exponent of the layer will be $\approx 0.213 \le \theta \le 0.295$. Linked to the growth rate is the mixedness of the layer, where generally a higher $\theta$ implies a less well mixed layer. Mixedness is particularly challenging to resolve since it depends on resolving a large enough portion of the sub-inertial range such that sub-cell variations in concentration are negligible (without a sub-cell representation of scalar variance). Finally, previous studies have highlighted significant asymmetry in the self-similar turbulent kinetic energy components, however late time self-similar anisotropy has not yet been conclusively demonstrated. 

Here a new initial condition is proposed which combines the most favourable features of prior studies such that simulations may be run using a wide range of algorithms ranging from Arbitrary-Lagrangian-Eulerian and Godunov methods through to algorithms based on compact differences. Simulations are then run with eight independent algorithms for an initial condition for which a grid converged solution may be achieved. 

Comparing results from many different codes simulating the same mixing flow has already proved fruitful in the past. The $\alpha$-group collaboration \cite{Dimonte2004} analysed the results of seven different codes in the context of Rayleigh--Taylor turbulent mixing. The conclusions were instrumental in understanding the effect of the initial perturbation spectrum on the value of the turbulent Rayleigh--Taylor growth parameter $\alpha$ (in $W(t)=\alpha\,{\cal A}\,g\,t^2$ where $W(t)$ is the mixing zone width, ${\cal A}$ is the Atwood number and $g$ is the acceleration of the interface).

A further set of simulations were then run for a more challenging condition which aimed to achieve later dimensionless times than have been presented to date. The results are then employed to give the current best-estimate of the key integral properties of the flow including the growth rate, integral mix measures and kinetic energy. The best estimates are accompanied by uncertainty estimates for each, which for the first time incorporate algorithmic uncertainty.

The layout of this paper is as follows. Section \ref{Problem} details the initial conditions, domain sizes, boundary conditions, algorithms, diagnostics, and the grid convergence study. Results for the standard case  are presented in Section \ref{standard} and the one-quarter length-scale case in Section \ref{quarter}. Finally, the key conclusions are summarised in Section \ref{concl}.

\section{Problem Description \label{Problem}}

\subsection{Standard Problem}

\begin{figure*}\centering
  \begin{overpic}[width=0.7\textwidth]{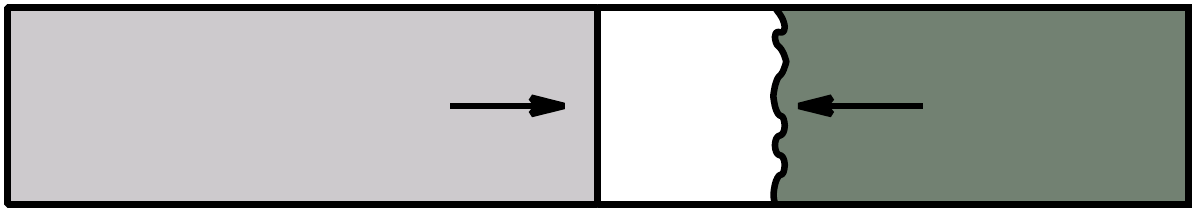}
    \small
    \put(20,10){Shocked}
    \put(22,6){heavy}
    \put(53,8){heavy}
    \put(80,8){light}
    \put(-1,-3){0}
    \put(48,-3){3.0}
    \put(62,-3){3.5}
    \put(97,-3){2.8$\pi$}
\end{overpic}
\caption{Schematic of the initial condition. \label{rmschematic}}
\end{figure*}
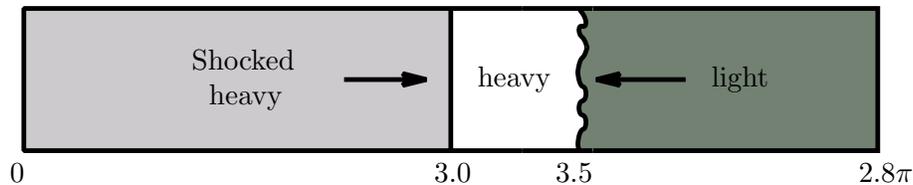

The objectives of the formulation of the initial conditions are to provide a platform from which an effective collaboration could be constructed to address gaps in understanding or regions of uncertainty, and to provide a reliable estimate of the growth exponent $\theta$ for a specific initial condition. Key to the formulation was to ensure that the initial condition could be run consistently in a range of numerical algorithms spanning Eulerian, Lagrange-remap through to compact differences. It should also be computationally feasible, i.e. a grid converged solution may be achieved at a resolution accessible by a wide range of codes. The choice of initial conditions build on previous studies \cite{Thornber2010,tritschler2014richtmyer,morgan2012late,morgan20132d,Olson2014} and have the following features:

\begin{itemize}
\item An initial range of length scales in the perturbation from $L/8$ to $L/4$ where $L$ is the cross-section. This should ensure that the layer growth rate is reasonably converged at $\approx 128^3$, and permits a future Direct Numerical Simulation (DNS) at an achievable resolution $\approx 1024^3$. 
\item A diffuse initial interface of error function form and thickness $L/32$. This leaves open a future option to run a DNS of this configuration, and is accessible for Large-eddy simulations (LES) which need an initial resolved interface thickness, whilst being narrow enough to not impact greatly the evolution of the instability.
\item A power spectrum with constant power at all initialised wavelengths and an overall amplitude of $0.1\lambda_{min}$. This form of initial conditions would provide more flexibility in later studies.
\item Mode amplitudes and phases defined using deterministic random numbers. These are the same at all grid resolutions such that a grid refinement study can be achieved with the same interface shape.
\item A moderate density ratio of $3:1$, $\gamma=5/3$ for both gases, and a Mach $1.84$ shock. 
\end{itemize}

The test case uses the initial conditions derived by \cite{Youngs2004}, which are shown schematically in Figure \ref{rmschematic}. The flow field consists of a heavy and light gas separated by a perturbed interface, where the perturbation satisfies a given power spectrum and mean amplitude. The heavy and light unshocked gases are at the same temperature in the undisturbed field, and the specific heats for each component ($C_{v1}$ and $C_{v2}$) are chosen to ensure this. 

The incident shock wave has a Mach number of $1.8439$, equivalent to a four-fold pressure increase. The initial conditions in the shocked domain, and left and right gases are

\begin{eqnarray}
0.0<x<3. \hspace{0.25cm} (\rho,u,p)&=&(6.37497 \,\rm{kg/m}^3,-61.48754\,\rm{m/s},399995.9\,\rm{Pa})\\
\rm{initial\,\, heavy}\hspace{0.25cm} (\rho,u,p)&=&(3.0\,\rm{kg/m}^3,-291.575\,\rm{m/s},100000\,\rm{Pa})\\
\rm{initial\,\, light} \hspace{0.25cm}(\rho,u,p)&=&(1.0\,\rm{kg/m}^3,-291.575\,\rm{m/s},100000\,\rm{Pa})
\end{eqnarray}

An initial velocity is given to the gas interface such that the centre of the interface is stationary after passage of the shock wave. The pre-shock densities of the heavy and light fluid are $3$ and $1$ kg/m$^{3}$ respectively, giving an Atwood number $A=0.5$. Both fluids have an initial pressure of 100 kPa. Post-shock densities were found to be $5.22$ and $1.8$ kg/m$^{3}$ for the heavy and light fluids respectively, giving a post-shock Atwood number of $At^+=0.487$. Assuming that the layer amplitude has been reduced by compression $a_0^+=(1-\Delta u/U_i)a_0$, where $U_i$ is the velocity of the incident shock,  gives the Richtmyer velocity $\dot a=k_{max} A^+ a^+_0 \Delta u=29.36$ m/s.

\begin{figure}
\begin{centering}
\includegraphics[width=0.49\textwidth]{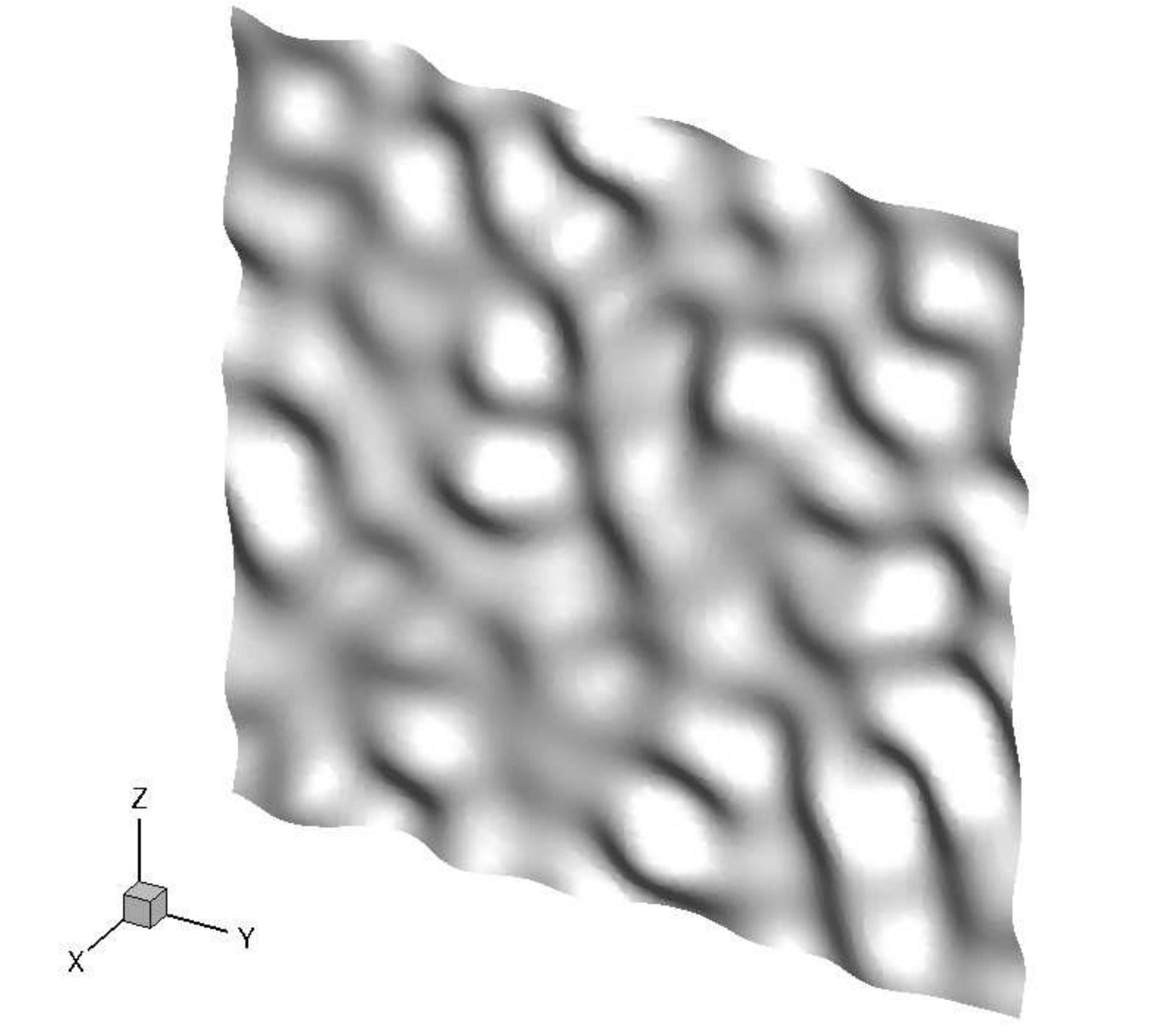}
\caption{Iso-surface of volume fraction $f_1=0.5$ at $180\times 128^2$ at $\tau=0$. \label{initvis}}
\end{centering}
\end{figure}

Here, the initial conditions use $\lambda_{min}=L/8$ and $\lambda_{max}=L/4$ and a constant power spectrum. The derivation of the perturbation in the initial condition is detailed in Appendix \ref{deriv}. A diffuse interface of gradient thickness of $\delta=L/32$ is required, thus the final form of the initial conditions is:

\begin{equation}
f_1(y,z)=\frac{1}{2} {\rm erfc} \left(\frac{\sqrt{\pi} (x-S(y,z))}{\delta}\right)
\end{equation}

\noindent where $S(y,z)=3.5+A(y,z)$, $A(y,z)$ is defined in Appendix \ref{deriv}, and $f_1$ is the volume fraction of the heavy gas. To improve the quality of the initial condition on coarse meshes, the initial condition is subsampled in $4 \times 4 \times 4$ sub-cells for each computational cell. This was deemed to be simpler than integrating the initial condition and gives satisfactorily converged results. A visualisation of the isosurface of $f_1=0.5$ is shown in Figure \ref{initvis} for the $180\times 128^2$ resolution. The diffuse initial condition was compared to a sharp interface in the Flamenco code, and it was verified that it does not have a large impact on the overall behaviour of the mixing layer, marginally increasing the growth rate and decreasing the mix parameter $\Theta$. Simulations using double the domain length highlighted that the spikes are expected to advect out of the domain at $t\approx 0.5$ s; thus, simulations which are run to $t=0.5$ s should give a reasonable guide to the asymptotic mix parameters and be reasonably independent of the boundaries (for lower order parameters). At the latest time, the integral width is less than 10\% of the domain size, and it should be reasonably statistically converged and not be box constrained \cite{Thornber2016}. However, it is important to emphasize that it may not be self-similar.

The computational domain is Cartesian as shown in Figure \ref{rmschematic} and measures $x\times y \times z=2.8\pi \times 2\pi \times 2\pi$ m$^3$. All codes but one (Hesione) in this study undertook a grid convergence study including mesh sizes $180\times 128^2$, $360 \times 256^2$, $720 \times 512^2$, and TURMOIL and Triclade were also run at a grid resolution of $1440\times 1024^2$. The boundary conditions are periodic in the $y$ and $z$ directions and transmissive in the $x$ direction. It should be noted that the boundary conditions in the direction of shock propagation vary considerably between the implementations, as it was decided that this should employ the `best practise' for each code. For example, the ALE algorithms would use a moving mesh option, and some Eulerian codes included an extended `buffer' region to damp out unphysical waves reflected from the outflow and inflow boundaries.  As detailed previously, simulations were run from $t=0$s to $t=0.5$s which is equivalent to a dimensionless time $\tau=6.15$, where the basis for non-dimensionalisation is detailed in Section \ref{mmricht}. Note that the computation is intended as a test of the high Reynolds number limit of turbulent mixing \cite{Zhou2016,Zhou2014}. Thus, the effects of physical viscosity on the resolved scales is assumed to be zero and all simulations presented here are inviscid. 

\subsection{Quarter-Scale Problem}

The grid convergence study demonstrated that several algorithms had  integral width $W$ with variations $<1\%$ between the lowest and highest grid resolutions on the standard problem. Quantitatively, the change in predicted integral widths at $t=0.5$ between the lowest and highest grid resolutions for each algorithm were: Ares (5.8\%), Flamenco (0.5\%), Flash (5.1\%), Miranda (3.8\%), NUT3D (16\%), Triclade (0.2\%) and TURMOIL (1.6\%). This indicated that several of these algorithms could reliably compute a converged integral width for a problem  where all physical length scales were divided by four, with the computational domain size maintained, at the highest grid resolution employed in the standard problem. This new case would have improved statistical fidelity, may be run to much later dimensionless times ($\tau \approx 125$) and thus have a greater chance of achieving self-similarity. In this problem, the initial range of physical length scales in the perturbation was modified to lie between $L/32$ to $L/16$ where $L$ is the cross-section, and the diffuse layer has an initial thickness $L/128$.   Note that one algorithm (NUT3D) ran a `half-scale' problem, where the initial length-scales were all scaled by a factor of two, instead of four. 

To facilitate the use of this test case in future algorithmic development and validation, Fortran and C implementations of the initial conditions may be made available to other research groups by contacting the lead author.

 \subsection{Initial Impulse and Non-Dimensionalisations \label{mmricht}}

In this subsection an approach is detailed which can estimate the initial growth rate for a narrowband multimode surface perturbation. This initial growth rate provides a consistent non-dimensionalisation for both problems explored in this paper. 

A single mode RMI at low to medium Mach numbers will have an initial growth rate $\dot {a}= k At^+ a_0^+ \Delta u$, where  $a_0^+=Ca_0=(1-\Delta u/U_i)a_0$. The current initial condition consists of a random multimode perturbation represented by a sum of Fourier modes with random phase $\phi$:

\begin{equation}
A(y,z)=\sum_{k_y,k_z} a_0 \cos(k_y y+k_z z+\phi).
\end{equation}

According to linear theory, after shock passage the perturbation is given by

\begin{eqnarray}
A(y,z)&=&\sum_{k_y,k_z} C\, a_0\,(1+k\, At^+\, \Delta u\, t) \cos(k_y y+k_z z+\phi)\\
&\approx& \sum_{k_y,k_z} C\, a_0\, k\, At^+\, \Delta u\, t \cos(k_y y+k_z z+\phi),
\end{eqnarray}

\noindent where the second line is valid towards the end of the linear phase, assuming an initial perturbation $a_0 \ll 2\pi/k$. It is assumed that the amplitude compression factor does not depend on the wavenumber $k=\sqrt{k_y^2+k_z^2}$. The initial variance of the perturbation is given by $\sigma^2(0)=\sigma^2_0=\sum_{k_y,k_z} a_0^2/2=\int_0^\infty P(k) dk$, where $P(k)$ is the power spectrum of the initial perturbation, so that

\begin{equation}
\sigma^2(t)\approx \sum_k \frac{1}{2} \left(C\, a_0\, k\, At^+\, \Delta u\, t\right)^2 \equiv \left(C\, At^+\,\Delta u\, t\right)^2 \int_0^\infty k^2 P(k) dk,
\end{equation}

\noindent which implies that $\dot \sigma=\bar{k}At^+\sigma_0^+ \Delta u$ with $\sigma_0^+=C\sigma_0$ and

\begin{equation}
\bar k=\sqrt{\frac{\int_0^\infty k^2 P(k)dk}{\int_0^\infty P(k)dk}}.
\end{equation}

\noindent For the narrowband perturbation utilised in this paper, $P(k)={\rm const.}$ for $k_{{\rm max}}/2 \le k \le k_{\rm max}$, giving $\bar k=\sqrt{7/12}k_{\rm max}$. 

The integral width $W=\int^{L_x}_0 \langle f_1\rangle \langle f_2\rangle dx$ is used to measure the growth of the layer, and must be related to the variance $\sigma$. For the type of perturbation chosen the height distribution $p(A)$ should be Gaussian:

\begin{equation}
p(A)=\frac{1}{\sigma \sqrt{2\pi}}\exp\left(-\frac{A^2}{2\sigma^2}\right).
\end{equation}

\noindent The dense fluid plane-averaged volume fraction is then

\begin{eqnarray}
\bar f_1(x)&=&{\rm prob}(A>x)=\int_x^\infty p(A)dA=\int_0^\infty p(A)dA-\int_0^x p(A)dA \nonumber\\
&=&\frac{1}{2} \left[1-{\rm erf}\left(\frac{x}{\sqrt{2}\sigma}\right)\right]
\end{eqnarray}

\noindent The integral width is therefore 

\begin{eqnarray}
W&=&\int_{-\infty}^\infty \frac{1}{4}\left[1-{\rm erf}^2\left(\frac{x}{\sqrt{2}\sigma}\right)\right] dx=\frac{\sqrt{2}}{4}\sigma \int_{-\infty}^\infty \left[1-{\rm erf}^2(t)\right]dt,\\
&\approx&0.564 \sigma
\end{eqnarray}

\noindent This result is close to that obtained with an assumed linear volume fraction distribution with bubble height equalling spike height $h$, which gives $W=h/3$, $\sigma=h/\sqrt{3}$, and thus $W= 0.577\sigma$. Thus the predicted initial growth rate is:

\begin{equation}
\dot W_0=0.564 \sqrt{\frac{7}{12}}\, k_{\rm max} At^+ \sigma^+ \Delta u
\label{wdotpred}
\end{equation}

\noindent For the initial conditions employed within both the standard and quarter scale case, $\dot W_0=12.649$m/s, from which $W=W_0^++\dot W_0 (t-t_{\mbox{shock time}})$ for the linear phase and $t-t_{\mbox{shock time}}\ge 0$.

\noindent All non-dimensional quantities within this paper are non-dimensionalised by $\dot W_0$, $\bar \lambda=2\pi/\bar k$, cross-sectional area $4\pi^2$ and $\rho_L=1$. For example, dimensionless time $\tau=t\dot W_0/\bar \lambda$.

\subsection{Quantities of Interest \label{quant}}

This study focuses on the integral properties of the mixing layer including width measures, mixing fractions, total kinetic energy in each direction, and kinetic energy spectra.

The time dependent evolution of integral mixing width $W$, molecular mixing fraction $\Theta$ and the mixing parameter $\Xi$ are defined as (see, for example \cite{Youngs1991,Youngs1994,Cook2002,Latini2007,Thornber2010}) 

\begin{equation}
W(t)=\int^{L_x}_0 \langle f_1\rangle \langle f_2\rangle dx.
\label{intwidtheq}
\end{equation}

\begin{equation}
\Theta(t)=\frac{\int \langle f_1 f_2 \rangle dx}{\int
\langle f_1 \rangle \langle f_2 \rangle dx},  \hspace{0.25cm} \Xi(t)=\frac{\int \langle \min(f_1,f_2) \rangle dx}{\int
\min(\langle f_1 \rangle,\langle f_2 \rangle) dx},
\end{equation}

\noindent where $\langle f_{1,2} \rangle$ indicates the $y-z$ plane averaged volume
fraction of species $1$, $2$ where species $1$ is the heavy gas. For codes which use mass fractions to track the species, volume fractions are derived from mass fractions assuming local pressure and temperature equilibrium.

Planar averaged kinetic energy in the $x$, $y$ and $z$ direction have been computed as the difference of the actual velocities minus the plane averaged velocity, summed over the entire mixing layer, 

\begin{eqnarray}
\rm{TKX}&=&\sum_{xyz}\frac{1}{2}\rho(u-\tilde{u})^2 dV,\,\,\, \tilde{u}=\frac{\sum_{yz} \rho\, u\, dV}{\sum_{yz} \rho\, dV},\\
\rm{TKY}&=&\sum_{xyz}\frac{1}{2}\rho v^2 dV,\\
\rm{TKZ}&=&\sum_{xyz}\frac{1}{2}\rho w^2 dV.
\label{eqntke2}
\end{eqnarray}

Finally, the two dimensional variable density spectra taken at the mixing layer centre (defined by $\langle f_1 \rangle=0.5$)\cite{Cook2002,Thornber2012b}:

\begin{equation}
E(k)=\frac{1}{2} \hat \nu_i^*\hat \nu_i,
\end{equation}

\noindent where $\nu_i=\rho^{1/2}u_i$, $\hat{(.)}$ indicates the Fourier transform of a quantity, and $\hat{(.)}^*$ is the complex conjugate of the transform.

\subsection{Numerical Methods}

Eight independent algorithms have been employed in this study which are briefly described here with references to publications with more complete information. 

{\bf Ares} is an Arbitrary Lagrangian--Eulerian (ALE) code developed at Lawrence Livermore National Laboratory (LLNL). The Lagrange time step uses a second order predictor-corrector method. The Gauss divergence theorem is used to solve the discrete finite difference equations of the compressible multi-component Navier--Stokes equations. Spatial derivatives are approximated using a second-order finite difference method. Artificial viscosity is applied to damp out spurious, high frequency oscillations which are generated near shocks and contact discontinuities. The maximum grid resolution run here is $720\times 512^2$.

{\bf Flamenco} is a Godunov-type solver which utilises a nominally fifth order reconstruction stencil (in 1D) \cite{Kim2005}, a HLLC Riemann solver \cite{HLLC}, coupled with a low Mach correction \cite{Thornber2007b,Thornber2007c,Thornber2007d}, and second order TVD time stepping \cite{Spiteri2002}. The governing equations are based on Navier--Stokes plus volume fraction equations \cite{Allaire2002}.  The maximum grid resolution run here is $720\times 512^2$.

{\bf FLASH} FLASH is a modular, massively parallel, open source code developed at the University of Chicago to investigate astrophysical flows \cite{fryxell2000flash}. The `Hydro module' is employed here to solve the compressible Euler equations on a block structured adaptively refined mesh. The Euler equations are solved using the Piecewise Parabolic Method (PPM) \cite{colella1984piecewise} complemented by the symmetric Monotized Central (MC) limiter \cite{van1977towards}. The maximum grid resolution run here is $720\times 512^2$.

{\bf Hesione}  is a finite difference Lagrange-remap code. In the Lagrange phase, it uses a second order accurate BBC scheme in space and time \cite{WOODWARD1984115}. For multimaterial flow, the remap phase requires an interface reconstruction (IR) based upon Youngs method and can operate with mixed meshes \cite{youngs1984interface}. A key difference in this numerical algorithm is that interface reconstruction  inhibits species diffusion, as if the two fluids are immiscible. That means the mixing is heterogeneous (with IR) and homogeneous with all other codes used by the collaboration.  The maximum grid resolution run here is $180\times 128^2$.

{\bf Miranda} uses a tenth--order compact differencing scheme for
spatial derivatives and a five--stage, fourth--order Runge--Kutta
scheme for temporal integration.  Full details of the numerical method are
given by Cook \cite{cook2007artificial}.  For numerical regularization of the
sharp, unresolved gradients in the flow, artificial fluid properties are
used to locally damp structures which exist on the length scales of the
computational mesh. The maximum grid resolution run here is $720\times 512^2$.

{\bf NUT3D}  solves the Euler equations augmented with mass fraction equations \cite{Tishkin1995,Lebo2006,Kuchugov2014}. The difference scheme follows the method proposed by  \cite{Vyaznikov1989}. A `sharp' second order approximation in space is coupled with the exact solution of the Riemann problem for the fluxes. Time integration is carried out by the `predictor-corrector' method. The maximum grid resolution run here is $720\times 512^2$.

{\bf Triclade} solves the  Navier--Stokes plus mass fraction equations using a  conservative finite difference method and is based on the wave propagation algorithm of Leveque \cite{LeVeque02} with high order accuracy provided by the corrections due to Daru and Tenaud \cite{Daru04}.  A fifth order time-space accuracy is chosen here, referred to as WP5 by Shanmuganathan {\it et al.} \cite{Shanmuganathan14} and a more detailed description of the method is given in appendix A.2 of that reference. The maximum grid resolution run here is $1440\times 1024^2$.

{\bf TURMOIL} solves the Euler equations as well as an equation for mass fraction using a Lagrange remap method \cite{Grinstein2007,Youngs1991}. An un-split, second-order finite difference method is used for the Lagrange phase and the monotonic advection method of van Leer \cite{vanLeer1977} in the directionally split remap phase. A higher-order artificial viscosity, similar to that proposed by Schultz \cite{Schultz1964} and Cook, Ulitsky and Miller \cite{Cook2013} is used in these calculations \cite{Williams2017}. The maximum grid resolution run here is $1440 \times 1024^2$.

Thus the eight algorithms include Godunov-type, ALE methods and compact differences. The algorithms are either Implicit LES (ILES) \cite{Youngs1991, Boris1992,Grinstein2007,Drikakis2009} or Artificial Fluid LES (AFLES) \cite{cook2007artificial}. In the case of ILES, it is assumed that there is sufficient separation between the integral length scales and the grid scale such that the growth of the integral length scale is independent of the exact dissipation mechanism. Furthermore, it is assumed that the Reynolds number is sufficiently high that the smallest scales represented on the grid are well mixed. This second assumption is the most restrictive, and poses a particular challenge in temporally transitional flows; however, the first assumption has been addressed in the formulation of the problem. 

\subsection{Grid Convergence}

Figures \ref{convW}, \ref{convTKX} and \ref{convTheta} in Appendix \ref{gc} show the convergence of $W$, $\rm{TKX}$ and $\Theta$ for each of the seven codes that provided convergence data. 

\noindent For the standard problem,  all algorithms are grid converged with respect to integral width $W$ at the highest grid resolution. Triclade, TURMOIL and Flamenco are converged with  $<$2\% variation at $180\times 128^2$ resolution. Miranda, NUT3D, Ares and FLASH converge at $360\times 256^2$ grid resolution, however it should be noted that the Miranda and Ares simulations did not use sub-cell sampling to improve the representation of the initial condition on the mesh, which slowed the observed rate of convergence.

As a second measure of the large scales, the kinetic energy converges well for all algorithms, however most show a moderate increase in early time kinetic energy implying that the solutions are not converged in that region. There is potential for a further increase of up to $\approx 10$\% of the peak kinetic energy at $720\times 512^2$ grid resolution based on Richardson extrapolation of the current parameters. However, it should be noted that Triclade results at $1440\times 1024^2$ have $<$2\% variation from the $720\times 512^2$ case, lying between the  $720\times 512^2$ and $360\times 256^2$ resolutions. This indicates strongly that the kinetic energies are reasonably converged for the very high order algorithms at  $720\times 512^2$ resolution.

The mixing measures are challenging to resolve at early times, as the flow is undergoing temporal transition from a sum of linearly growing modes through to a self-similar layer. Thus at the early linear stages the measured mix parameters $\Theta$ and $\Xi$ are determined by the ability of the numerical scheme to resolve an extremely stretched contact surface. However, at later times, turbulence acts to increase the amount of mixing, such that the gradients are again resolved on the given mesh. The target of this problem was to determine the late-time self-similar mixed state of the specified narrowband initial condition, where all modes have saturated and an inertial range established. At the final time, each individual algorithm has converged to within 3\%.

For the quarter scale problem convergence is not as clear. For integral width, based on the standard problem (which is converged at $180\times 128^2$ resolution), Flamenco, FLASH, Triclade, TURMOIL and Miranda should have converged at the maximum grid resolution. Examining the finest two resolutions ($360\times 256^2$ and $720\times 512^2$), Flamenco and TURMOIL vary by $<$3\% for the highest two resolutions, however for the other algorithms this difference is larger. This is not possible to demonstrate conclusively without running a higher grid resolution, which is beyond the computational resources available.  

For kinetic energies, most algorithms are showing a worst case variation on the order of 10--20\% between the highest two grid resolutions at the peak post-shock passage value. For mix $\Theta$, each algorithm shows very good convergence, with the greatest difference between the finest two resolutions of 5\%.

\section{Standard Problem: Results and Discussion \label{standard}}

\subsection{Visualisations \label{vis}}

\begin{figure*}
\begin{centering}
\subfigure[\hspace{0.1cm} Flamenco]{\includegraphics[width=0.4\textwidth]{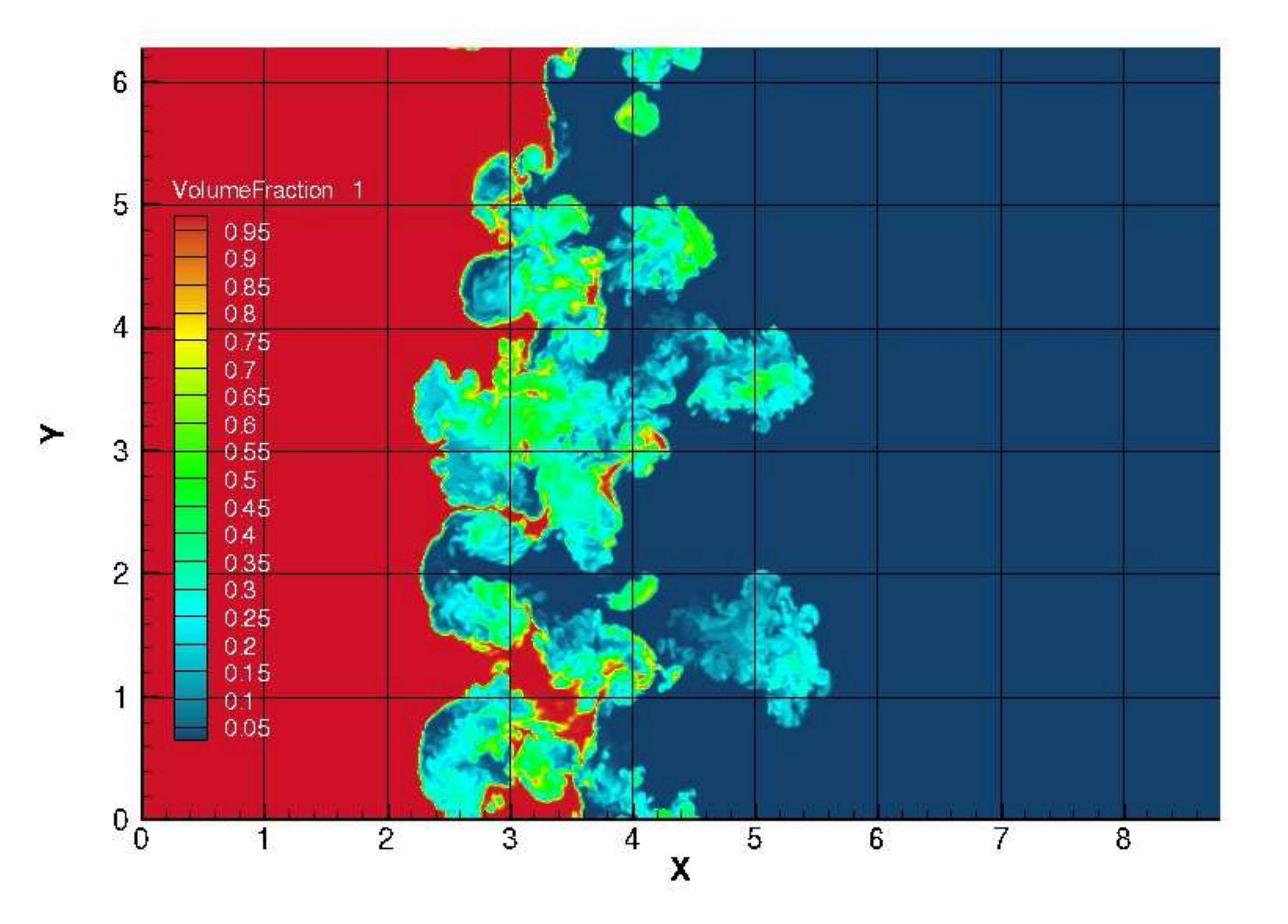}}
\subfigure[\hspace{0.1cm} NUT3D]{\includegraphics[width=0.4\textwidth]{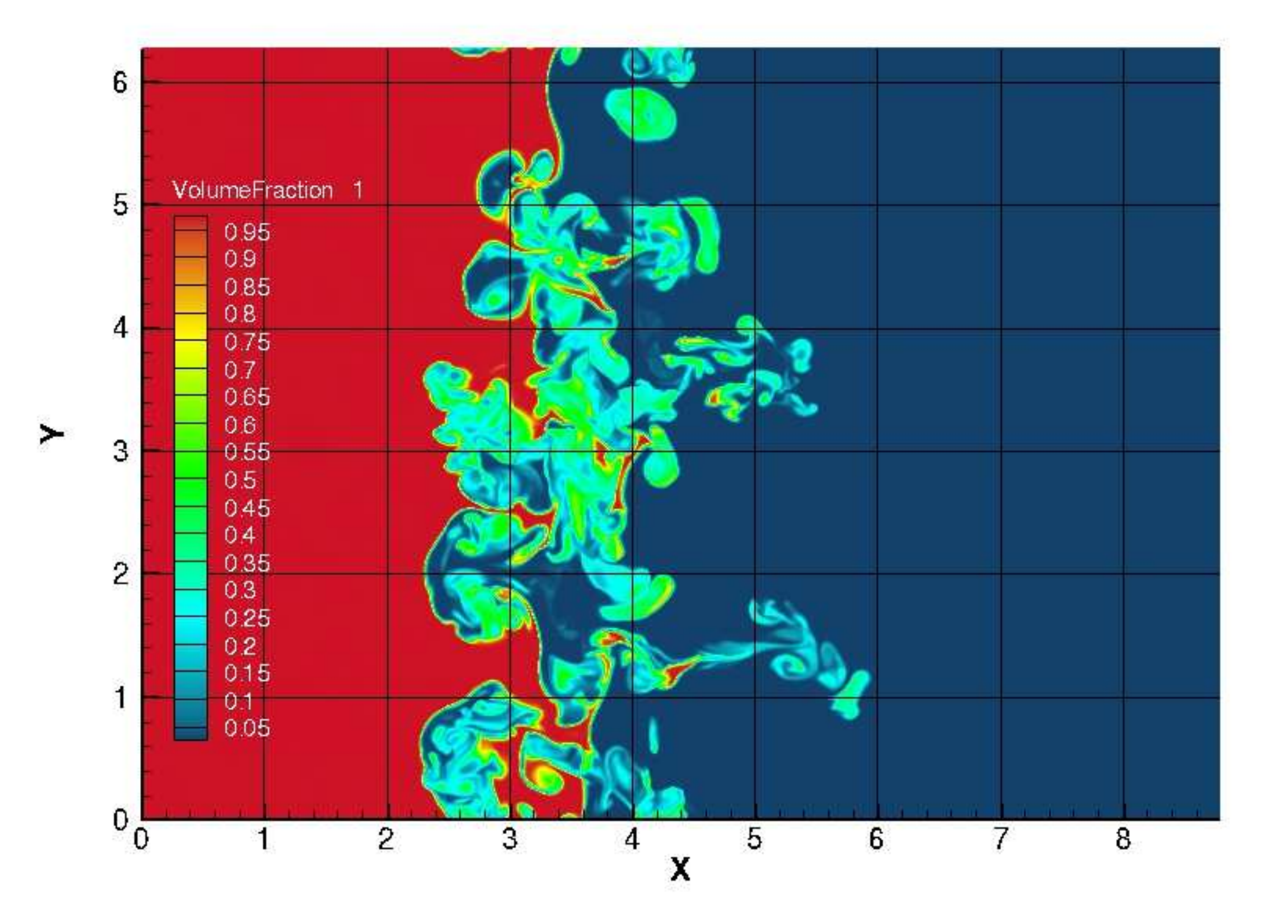}}
\subfigure[\hspace{0.1cm} Triclade]{\includegraphics[width=0.4\textwidth]{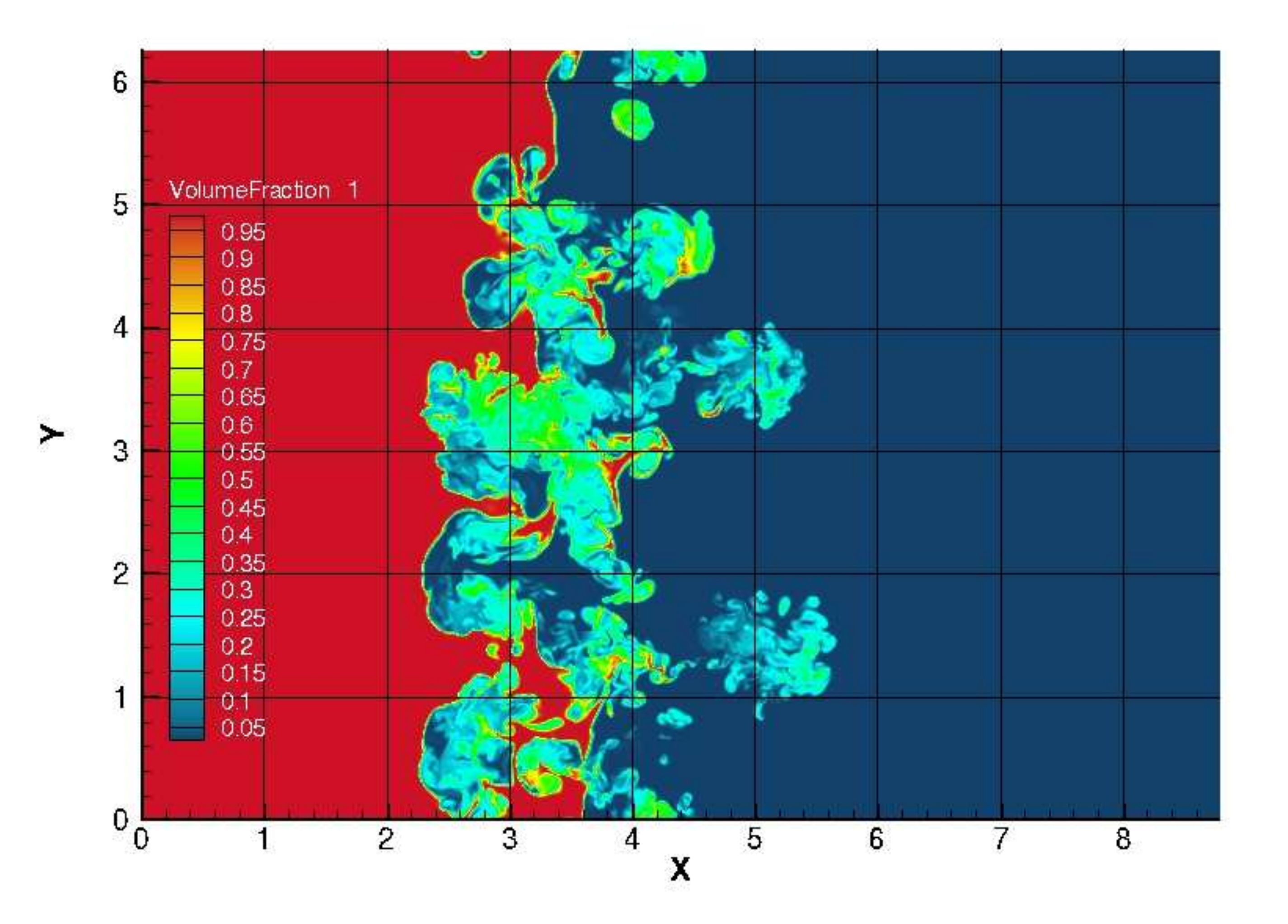}}
\subfigure[\hspace{0.1cm} TURMOIL]{\includegraphics[width=0.4\textwidth]{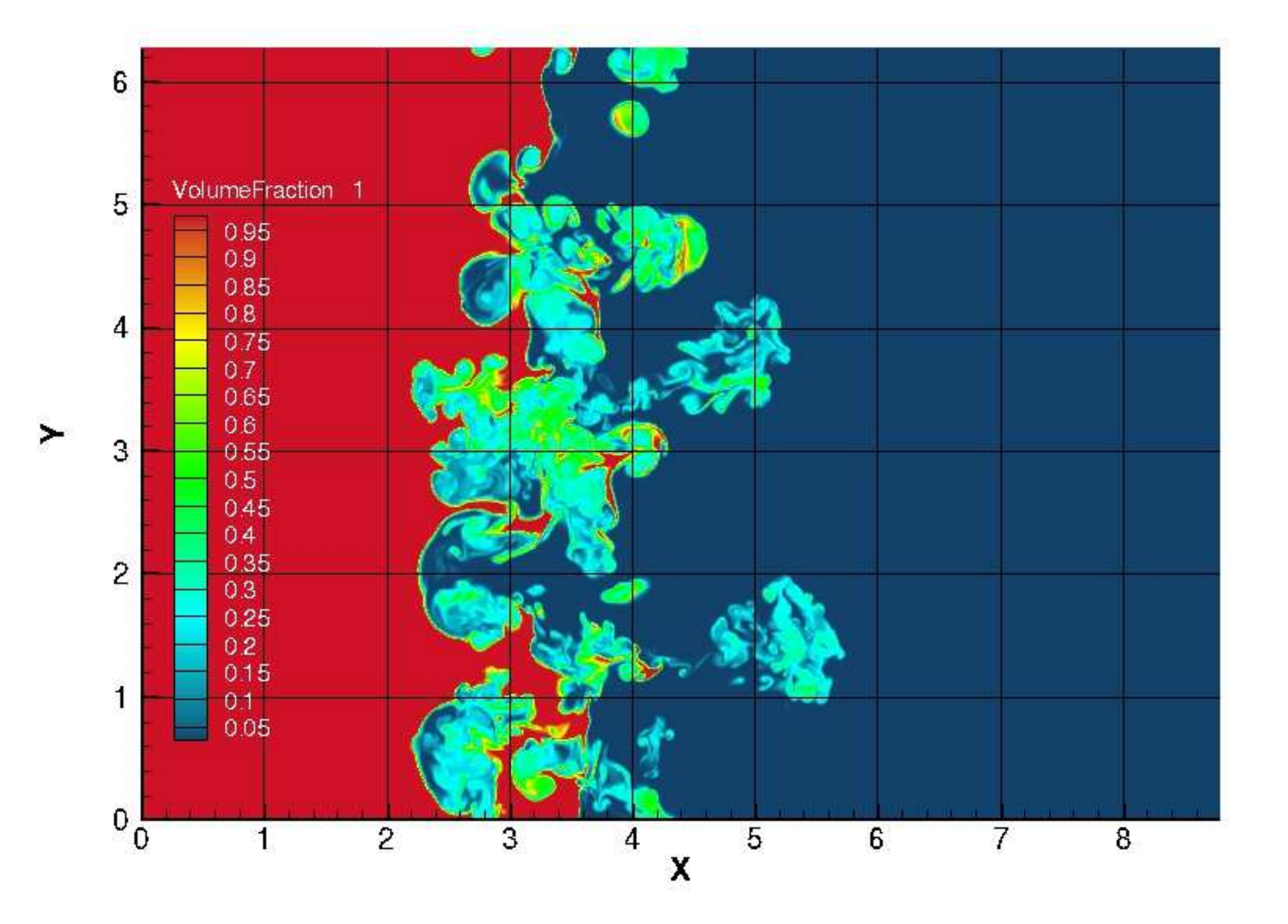}}
\subfigure[\hspace{0.1cm} Ares]{\includegraphics[width=0.4\textwidth]{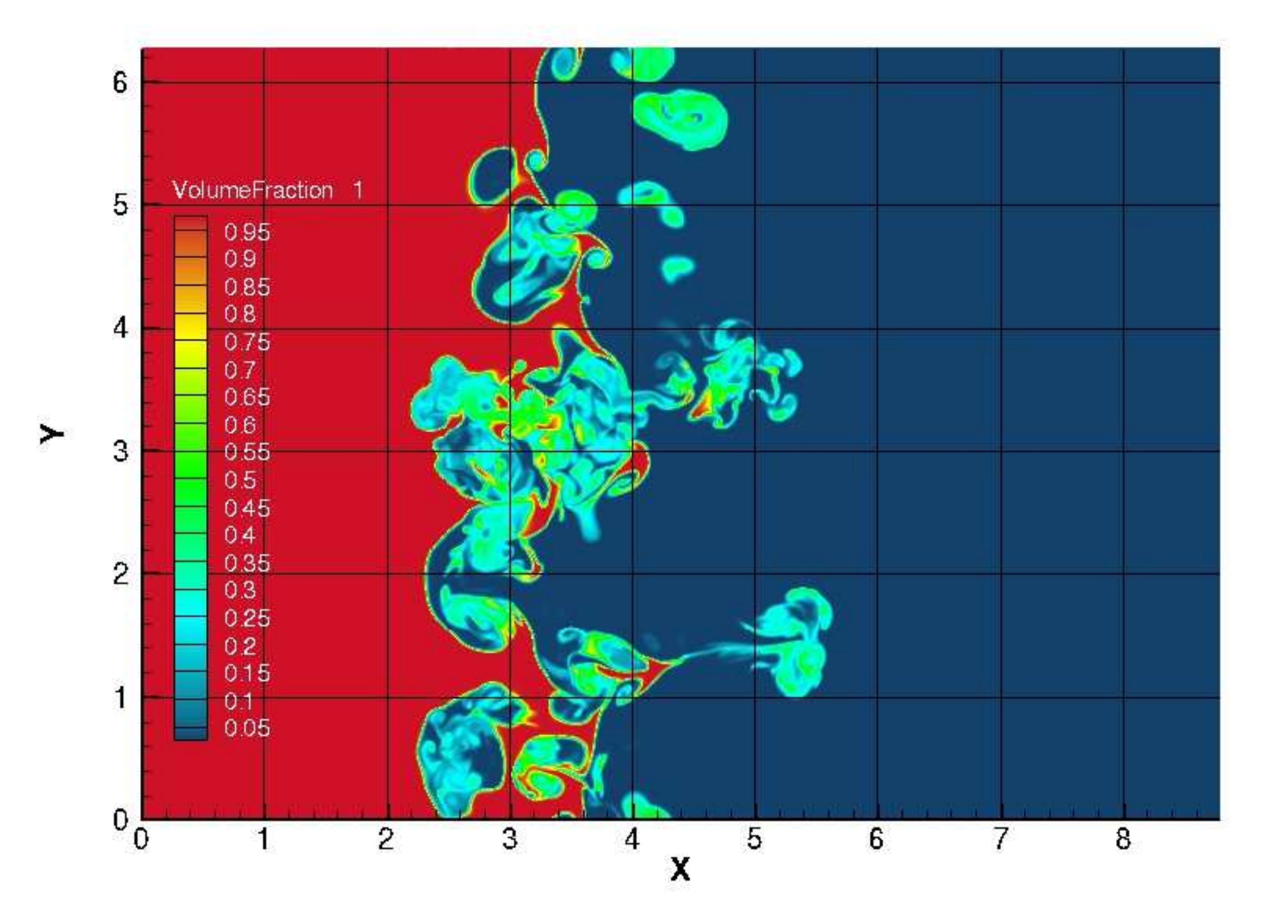}}
\subfigure[\hspace{0.1cm} Miranda]{\includegraphics[width=0.4\textwidth]{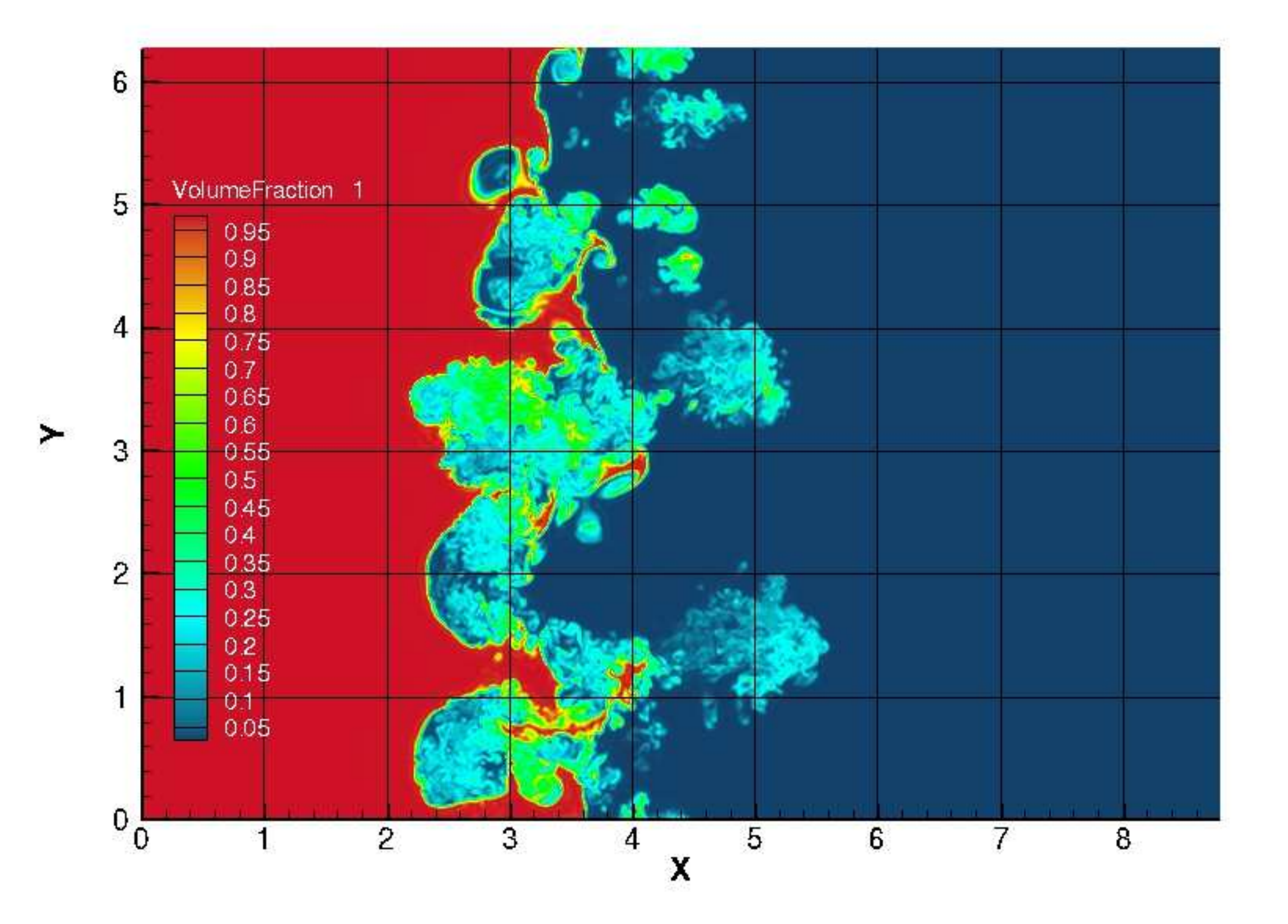}}
\subfigure[\hspace{0.1cm} FLASH]{\includegraphics[width=0.4\textwidth]{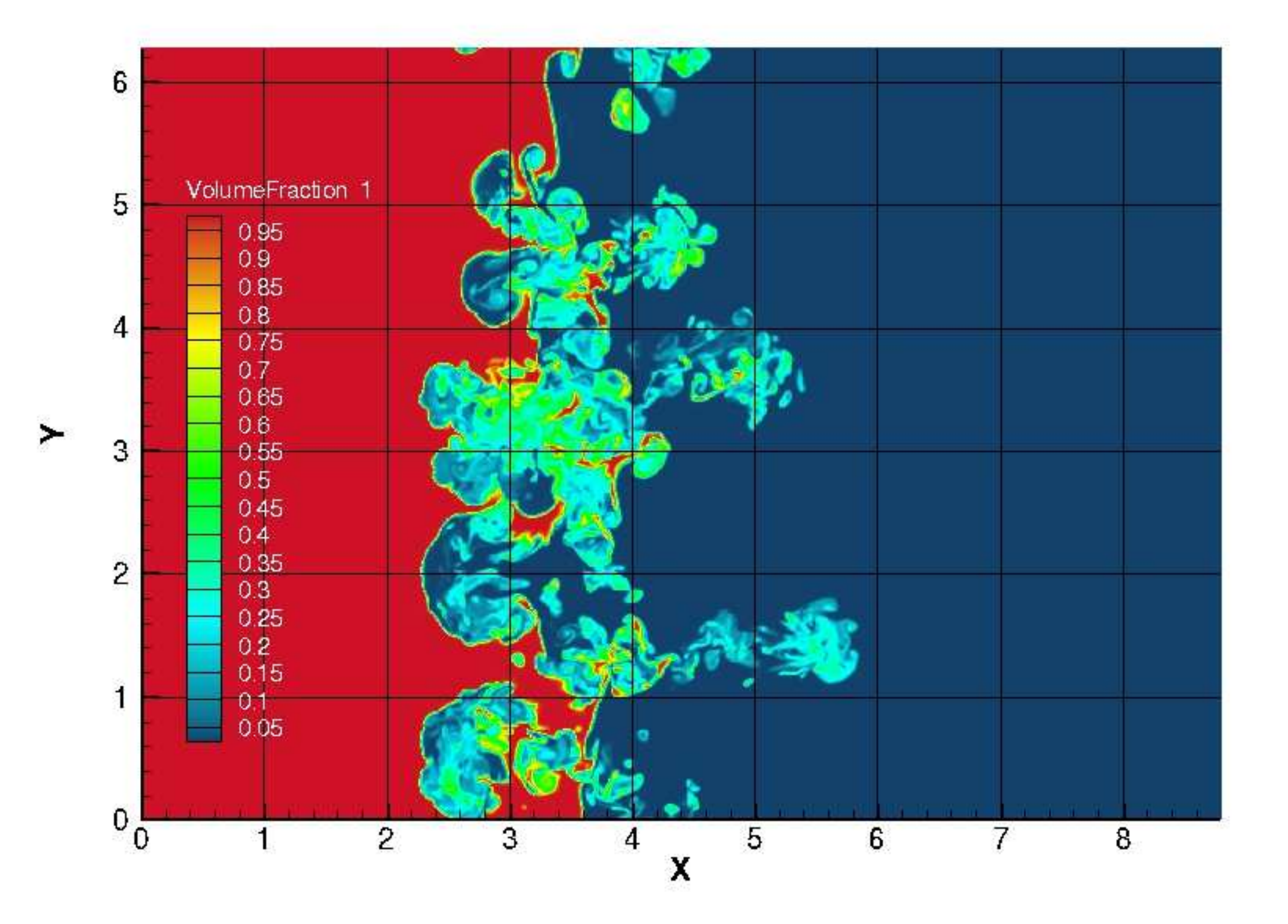}}
\subfigure[\hspace{0.1cm} Triclade $1440\times 1024^2$]{\includegraphics[width=0.4\textwidth]{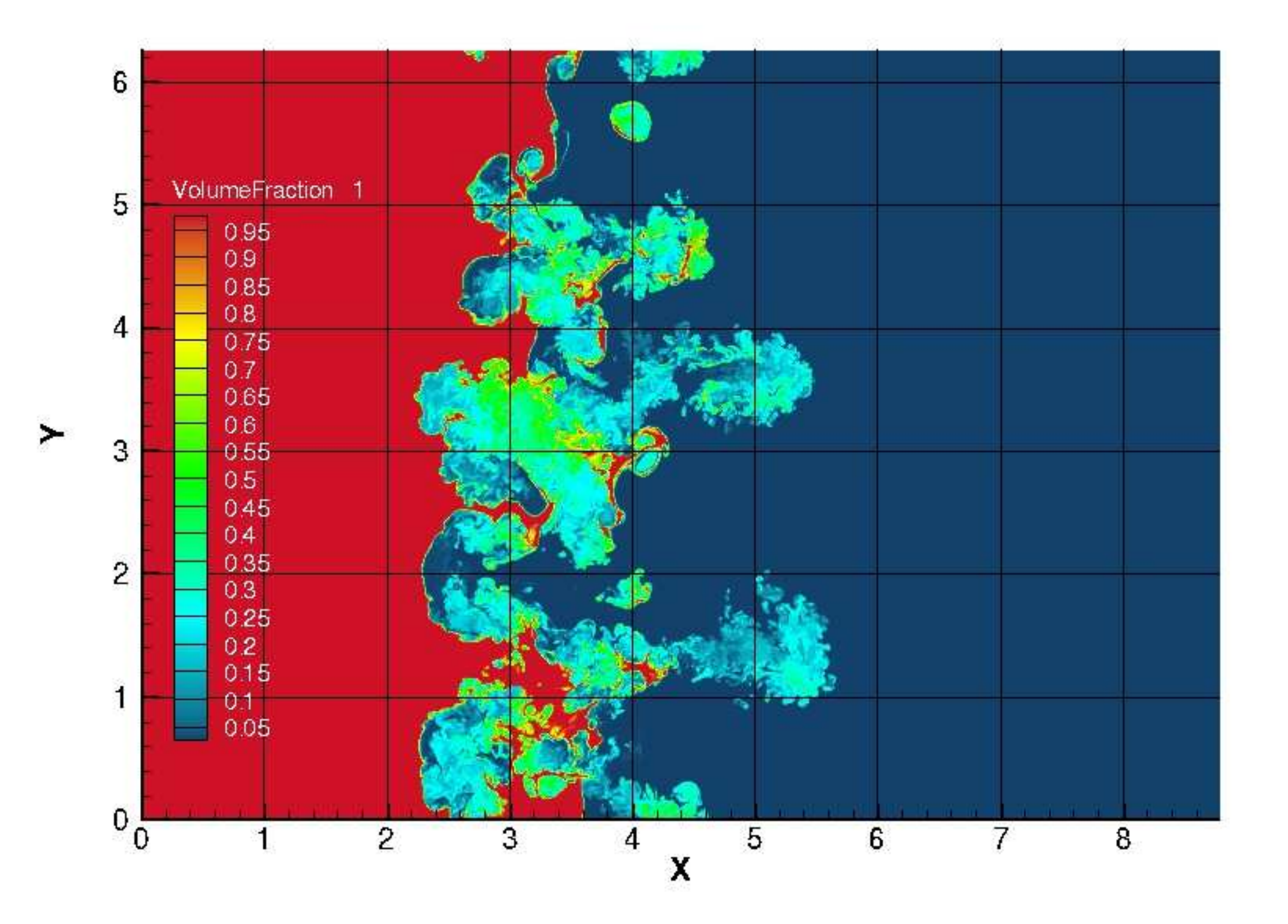}}
\caption{Visualisation at $\tau=1.23$ ($t=0.1$ s)  for the standard problem \label{vis01}}
\end{centering}
\end{figure*}

\begin{figure*}
\begin{centering}
\subfigure[\hspace{0.1cm}Flamenco]{\includegraphics[width=0.4\textwidth]{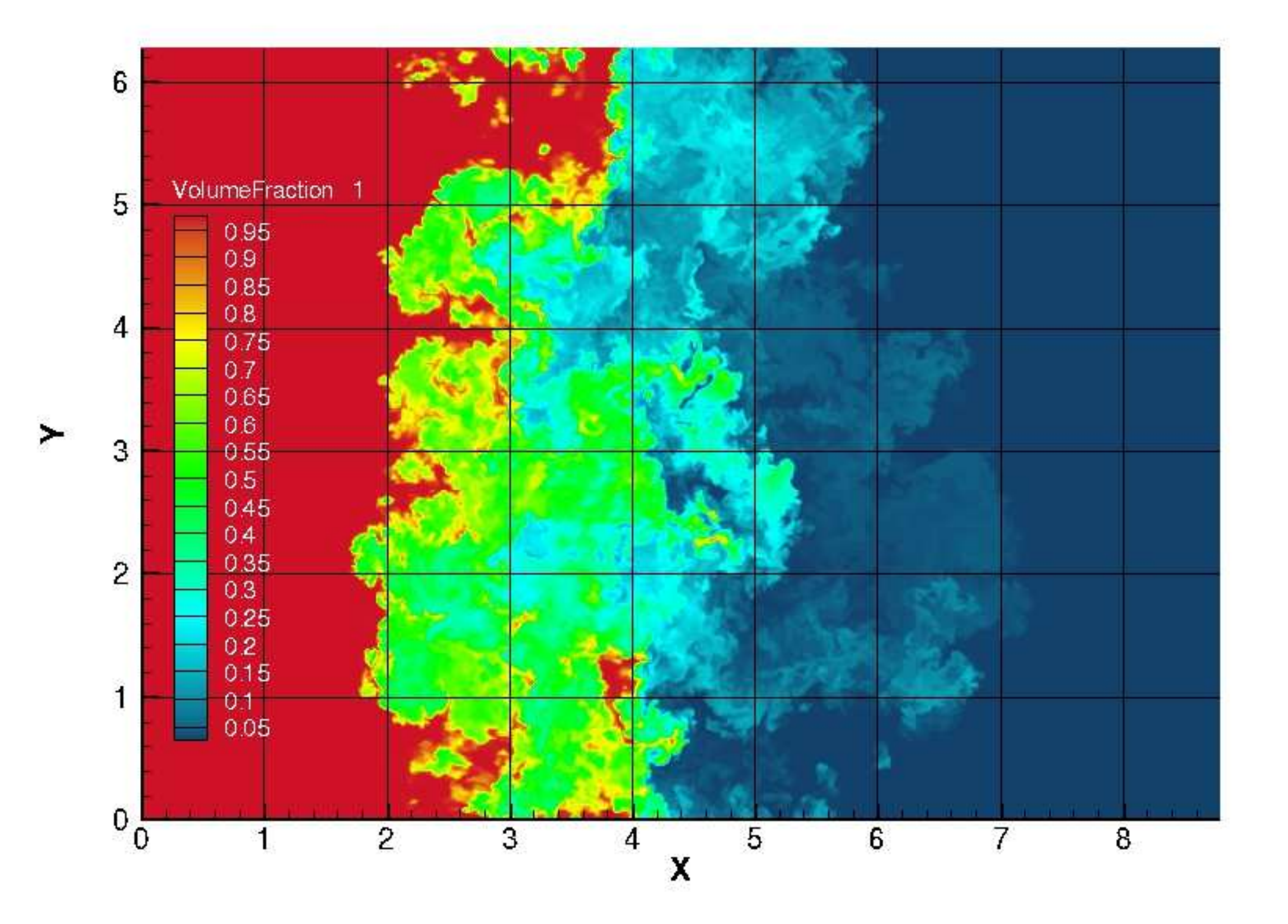}}
\subfigure[\hspace{0.1cm}NUT3D]{\includegraphics[width=0.4\textwidth]{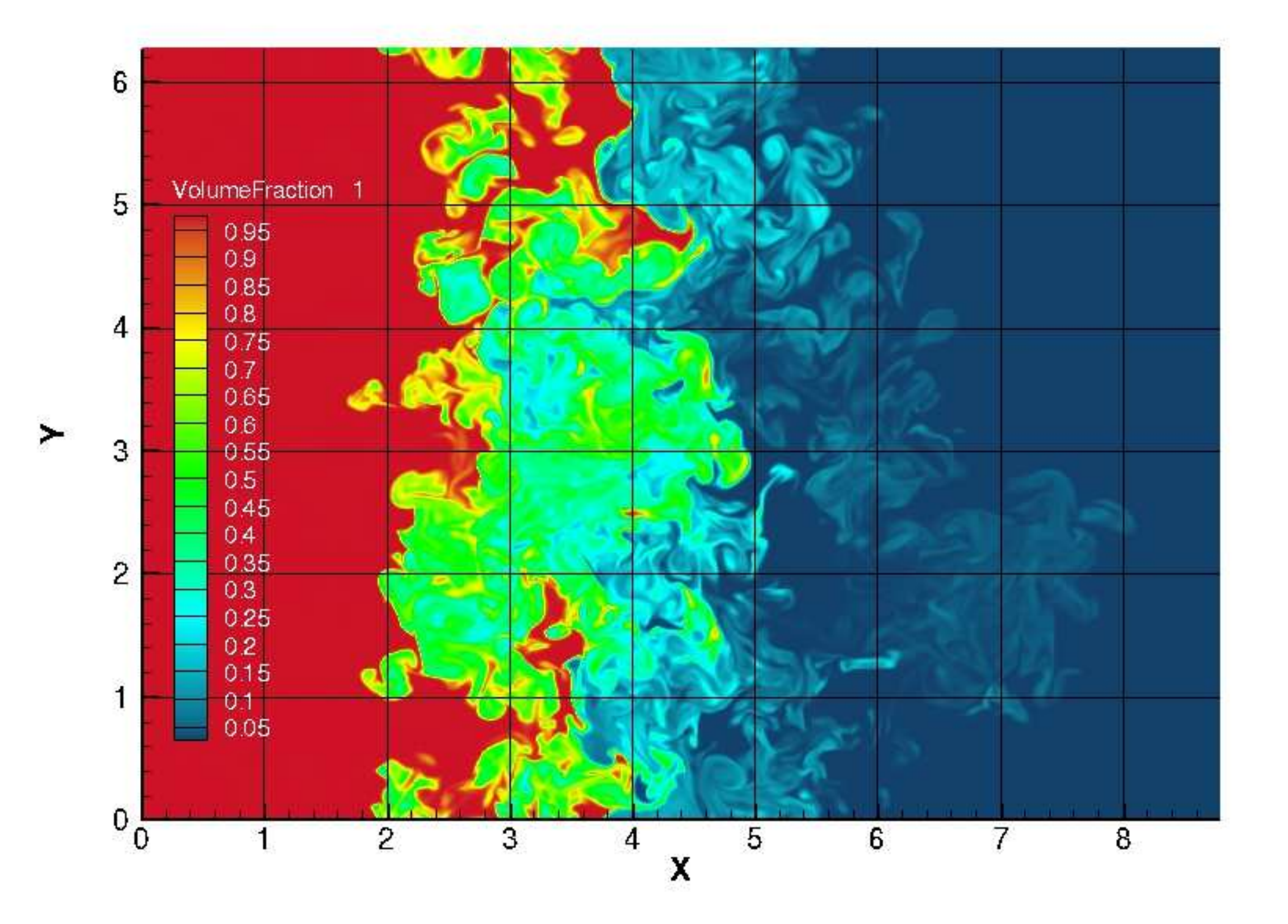}}
\subfigure[\hspace{0.1cm}Triclade]{\includegraphics[width=0.4\textwidth]{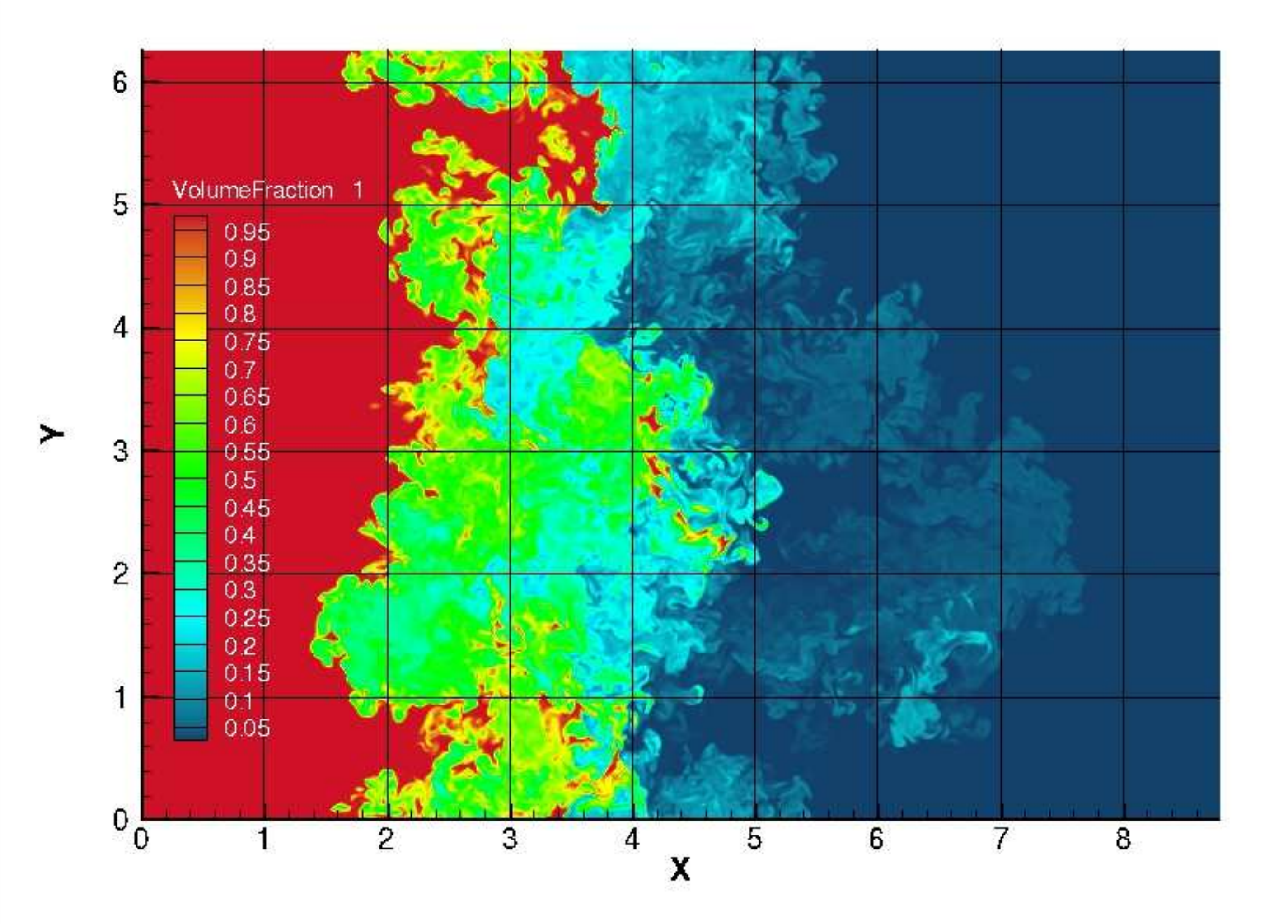}}
\subfigure[\hspace{0.1cm}TURMOIL]{\includegraphics[width=0.4\textwidth]{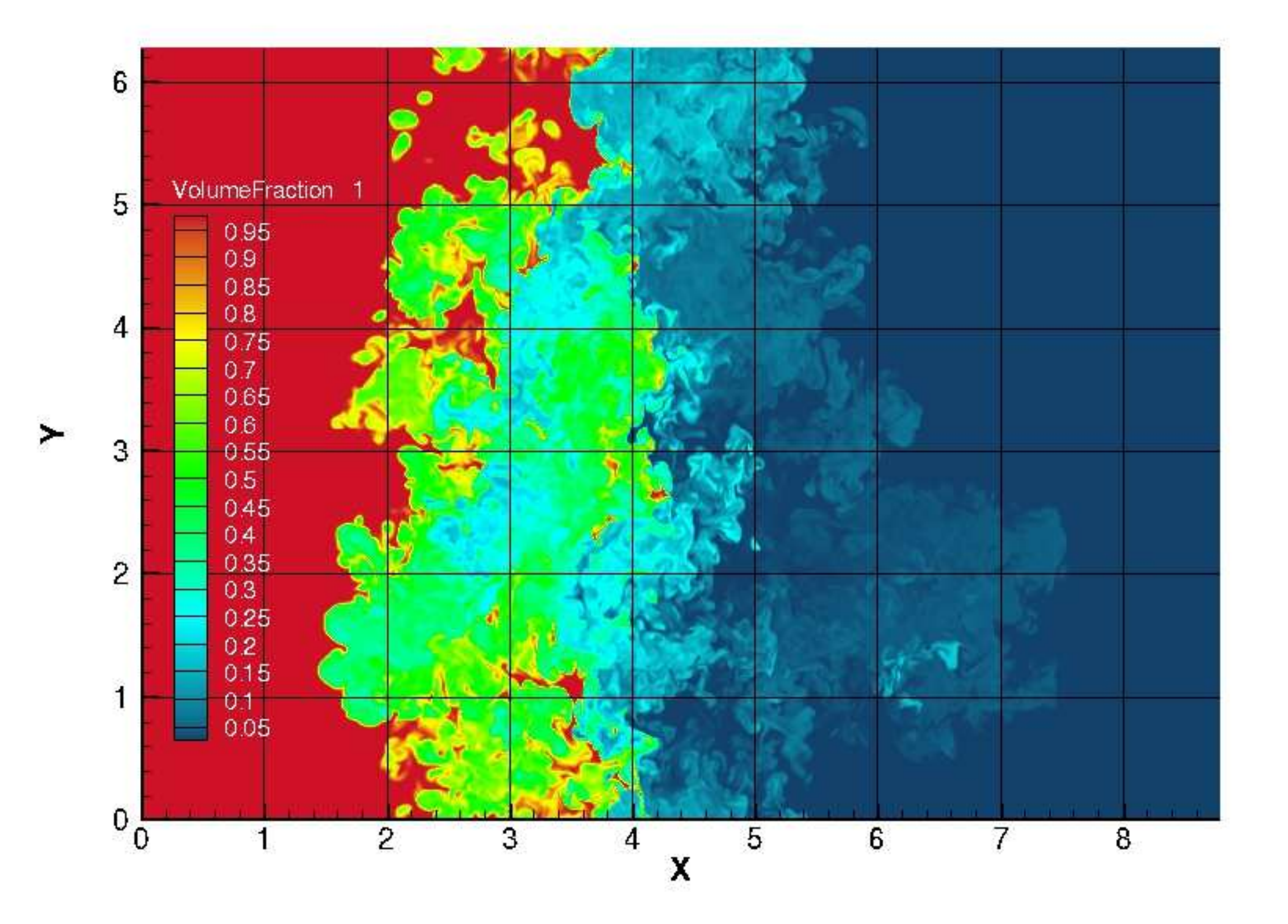}}
\subfigure[\hspace{0.1cm}Ares]{\includegraphics[width=0.4\textwidth]{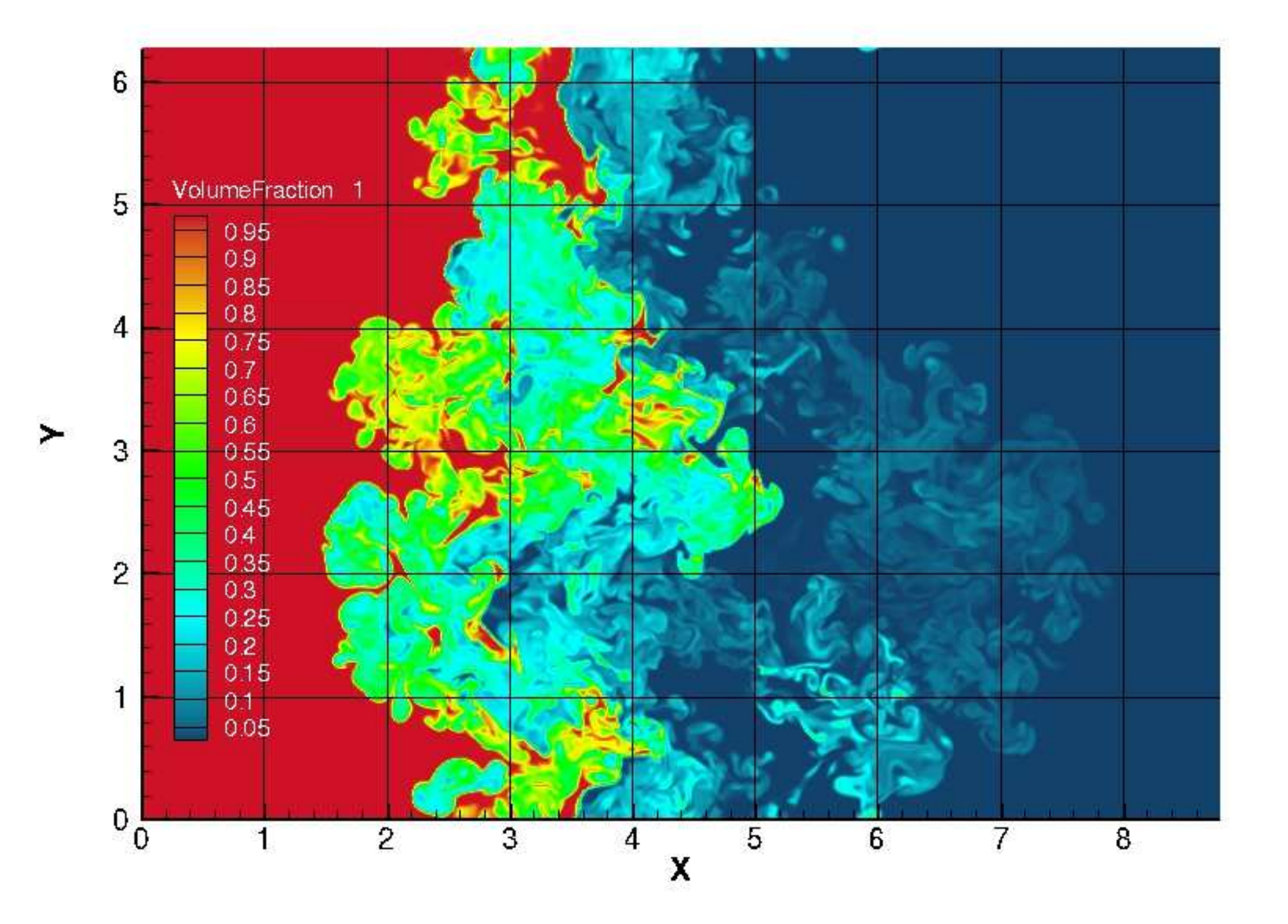}}
\subfigure[\hspace{0.1cm}Miranda]{\includegraphics[width=0.4\textwidth]{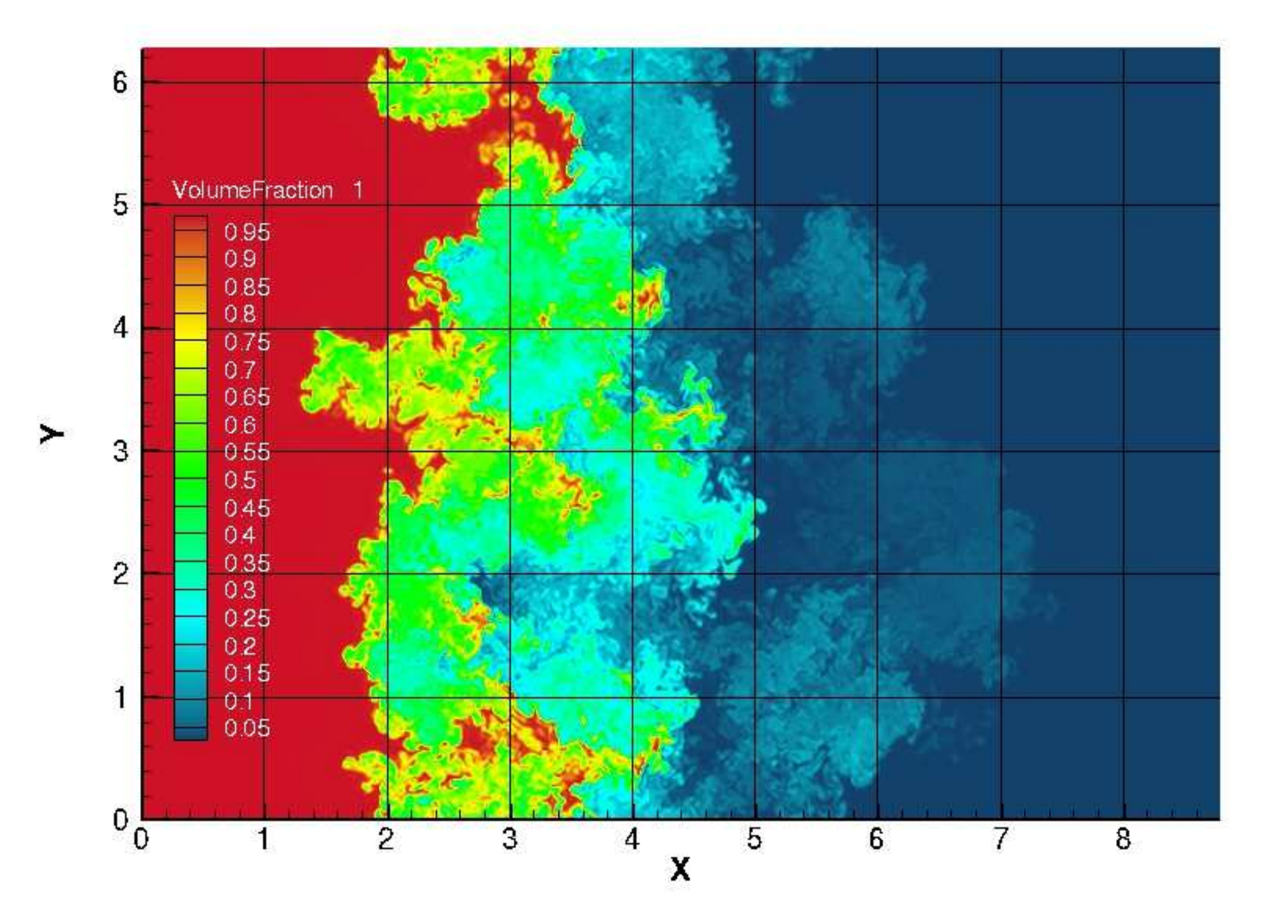}}
\subfigure[\hspace{0.1cm}FLASH]{\includegraphics[width=0.4\textwidth]{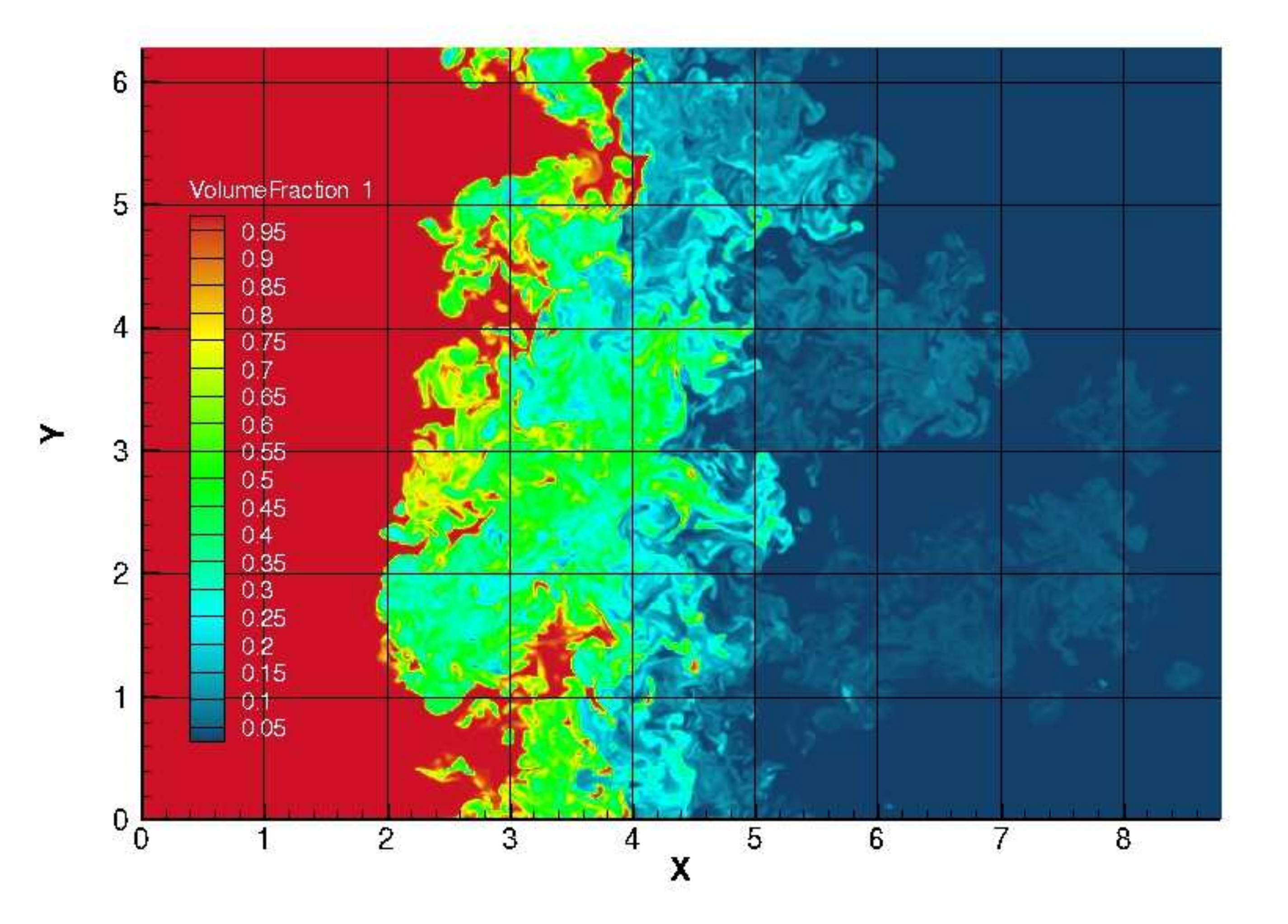}}
\subfigure[\hspace{0.1cm}Triclade $1440\times 1024^2$]{\includegraphics[width=0.4\textwidth]{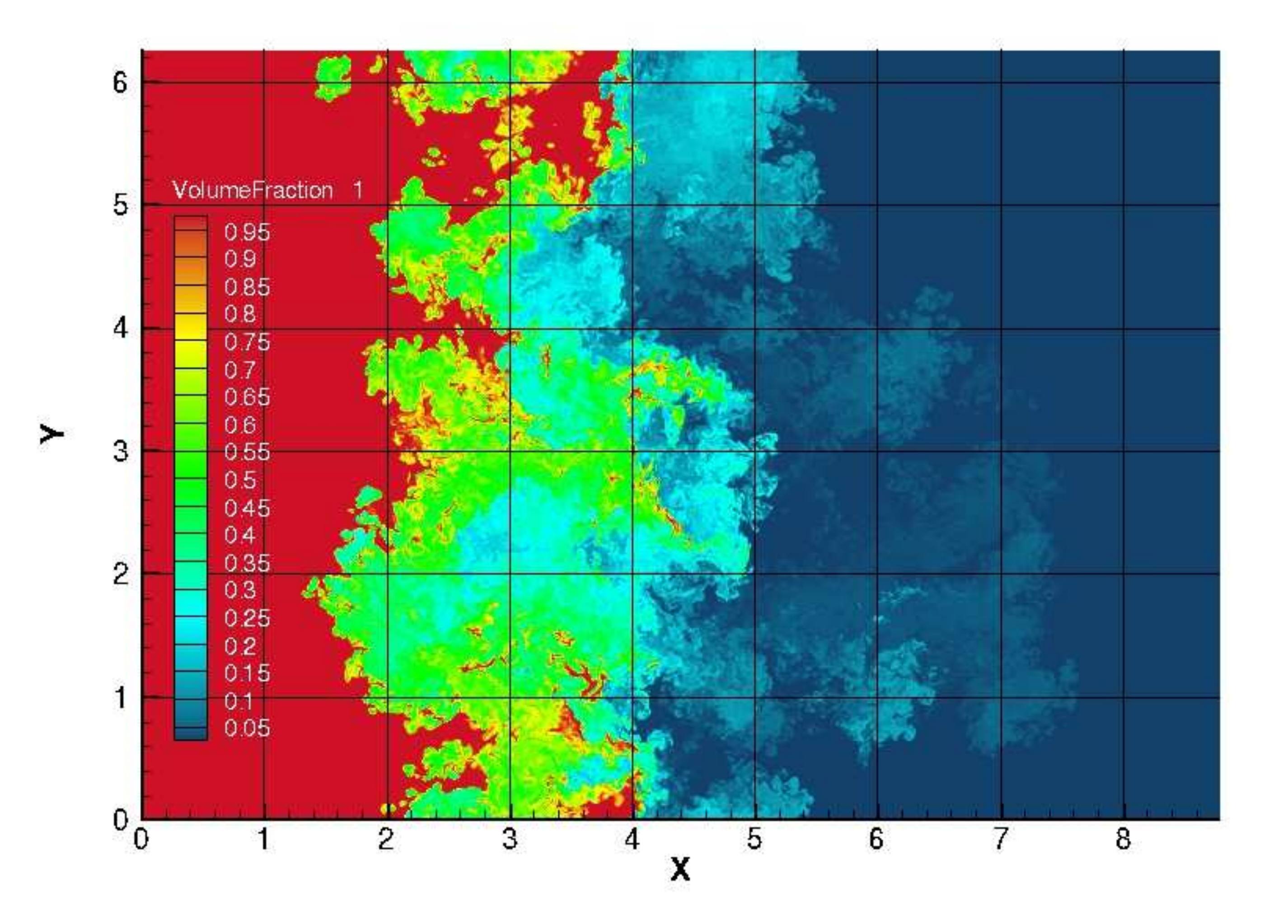}}
\caption{Visualisation at $\tau=6.15$ ($t=0.5$ s) for the standard problem. \label{vis05}}
\end{centering}
\end{figure*}

Visualisations of the standard test case at times $\tau=1.23$ ($t=0.1$ s) and $\tau=6.15$ ($t=0.5$ s) are shown in Figures \ref{vis01} and \ref{vis05} for each algorithm at $720 \times 512^2$, and an additional visualisation of the $1440\times 1024^2$ Triclade simulation. The visualisations at early time highlight the classical bubble and spike phenomenology associated with RMI at this Atwood number, showing narrow spikes penetrating into the lighter fluid and relatively wide bubbles. The large scale features are qualitatively similar for all algorithms; for example, the prominent spike which is located at $(x,y)\approx (5.5,1.3)$. However, there is a clear difference in resolution of the small scales, where the higher order algorithms permit the growth of finer scale fluctuations, as is again highlighted clearly at the aforementioned spike head. At the early time there are still coherent contact surfaces visible, linking the bubble and spikes in some regions of the flow, e.g. $(x,y)\approx (3.5,5.5)$.

The later time visualisations provide an indication of the wider range of length-scales present, implying that the flow is transitioning towards being a fully turbulent mixing layer with a smoother, well mixed, profile. The overall mixing layer width is qualitatively similar between the algorithms, and the positioning of the key bubbles and spikes. However, there are differences in the fine scale features as highlighted previously, which at later times have caused small differences in the position of these larger features. The computations of the standard problem were terminated at $\tau=6.15$ ($t=0.5$ s) as it is clear that the dominant wavelengths are approaching the domain size, thus impacting the statistical accuracy of the quantities of interest. 

\subsection{Mixing Layer Width and Mix Measures}

\begin{figure}
\begin{centering}
\includegraphics[width=0.49\textwidth]{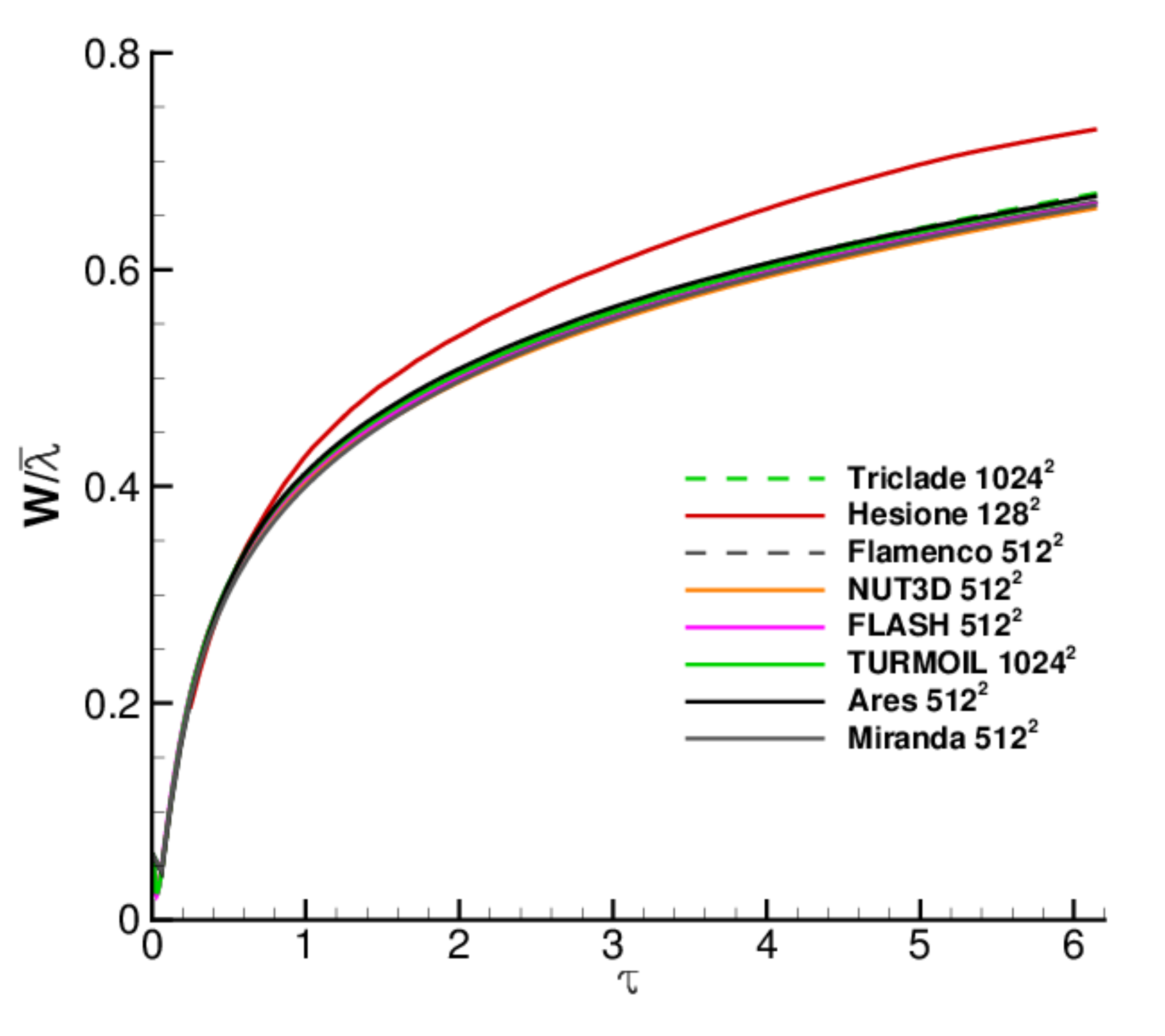}
\includegraphics[width=0.49\textwidth]{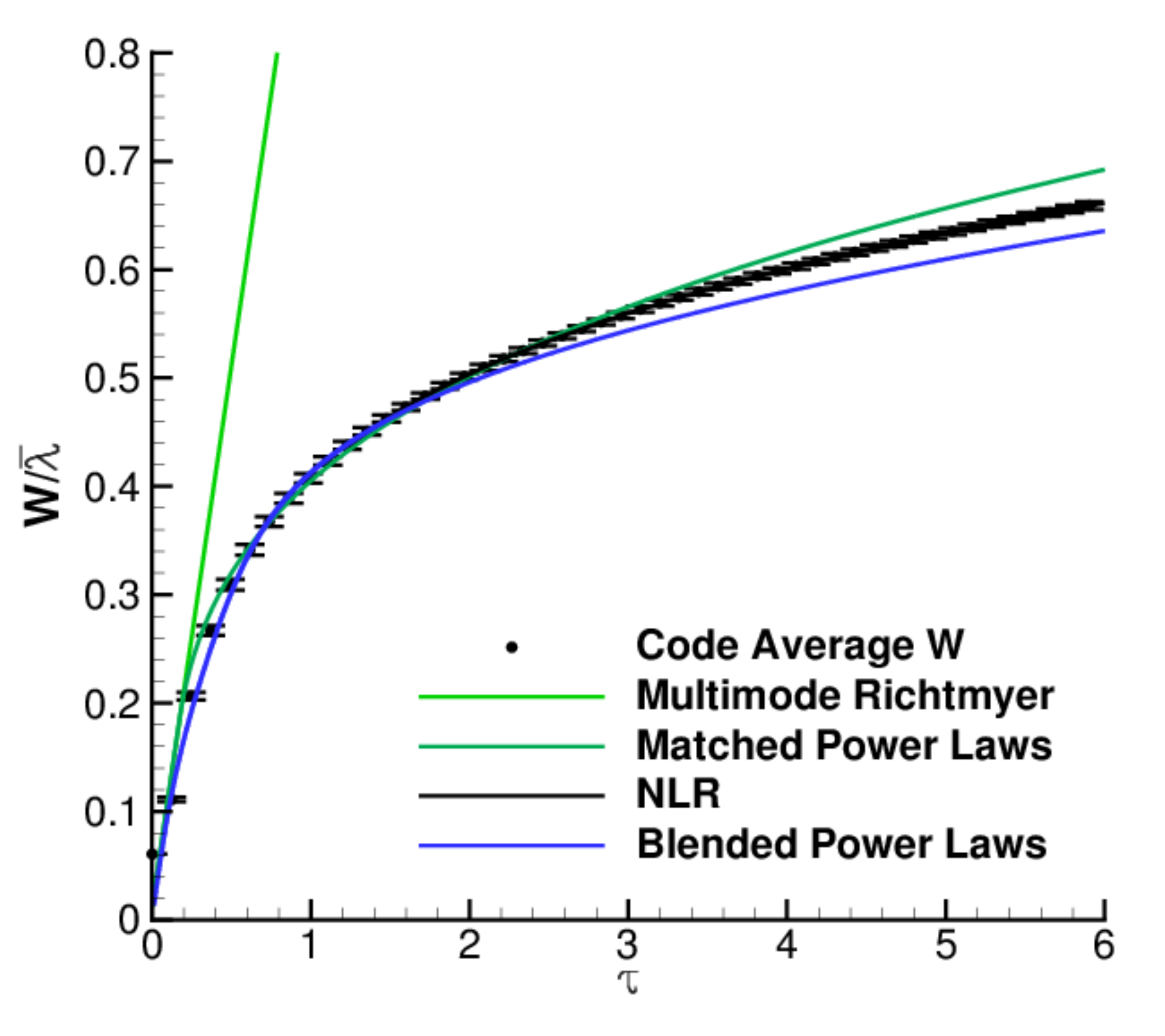}
\caption{Integral width $W$ for all algorithms at the highest grid resolution, and the code-averaged $W$ compared to the new linear model (Eqn. (\ref{wdotpred}), the matched power law (Eqn. (\ref{mpl})), the blended power law (Eqn. (\ref{bpl})) and the non-linear regression data fit for the standard problem (solid line passing through the symbols, starting at $\tau=1.23$.)\label{StandardW}}
\end{centering}
\end{figure}

\begin{figure}
\begin{centering}
\includegraphics[width=0.49\textwidth]{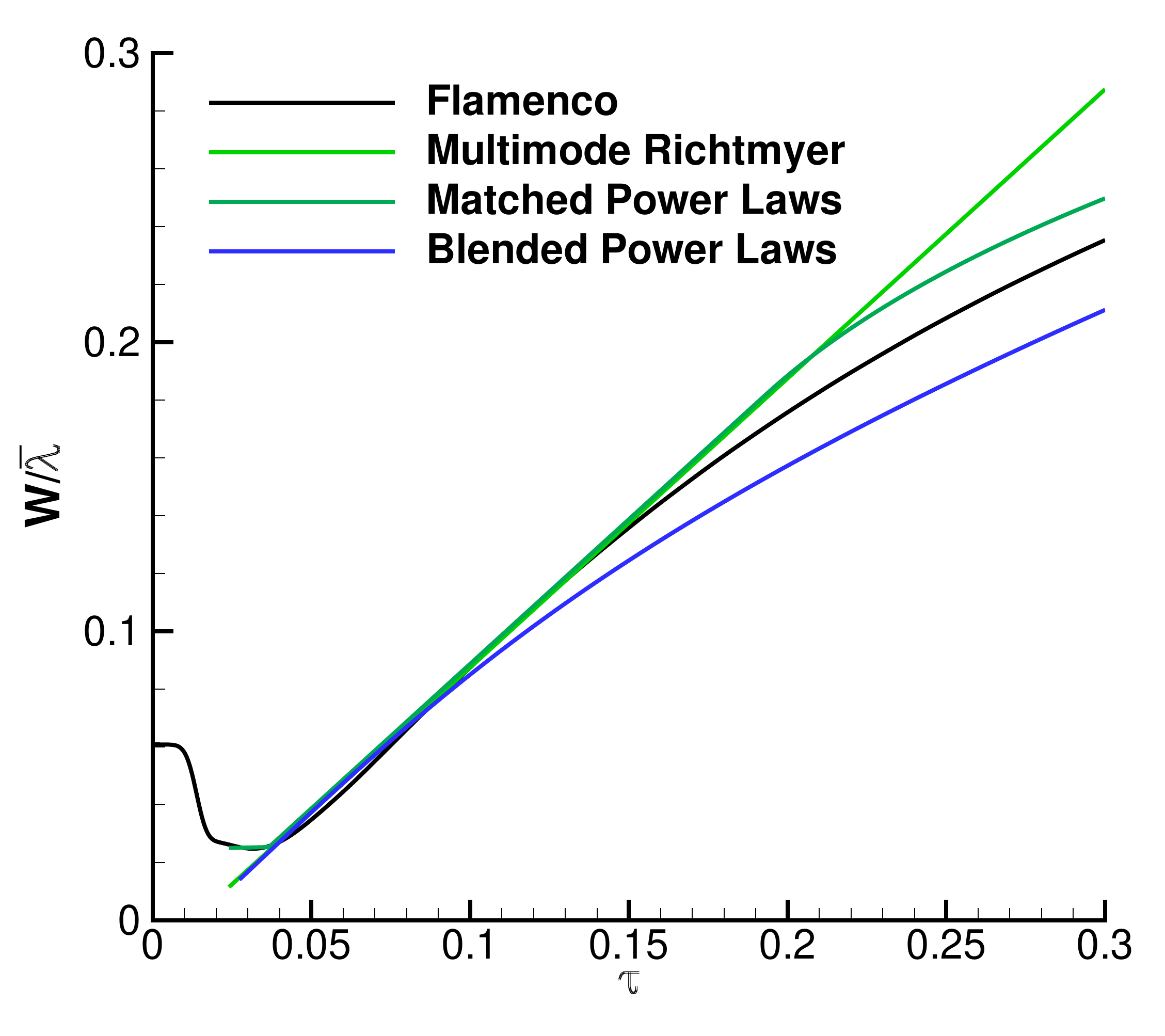}
\caption{The predicted integral width from the new linear model [Eqn. (\ref{wdotpred})]  and the two power law fits [Eqs. (\ref{mpl}) and (\ref{bpl})] compared to Flamenco data.\label{StandardCloseW}}
\end{centering}
\end{figure}

\begin{table}
\centering
\caption{Mixing layer width and mix measures as a function of time for the code-averaged data  for the standard problem.\label{dataresults2}}
\begin{tabular}{cccccccc}
$t$ (s) & $\tau$ & $W$ & $\sigma_{W}$ & $\Theta$ & $\sigma_{\Theta}$  & $\Xi$ & $\sigma_{\Xi}$ \tabularnewline
\hline
\hline
              0.05& 0.62 & 0.351& 0.0049&  0.253& 0.0334&  0.247& 0.0330\tabularnewline
              0.10& 1.23 & 0.450& 0.0039&  0.425& 0.0404&  0.403& 0.0375\tabularnewline
              0.15& 1.85 & 0.507& 0.0029&  0.541& 0.0435&  0.512& 0.0415\tabularnewline
              0.20& 2.46 & 0.547& 0.0028&  0.616& 0.0426&  0.589& 0.0424\tabularnewline
              0.25& 3.08 & 0.579& 0.0028&  0.667& 0.0395&  0.644& 0.0408\tabularnewline
              0.30& 3.69 & 0.606& 0.0028&  0.702& 0.0357&  0.683& 0.0379\tabularnewline
              0.35& 4.31 & 0.629& 0.0029&  0.726& 0.0317&  0.712& 0.0341\tabularnewline
              0.40& 4.92 & 0.649& 0.0031&  0.745& 0.0280&  0.735& 0.0300\tabularnewline
              0.45& 5.54 & 0.668& 0.0035&  0.759& 0.0249&  0.753& 0.0261\tabularnewline
              0.50& 6.15 & 0.684& 0.0040&  0.769& 0.0225&  0.767& 0.0229\tabularnewline
\end{tabular}
\end{table}

Code-averages have been taken of the seven non-interface reconstructing algorithms. Figure \ref{StandardW} plots the integral width for all algorithms, the code-average integral width and the standard deviation plotted as error bars. The code-averaged results are also compared to three theoretical models, that developed in Section \ref{mmricht} for the linear phase, a second inspired by the approach of Mikaelian \cite{Mikaelian2015}, and a third which utilises a straight error function blending of the linear and non-linear stages. 

Mikaelian's model combines growth of the width which is linear in time at early time, matched to a nonlinear growth at late time. This approach is designed to match a post-shock fully turbulent layer width which is linear in time initially with a late time decay through a power law. Here the model is adapted such that the linear phase, given by $W=W^+_0+4\alpha At^+ \Delta Vt$, is calibrated to match the initial linear growth $\dot W_0$ defined in Section \ref{mmricht} by an appropriate choice of $\alpha$. The linear phase is then joined to an assumed power law behaviour at the temporal location where the growth rates predicted from both models would be equal, giving,

\begin{equation}
W=W^*\left(1+\frac{\dot{W}^*}{\theta W^*}(t-t^*)\right)^\theta,
\label{mpl}
\end{equation}

\noindent where the layer width $W=W^b+W^s$,  $\theta=\theta^b,\theta^s$, $\alpha=\alpha^s,\alpha^b$, $W^*$ is the width at the transition between the two power laws, and $W^+_0$ is the initial layer width. The superscripts $(.)^s$ and $(.)^b$ denote the spike and bubble values which may be allowed to vary. The current curve uses the following parameters recommended by Mikaelian, except for $\beta$ which is chosen as the dimensionless time where $W$ varies by more than 5\% from that predicted by the theory in Section \ref{mmricht}, and $\alpha^{b,s}$ which have been calibrated to match the new linear theory as follows:

\begin{equation}
\alpha^b=\alpha^s=\frac{0.564 \sqrt{\frac{7}{12}}\, k_{\rm max}\, \sigma^+ }{4}=0.022, \,\,\, \theta^b=0.25, \,\,\, \theta^s=0.31
\end{equation}
\begin{equation}
\beta=32, \,\,\,At^+=0.487, \,\,\,U_i=434\,{\rm m/s}, \,\,\,t^*=3.41\times 10^{-3}\,{\rm s}.
\label{kmdets}
\end{equation}

\noindent Both $\alpha^{b,s}$ have been set to the same value but could be set to different values as long as the sum remains unchanged.  In this way the coefficients use only quantities which can be estimated `a priori'.  This curve is labelled `Matched Power Laws' in Figure \ref{StandardW}.

The third curve labelled `Blended Power Laws' writes the dimensionless time $\tau$ as a function of dimensionless integral width $\tilde{W}$ as:

\begin{equation}
\tau=\left(\tilde{W}/\tilde{A}\right)^{1/\theta}+\tau_0 \mathcal{Z}+\tau_{\mbox {shock time}}+\left(\tilde{W}-\tilde{W_0^+}\right)\left(1-\mathcal{Z}\right)
\label{bpl}
\end{equation}

\begin{equation}
\mathcal{Z}=\frac{1}{2} \left\{1+\mbox{erf}\left[\mathcal{C}\left(\tilde{W}-\tilde{W_0^+}-\mathcal{B}\right)\right]\right\},
\end{equation}

\noindent where all $\tilde{(.)}$ are dimensionless quantities, and $ \tilde{W_0^+}$ is the dimensionless post-shock integral width from theory. Here $\mathcal{Z}$ acts to blend in the zeroth order term $\tau_0$ in the data fit for the late time nonlinear behaviour, in the form $\tilde{W}=\tilde{A}\left(\tau-\tau_0\right)^\theta$, with a simple linear initial growth $\tilde{W}=\tau$. The dimensionless parameters in the non-linear fit are defined as 
\begin{eqnarray}
\tilde{A}=\frac{A}{\bar{\lambda}^{1-\theta}\,\dot W_0^{\theta}}=0.3438,\,\,\, \tau_0=-2.075,\,\,\, \theta=0.296,\,\,\,\tau_{\mbox {shock time}}=0.0011\frac{ \dot W_0}{\bar \lambda}+0.015.
\end{eqnarray}

\noindent Note that $\tau_{\mbox {shock time}}$ takes into account the time taken for the shock to reach the interface ($0.0011$ s), and a delay in initial growth due to the inversion process ($\Delta \tau \approx 0.015$). These have been determined from the quarter scale problem, as will be discussed in Section \ref{quarter}. $\mathcal{B}=0.395$ and $\mathcal{C}=5.8$ are dimensionless free parameters which are set `by eye' as non-linear regression in Mathematica \textregistered$\,$ did not perform well in optimising these coefficients. Assuming that the dimensionless parameters $\mathcal{B}$ and $\mathcal{C}$, and exponent $\theta$ are the same for all narrowband cases then all other quantities can be evaluated \`{a} priori. 

Examining  Figure \ref{StandardW} and the close up shown in Figure \ref{StandardCloseW}, the new theory predicting the linear growth rate of a multimode narrowband perturbation shows excellent agreement with the code-averaged data, and the Flamenco data which is plotted as the sampling rate is much higher than the code-averaged data in the linear phase. Regarding the `blended' and `matched' power laws, both of the models show very good agreement within their regions of applicability and underlying assumptions. The `Matched' power laws overestimate late time behaviour as the $\theta$ prescribed is higher than that computed from the simulations (and no attempt was made to calibrate the $\theta$). The prediction from the `Blended Power Laws' is better, however does not match as well as the non-linear regression since this formula is calibrated to match the very late time behaviour simulated in the subsequent quarter scale case. For future modelling and code validation, the dimensional data is tabulated in Table \ref{dataresults2}.

There is excellent agreement between all algorithms, which is expected given that this is a measure of the large scales and should be relatively algorithm-independent. The standard deviation of the code-averaged integral width is $\sigma_W \le 0.7$\%. As expected, the interface reconstruction code Hesione differs from the diffuse-interface algorithms as the interface reconstruction method suppresses the dissipation of density fluctuations. The predicted integral width for Hesione is 3.7\% higher at the latest time when compared to Ares at the same grid resolution, which uses a similar method but without IR.

\begin{table}
\caption{$\theta$ for each algorithm, along with Richardson extrapolated values of $\Theta$ and $\Xi$ at $\tau=6.15$ for the standard problem. * indicates that the measure is not uniformly converging, thus the Richardson extrapolated value is replaced by the value at the highest grid resolution. Note that Richardson extrapolation is not necessarily robust, for example the extrapolated values for Flamenco and Ares would not be considered to be a realistic bound. \label{thetas}}
\begin{tabular}{ccccccc}
Code &  $\theta (\tau>1.23)$& $\Theta_{R}$ & $\Xi_R$ \tabularnewline
\hline
\hline
Ares & 0.215 &  0.7827 & 0.9291\tabularnewline
Flamenco & 0.220 & 0.8717 & 0.6872\tabularnewline
Flash & 0.221 & 0.8446 & 0.8027\tabularnewline
Hesione &0.242 & 0.2740*  & 0.2708*\tabularnewline
Miranda & 0.228 & 0.8010* &  0.7922*\tabularnewline
NUT3D & 0.210 & 0.7677 & 0.7655\tabularnewline
Triclade & 0.218 & 0.7685 & 0.7842\tabularnewline
TURMOIL & 0.222 & 0.8312 & 0.7867*\tabularnewline
\end{tabular}
\end{table}

Two approaches have been employed to compute the simulated value of $\theta$, the first is to employ non-linear regression to fit the growth curve for $\tau>1.23$ ($t>0.1$ s) to a form $W=A(t-t_0)^\theta$ and the second is to utilise the derivatives of the integral width to directly determine  $\theta^{-1}=(1-\ddot{W} W/\dot{W}^2)$. This power law behavior of the mixing zone width ($W\propto t^\theta$) is a consequence of the interaction of large scale structures in the mixing zone (low-$k$ part of the spectrum) as proved by Poujade and Peybernes \cite{poujpeyb}. A Rayleigh--Taylor or Richtmyer--Meshkov mixing layer width will grow in the self-similar regime according to 
\begin{align}
\ddot{W}=C_b\,{\cal A}\,g-C_d\,\frac{\dot{W}^2}{W}
\end{align} which yields the power law solution for RMI (when $g=0$) with $\theta^{-1}= 1+C_d$. 

Using non-linear regression on the code-averaged integral width for  $\tau >1.23$ assuming $W=A(t-t_0)^{\theta}$ gives 

\begin{equation}
W=0.807(t-0.0309)^{0.219}, 
\end{equation}

\noindent in line with previous results for early time growth (not yet self similar) of the mixing layer \cite{Thornber2010,Lombardini2012, Youngs2004}. This is plotted in Figure \ref{StandardW} as the solid black line which passes through all data points. Note that $\theta$ increases at later time, for example at $t>0.4$, $\theta=0.228$. Table \ref{thetas} tabulates the values of $\theta$ for each individual code, illustrating that the range $0.210<\theta<0.228$ brackets all diffuse interface algorithms. 

\begin{figure}
\begin{centering}
\includegraphics[width=0.49\textwidth]{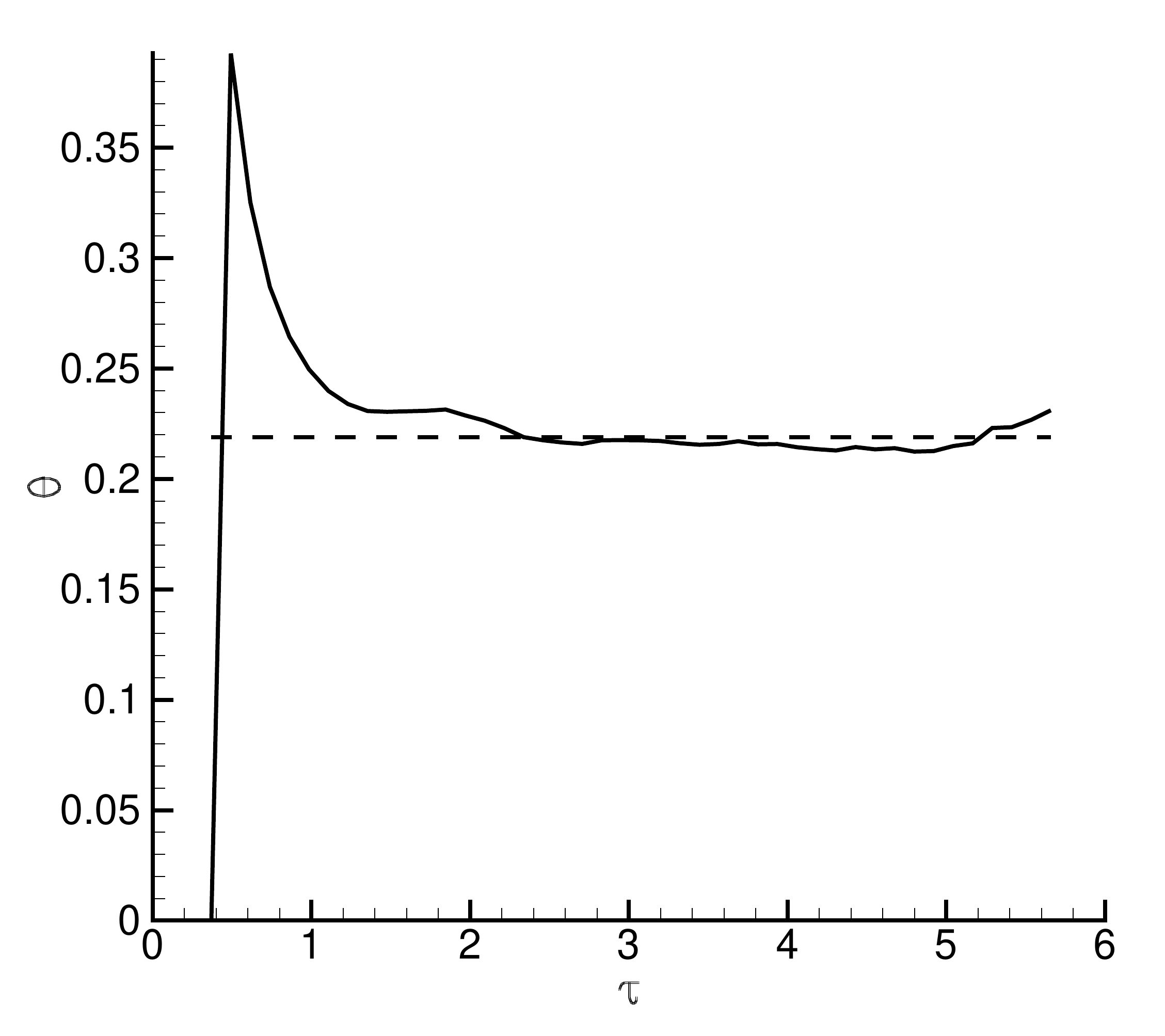}
\caption{The time dependent value of $\theta$ computed from $W$ and its derivatives (solid line) compared to non-linear regression (dashed line) from Flamenco for the standard problem \label{thetat}}
\end{centering}
\end{figure}

The time varying value of $\theta$ determined from the derivatives of $W$ is plotted in Figure \ref{thetat}, where the derivatives were evaluated using second--order central differences with temporal spacing of $\Delta t=0.04$. This shows that following the initial linear growth for $\tau < 0.8$, the value of $\theta$ moderately decreases from $\tau> 1.23$ to $\tau < 4.8$, followed by an increase at late time. 

\begin{figure*}
\begin{centering}
\includegraphics[width=0.49\textwidth]{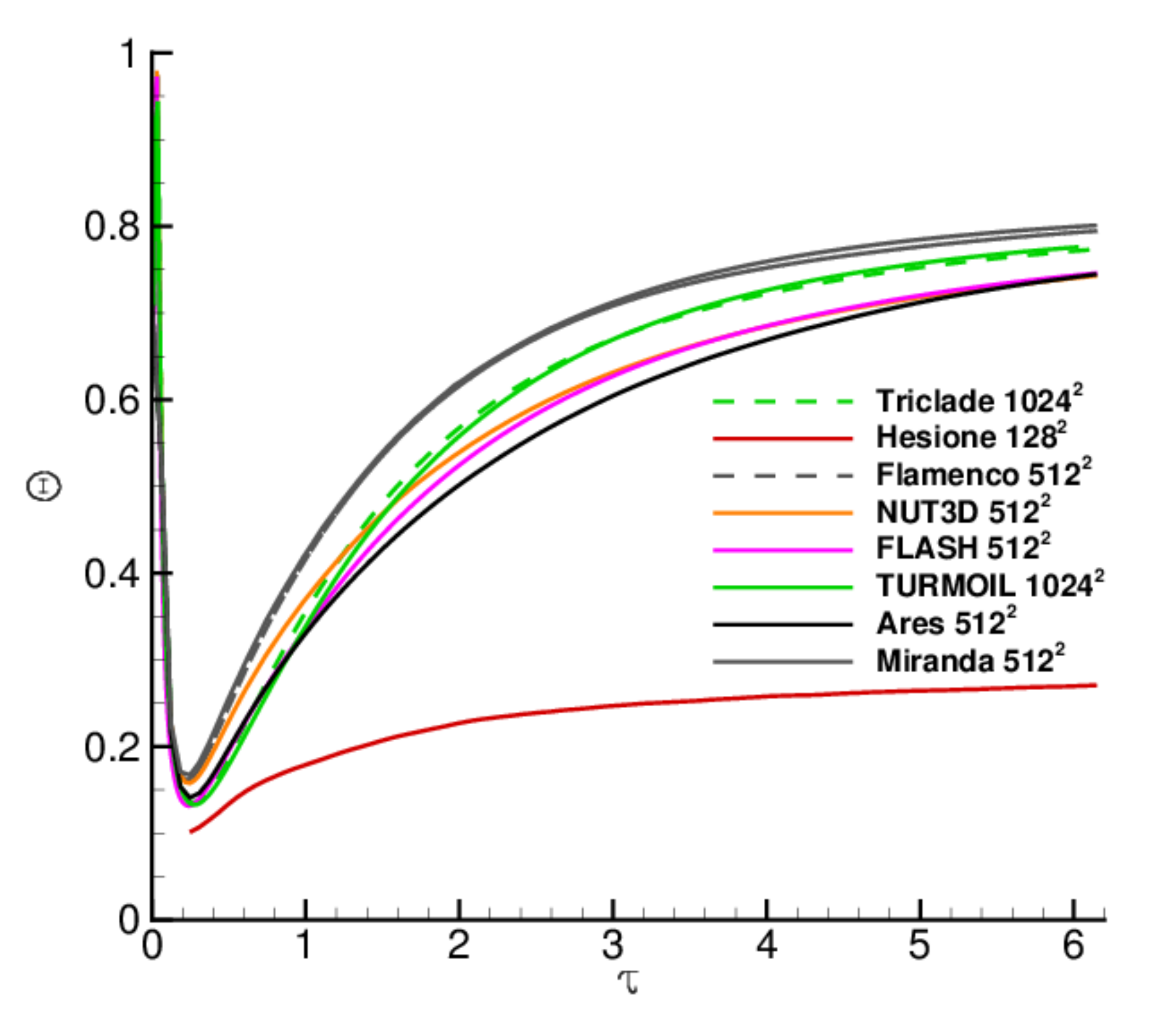}
\includegraphics[width=0.49\textwidth]{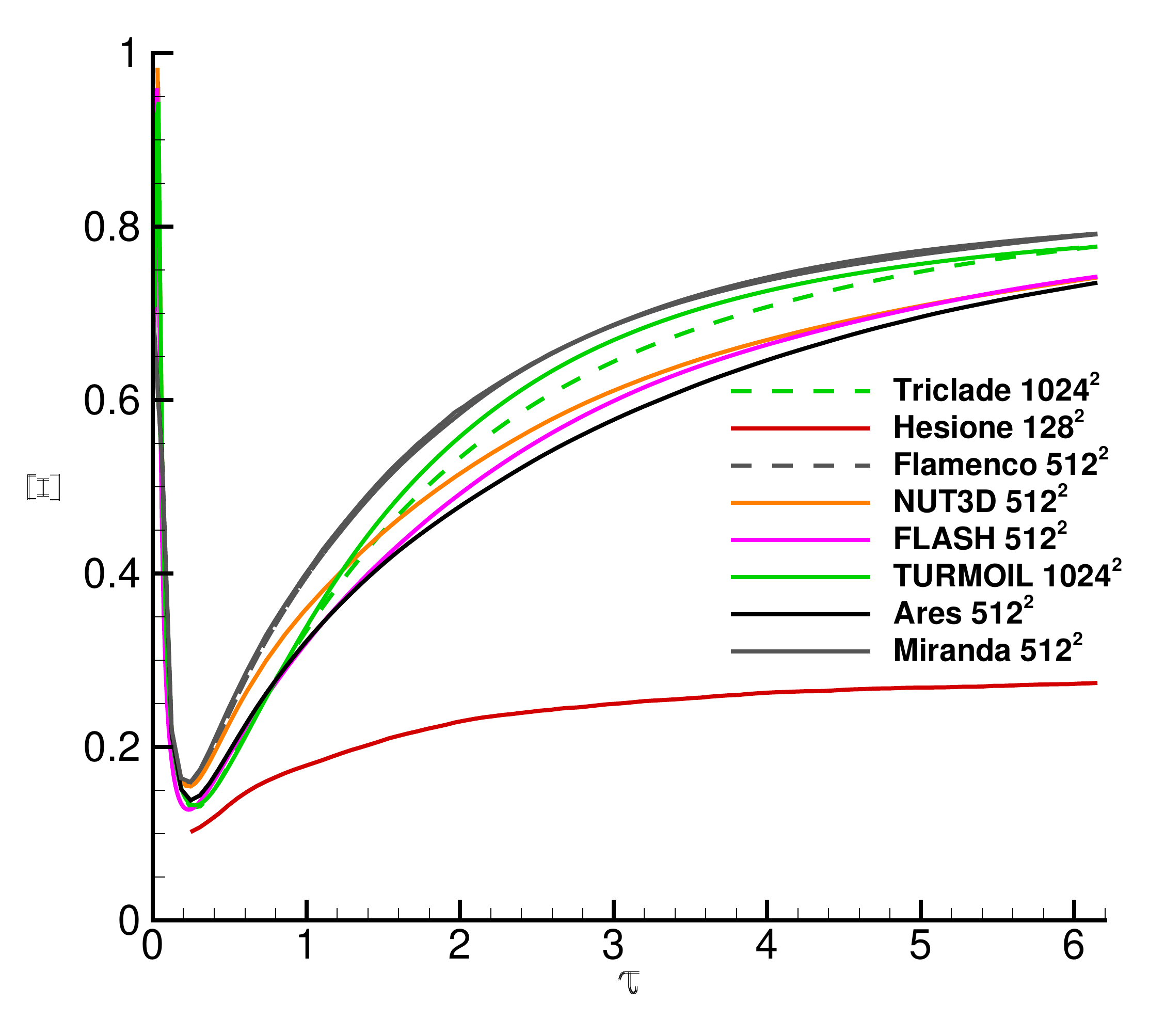}
\includegraphics[width=0.49\textwidth]{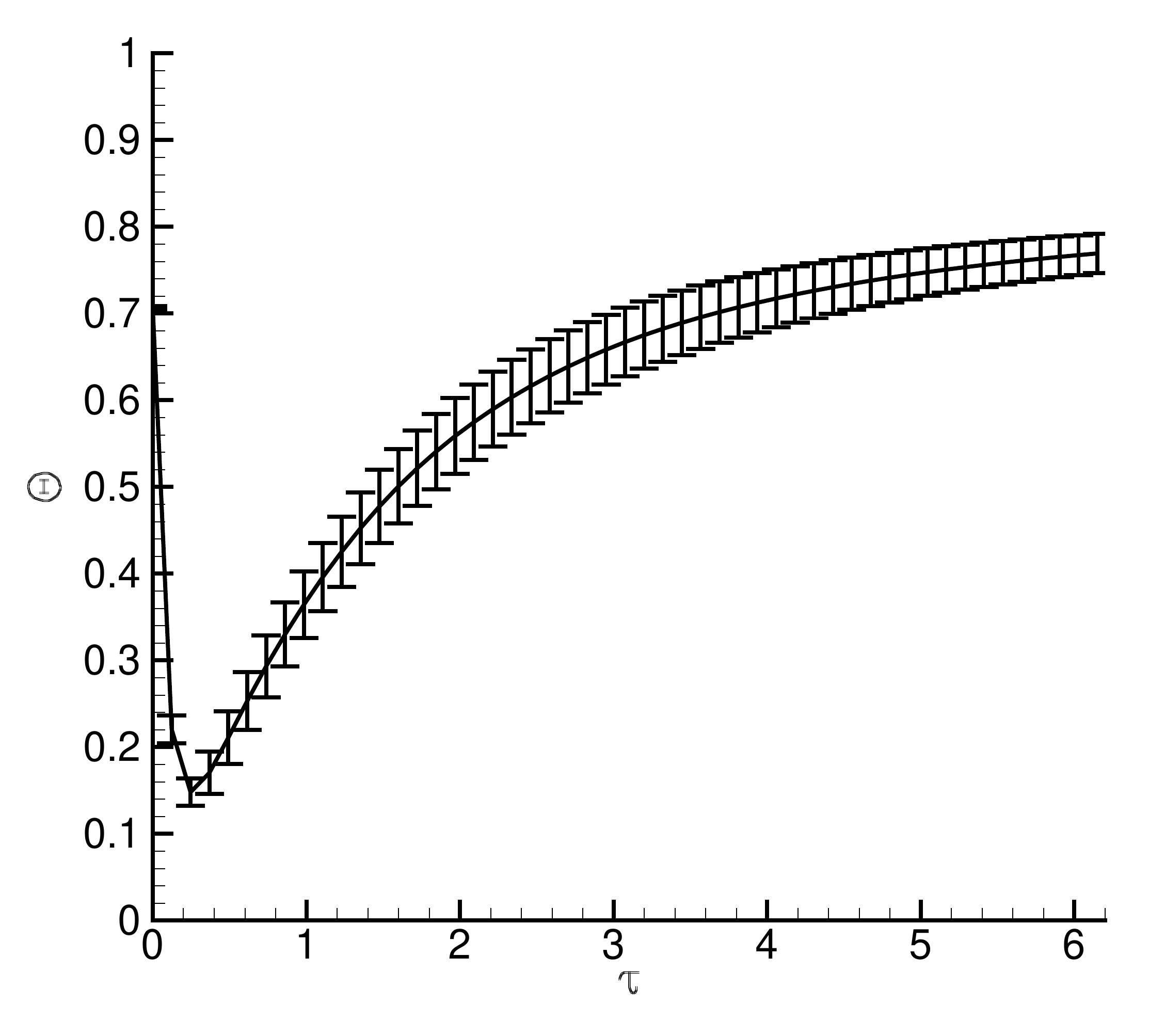}
\includegraphics[width=0.49\textwidth]{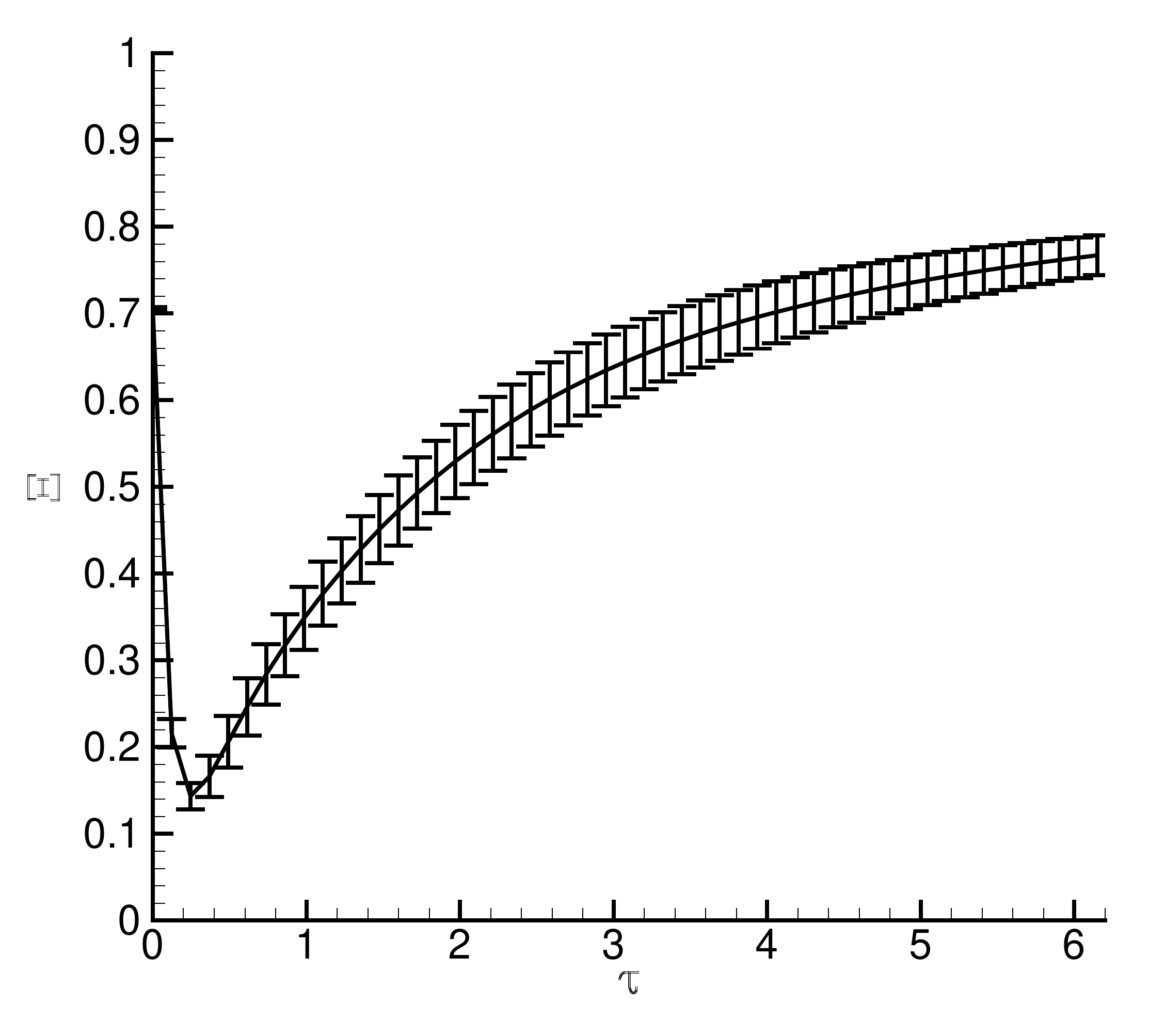}
\caption{Mix parameters $\Theta$ and $\Xi$ for all algorithms  for the standard problem. \label{StandardMix}}
\end{centering}
\end{figure*}

The integral mix measures $\Theta$ and $\Xi$ are plotted in Figure \ref{StandardMix}, including the code-averaged data (not including Hesione) and the standard deviation is plotted as error bars. The data is also tabulated in Table \ref{dataresults2}. For Hesione, interface reconstruction (IR) inhibits numerical diffusion so that the two fluids are treated as immiscible. That means the mixing is heterogeneous  which leads to a predicted $\Theta \approx 0.24$ if mixed cells are homogenised in contrast to $0.7-0.8$ for other algorithms. The $\alpha$ collaboration \cite{Dimonte2004} reported the same difference using HYDRA and ALEGRA in their interface reconstruction (IR) version.

Referring specifically to the diffuse interface algorithms, there is substantially greater variation in integral mix measures compared to the integral width. The early time quasi-linear phase shows good agreement between all codes, where the initialised diffuse interface is thinned by the extension of the contact surface as the fluids inter-penetrate. As time progresses, the energy containing scales begin to break down, transferring energy to smaller scales which then increase the efficiency of the mixing by stirring the fluid, thus increasing the cross-sectional area. 

Employing the time-to-transition criteria of Zhou \cite{Zhou2007} and taking $\delta =L/4$ and $u=\sqrt{2/3({\rm TKX}+{\rm TKY}+{\rm TKZ})}$ evaluated at $t=0.03$s implies that the transition time $\tau^{tr} \approx 0.18$ ($t^{tr}\approx 0.015$ s). However, at such an early stage the components of kinetic energy are strongly biased towards the shock-direction ($x$), using only the TKY and TKZ components gives $\tau^{tr} \approx 0.42$ ($t^{tr} \approx 0.034$ s). This is approximately the time of minimum mix parameters, i.e. where mixing begins to accelerate. The time of minimum mixing measures is not very sensitive to algorithm choice. As time increases, the layer tends towards a fully established mixing layer in which the mix measures increase. Should perfect self-similarity be achieve, these integral mix measures should tend towards a constant value. 

From the mix parameters it is clear that (i) the layer has not achieved self-similarity in the prescribed period and (ii) uncertainty in mix is greatest at the transition period. This can be seen clearly in the error-bars plotted for the code-averaged results, where the maximum standard deviation is $\sigma_\Theta=12$\% at $\tau=0.37$ ($t=0.03$ s), moves below 10\% at $\tau=1.1$ ($t=0.09$ s) and is 2.8\% at $\tau=6.15$ ($t=0.5$ s). This uncertainty in the results from differing algorithms is determined by two numerical factors. Firstly, at an early time the contact surface is stretched and becomes thinner than the local grid scale. Thus, the computed mix measures are dependent on the minimum thickness contact surface that each algorithm is able to represent, since once the contact surface is smaller than that, the layer can no longer thin. The second factor is that each algorithm has a different minimum number of points required to resolve the growth of a single vortex, which again is dependent on the dissipative properties of the algorithm (strongly linked to order of accuracy). The mixing layer must be resolved with a sufficient number of points to be able to simulate the physically present range of the length scales in the problem. It should be noted that in the high Re limit, and given sufficient grid resolution, the species variance should be sufficiently resolved using a Large--Eddy Simulation, and thus all algorithms should converge. 

Combining the complicated physical and numerical factors, it is encouraging that the standard deviation of the code-averaged results reduces considerably at late time, indicating that although self-similarity has not yet been achieved, the simulations appear to be tending to a much narrower uncertainty band at late times. At the latest time, $\Theta=0.769\pm 0.0225$ and $\Xi=0.767\pm 0.0229$ which is consistent with previous studies of narrowband and other closely related perturbations \cite{Lombardini2012,Thornber2010,Oggian2015,Zhou2016,Thornber2016}, and close to the experimentally measured results of Krivets and Jacobs \cite{Krivets2017}, but lower than those by Orlicz {\it et al.} and Weber {\it et al.} \cite{Orlicz2013,Weber2012}. 

Table \ref{thetas} gives the values of $\Xi$ and $\Theta$ for each algorithm at the latest time. Where convergence is uniform, asymptotic values have been computed using Richardson extrapolation of the three provided grid levels. Note that the Richardson extrapolated values are not expected to be realistic (particularly for Ares and Flamenco), even though their convergence was relatively uniform. The time of minimum mixing measures is not very sensitive to algorithm choice.

\subsection{Kinetic Energy and Anisotropy}

\begin{figure*}
\begin{centering}
\includegraphics[width=0.49\textwidth]{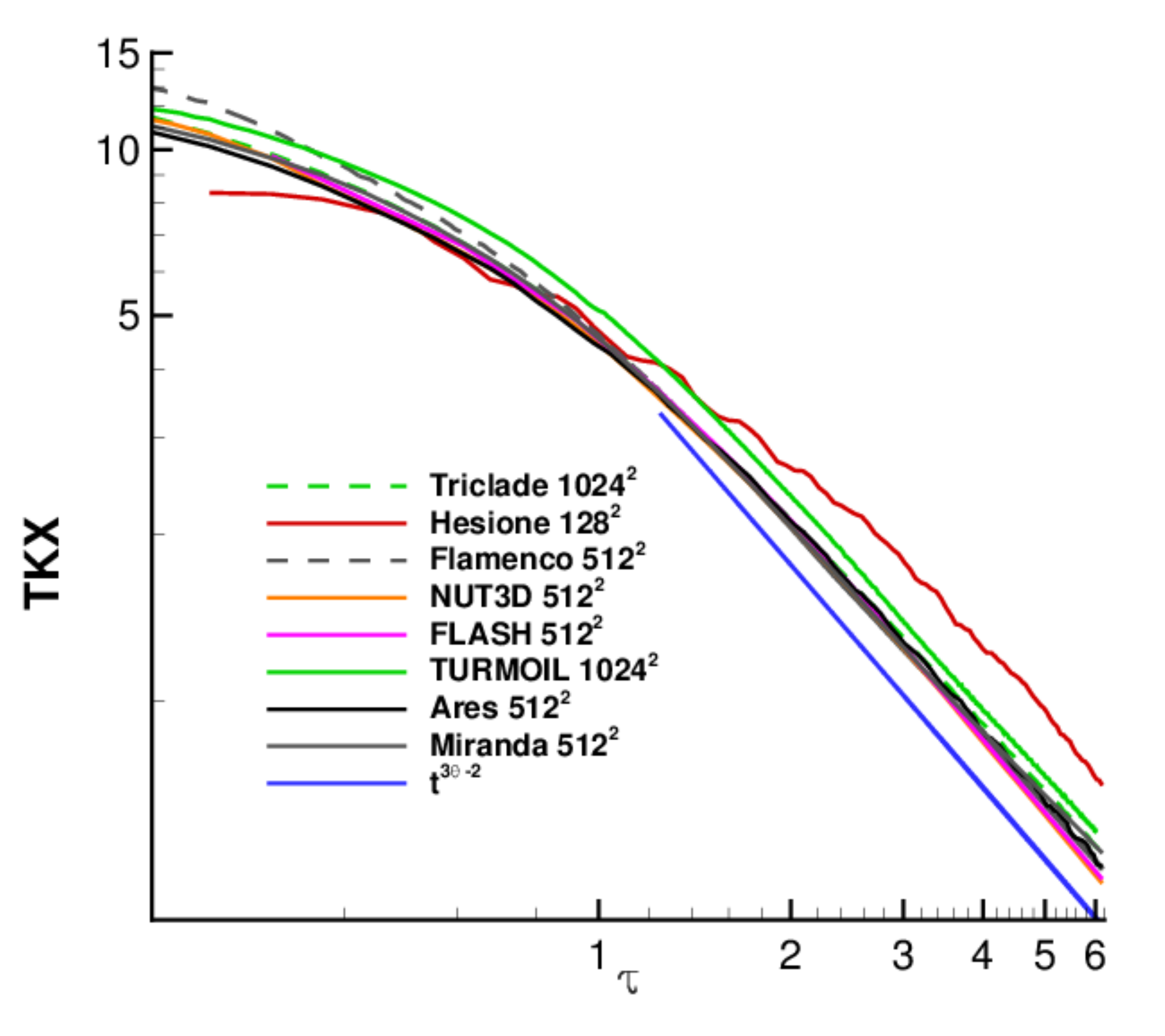}
\includegraphics[width=0.49\textwidth]{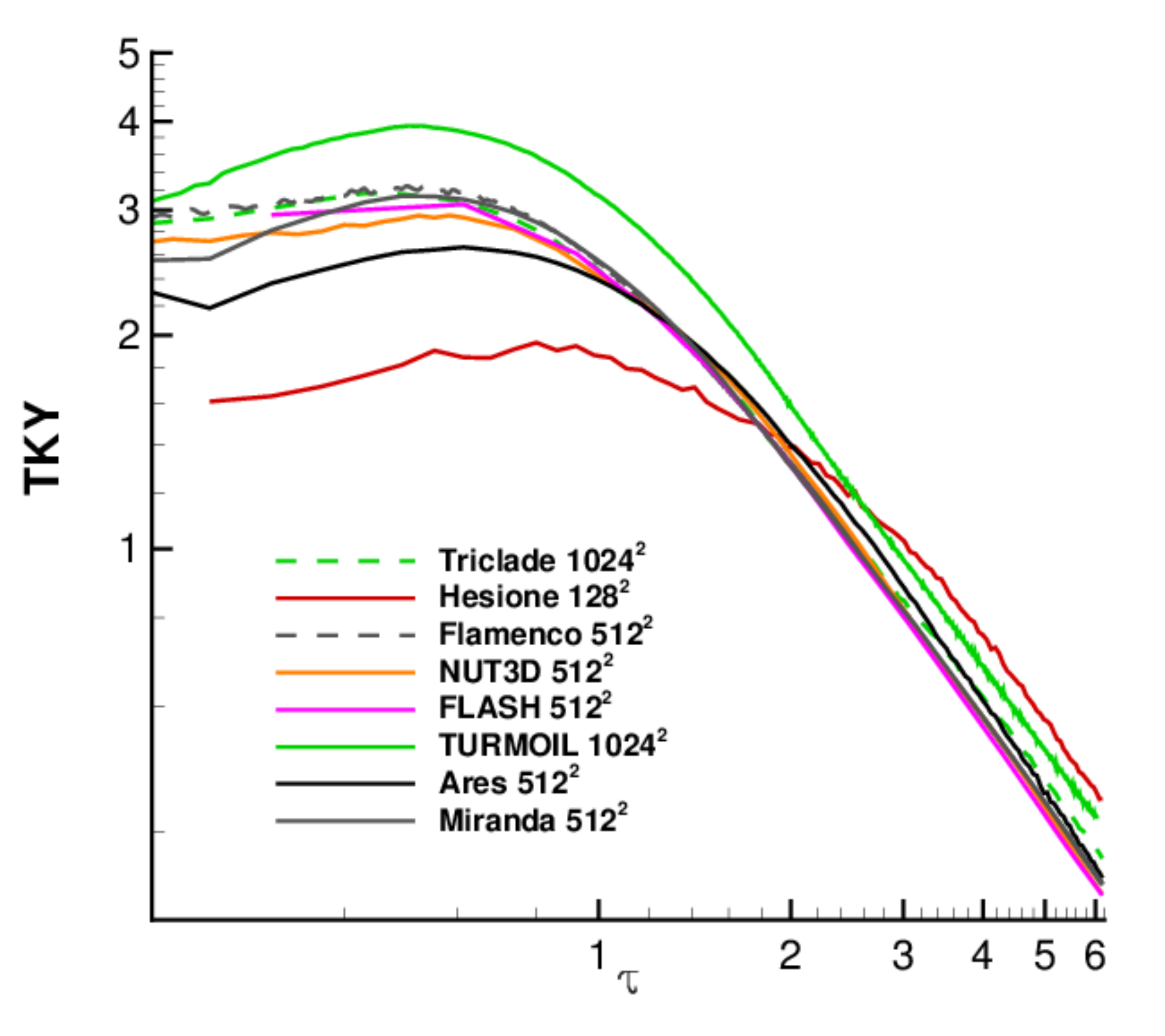}
\includegraphics[width=0.49\textwidth]{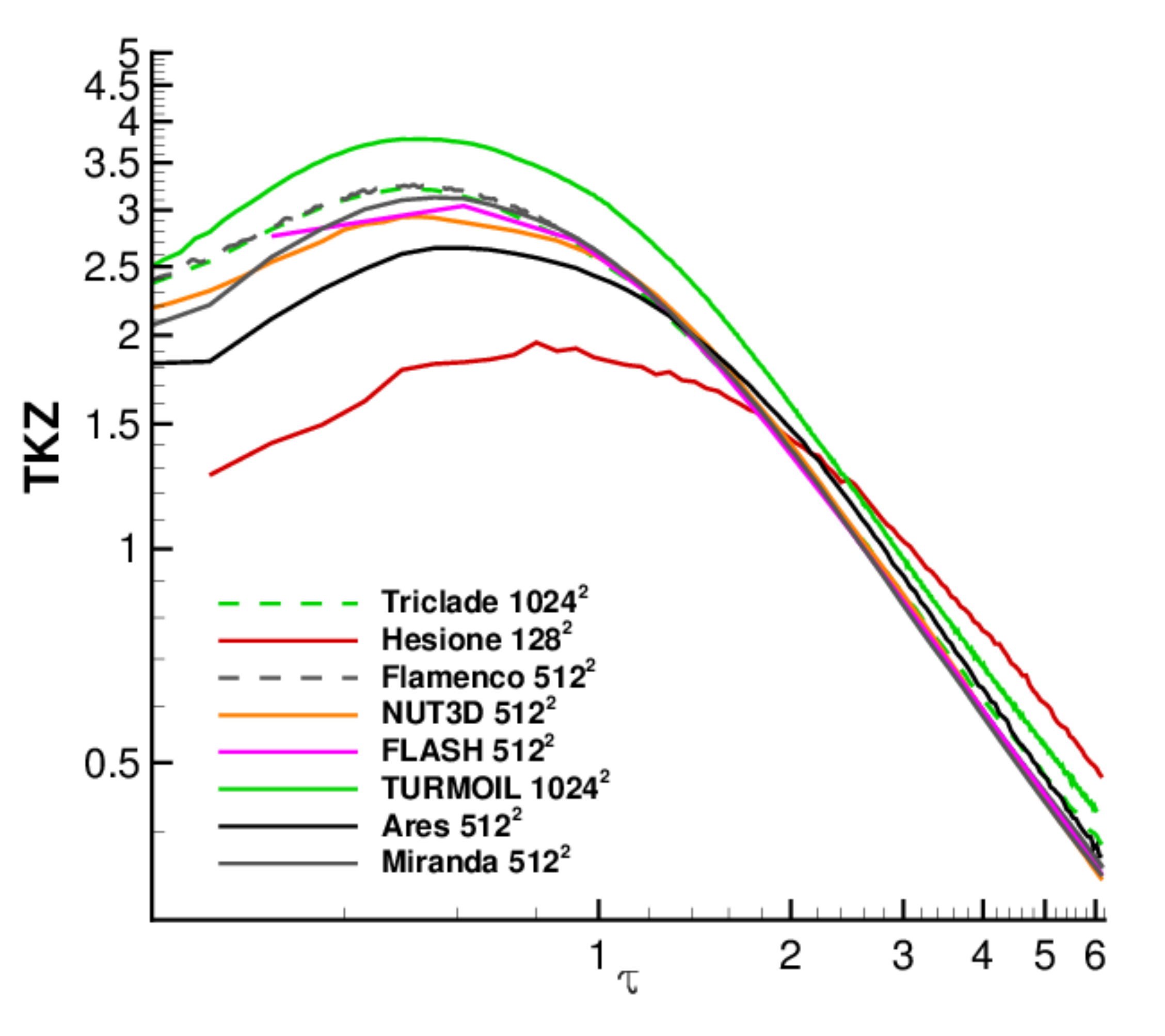}
\includegraphics[width=0.49\textwidth]{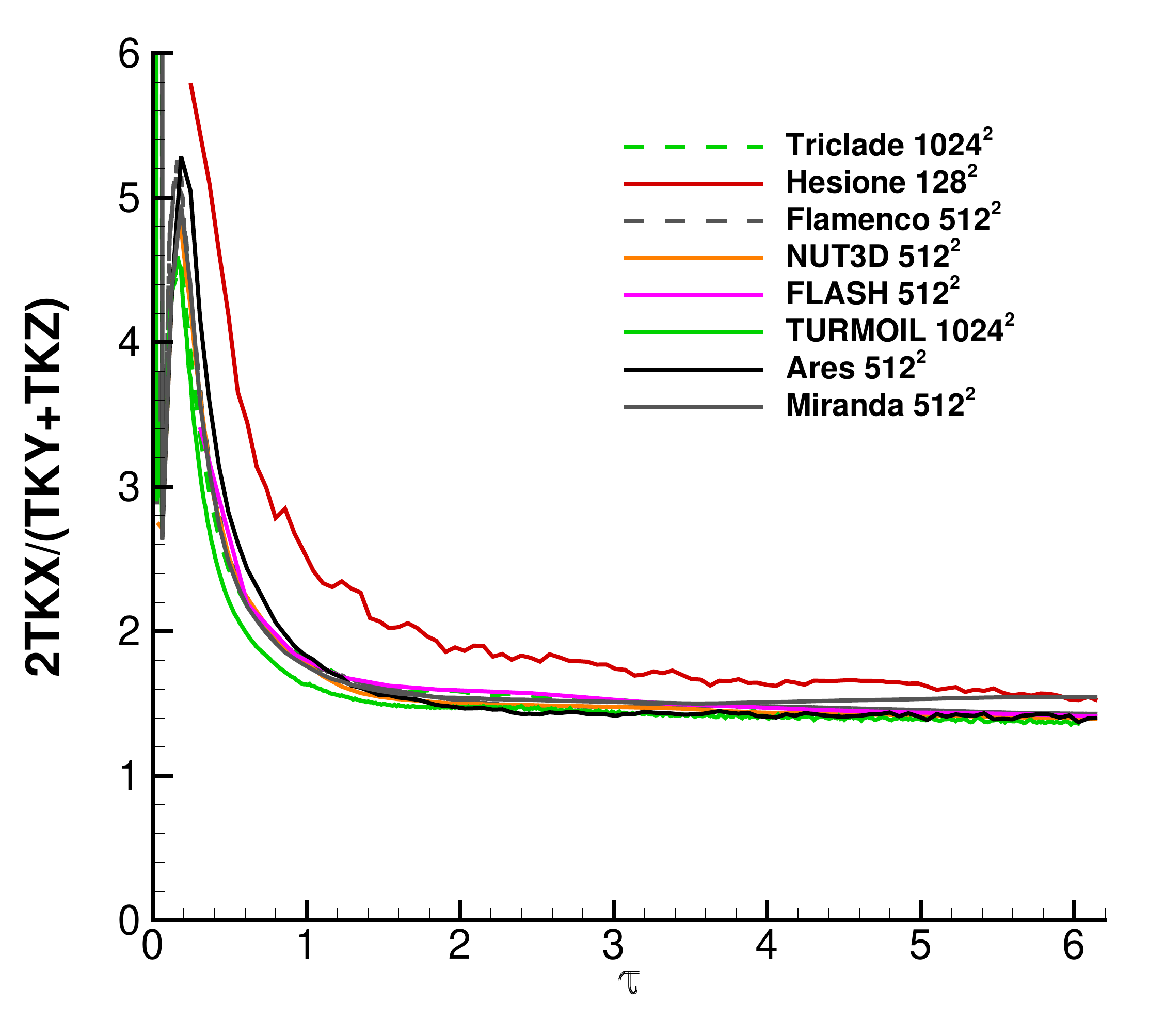}
\caption{Turbulent kinetic energy components and anisotropy ratio $2{\rm TKX}/({\rm TKY}+{\rm TKZ})$ for the standard problem. \label{TKE}}
\end{centering}
\end{figure*}

\begin{figure}
\begin{centering}
\includegraphics[width=0.49\textwidth]{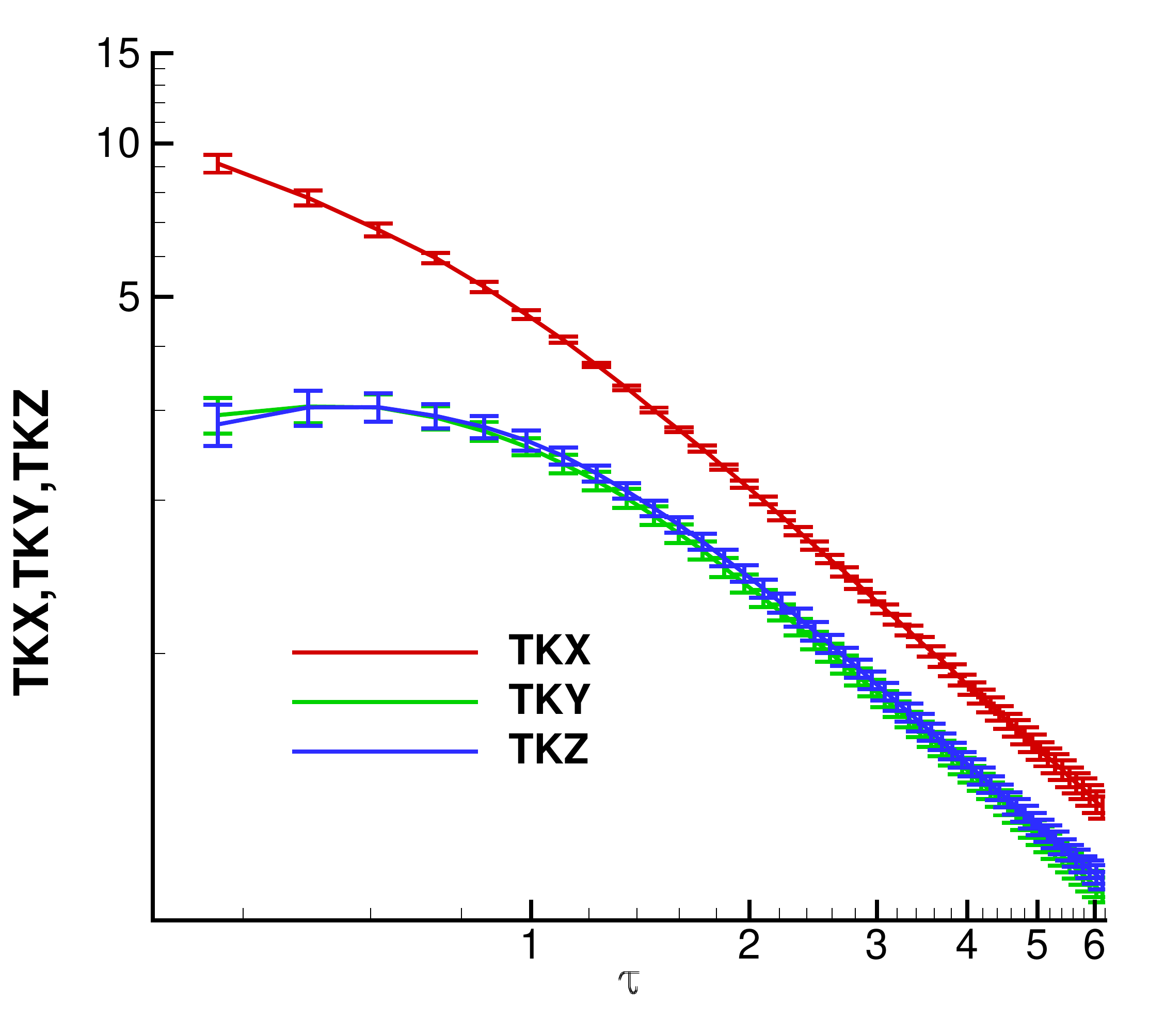}
\includegraphics[width=0.49\textwidth]{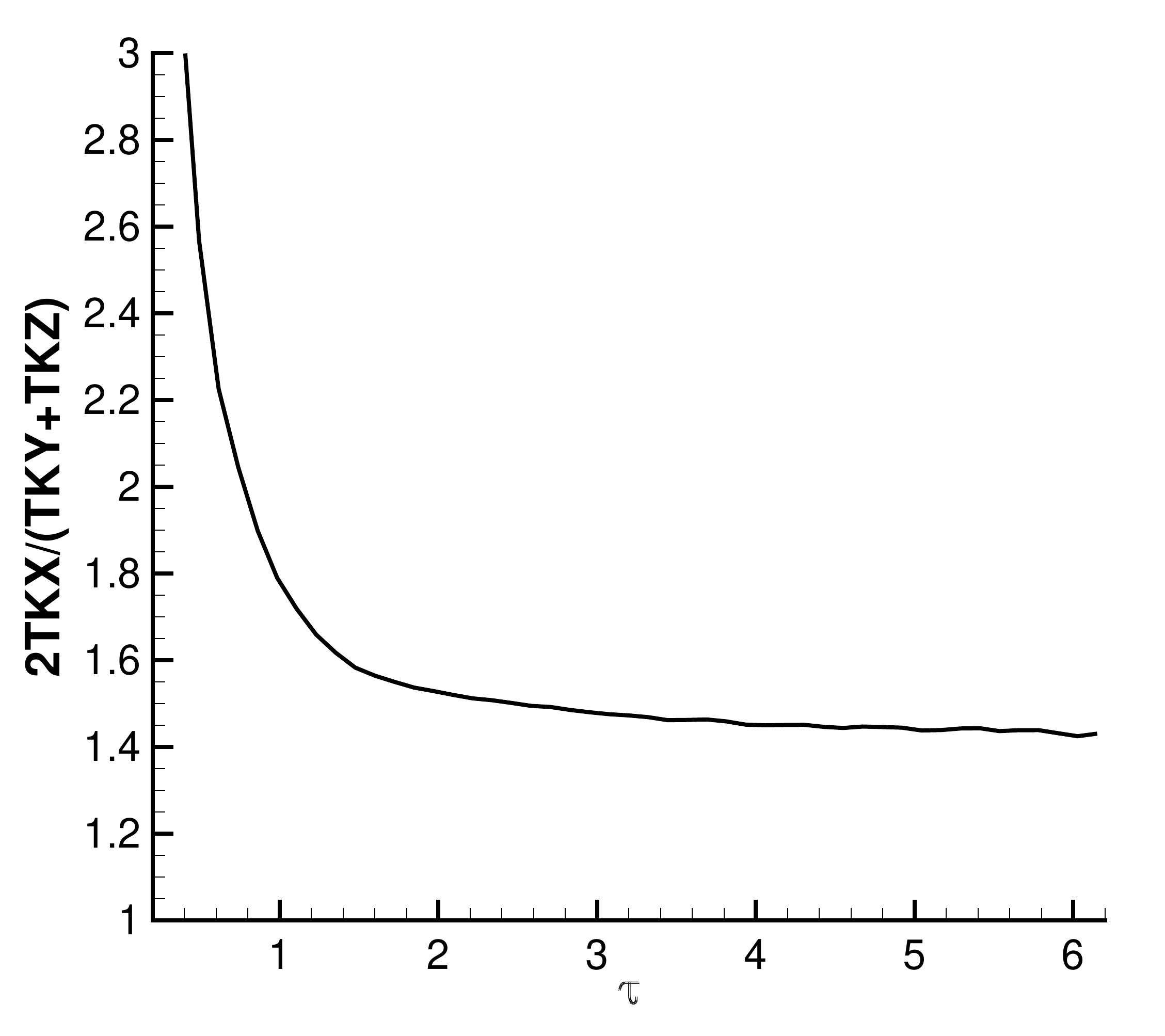}
\caption{Code-averaged dimensional kinetic energies and anisotropy ratio for the standard problem. \label{codeavs}}
\end{centering}
\end{figure}

\begin{table}
\caption{Dimensional kinetic energies and the anisotropy measure as a function of time and standard deviations for the code-averaged data for the standard problem. \label{datatkxstandard}}
\small{
\begin{tabular}{ccccccccc}
$t$ (s) & $\tau$ & TKX & $\sigma_{{\rm TKX}}$ & TKY & $\sigma_{{\rm TKY}}$  &TKZ & $\sigma_{{\rm TKZ}}$ & Aniso. \tabularnewline
\hline
\hline
              0.05& 0.62 &10983.94&  318.25& 4930.47&  305.09& 4936.73&  316.48 & 2.23\tabularnewline
              0.10& 1.23 &  5978.18&   55.41& 3541.15&  144.92& 3664.06&  132.97 & 1.66\tabularnewline
              0.15& 1.85 &3766.78&   45.97& 2396.90&  101.51& 2502.81&   90.99 & 1.54\tabularnewline
              0.20& 2.46 &  2644.78&   47.58& 1724.41&   67.77& 1797.25&   76.27 & 1.50\tabularnewline
              0.25& 3.08 &1987.92&   41.08& 1323.75&   47.25& 1369.86&   54.28 & 1.48\tabularnewline
              0.30& 3.69 & 1574.77&   43.72& 1058.62&   36.75& 1092.80&   41.19 & 1.46\tabularnewline
              0.35& 4.31 & 1289.17&   43.06&  875.09&   32.92&  901.20&   33.67 & 1.45 \tabularnewline
              0.40& 4.92 & 1084.62&   41.88&  736.59&   28.71&  764.62&   26.72 & 1.44\tabularnewline
              0.45& 5.54 &  930.11&   41.10&  632.30&   26.58&  662.20&   23.22 & 1.44 \tabularnewline
              0.50& 6.15 & 812.86&   40.63&  552.11&   23.37&  583.82&   22.72 & 1.43 \tabularnewline
\end{tabular}}
\end{table}

Figure \ref{TKE} plots the fluctuating kinetic energy in each component direction, and a simple measure of anisotropy ($2{\rm TKX}/({\rm TKY}+{\rm TKZ})$) as a function of time, where in the plots all kinetic energies (and spectra) are non-dimensionalised by $\rho_L \dot W_0 ^2 \bar \lambda A/2$. Figure \ref{codeavs} plots the average of all of the algorithms except the two outliers, TURMOIL and Hesione. A sample of the same data is also tabulated in Table \ref{datatkxstandard}. 

As kinetic energy is primarily a measure of the large scales, it should converge well with increasing grid resolution. At early time there is a strong peak in fluctuating kinetic energy in all directions, as the measure also includes contributions from the non-planar shock. As the shock clears the mixing layer and straightens, the kinetic energy deposition by the shock wave dominates, and peaks at $\tau=0.12$ ($t=10^{-2}$ s) in the shock direction, and $\tau=0.62$ ($t\approx 5\times 10^{-2}$ s) in the in-plane directions. This delay is due to the transfer of energy from the shock-direction component into the in-plane components, and is visible in the anisotropy measure. This time scale matches well the time-to-transition estimate of  $\tau^{tr} \approx 0.42$ ($t^{tr}\approx 0.034$ s) in the previous section. The kinetic energy components then decay with an approximate power-law form, where an additional line is plotted in Figure \ref{TKE} representing the predicted self-similar decay rate based on $W \propto t^{\theta}$ which gives fluctuating kinetic energy $\propto t^{3\theta-2}$ (see, for example \cite{Thornber2010}).

Overall, six of the algorithms collapse nearly perfectly, which is an excellent result given that each algorithm has quite different order of accuracy and dissipation mechanism. This strongly suggests that a reasonable separation of energetic from dissipative scales has been achieved. There are two algorithms which give different solutions, Hesione and TURMOIL, which have consistently higher kinetic energy throughout the layer, the reason for which has not yet been identified but may be linked to the treatment of acoustic waves in the ALE method. The code averages have been taken of the six algorithms, neglecting these two outliers, as otherwise the standard deviations are effectively dominated by just the single algorithm. Taking this approach, the standard deviations are remarkably small, $\ll$1\% at early time and $\approx 5$\% at the latest time. 

Plotting both the anisotropy for each algorithm, and the code averaged kinetic energy anisotropy (2TKX/(TKY+TKZ)), it can be seen that for this initial condition and time period the results are not tending towards isotropy. At the final time, this anisotropy measure is 1.43 for the code-averaged data, where the anisotropy predicted by each individual algorithm ranges from 1.25 (TURMOIL) to 1.55 (Miranda). This lends substantial additional weight to prior observations that although the initial anisotropy reduces due to transfer from the shock direction to the in-plane directions, this transfer does not result in isotropy prior to the commencement of power-law decay \cite{Thornber2010,Oggian2015,Lombardini2012,Ristorcelli2013}.

\begin{figure}
\begin{centering}
\includegraphics[width=0.49\textwidth]{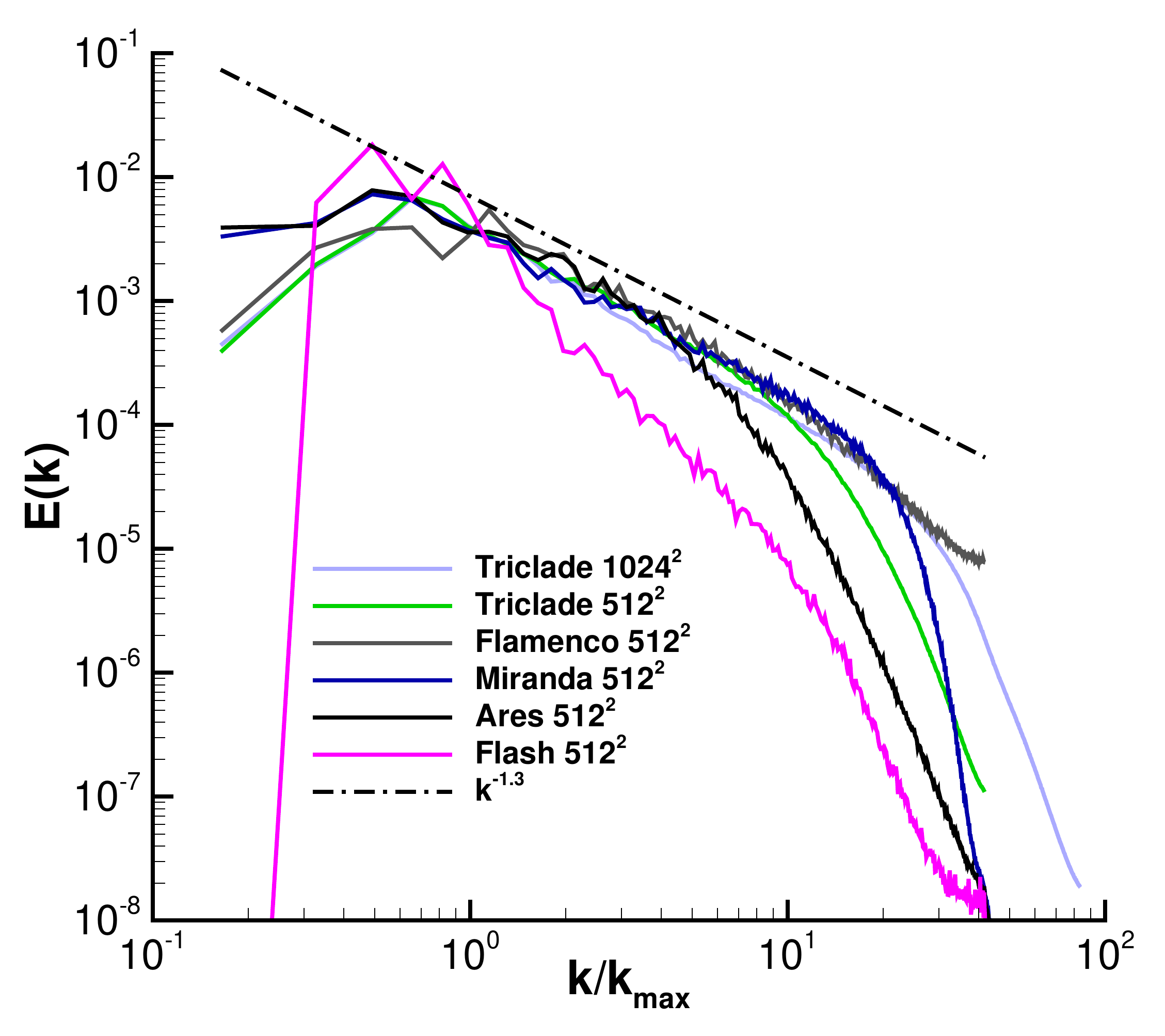}
\includegraphics[width=0.49\textwidth]{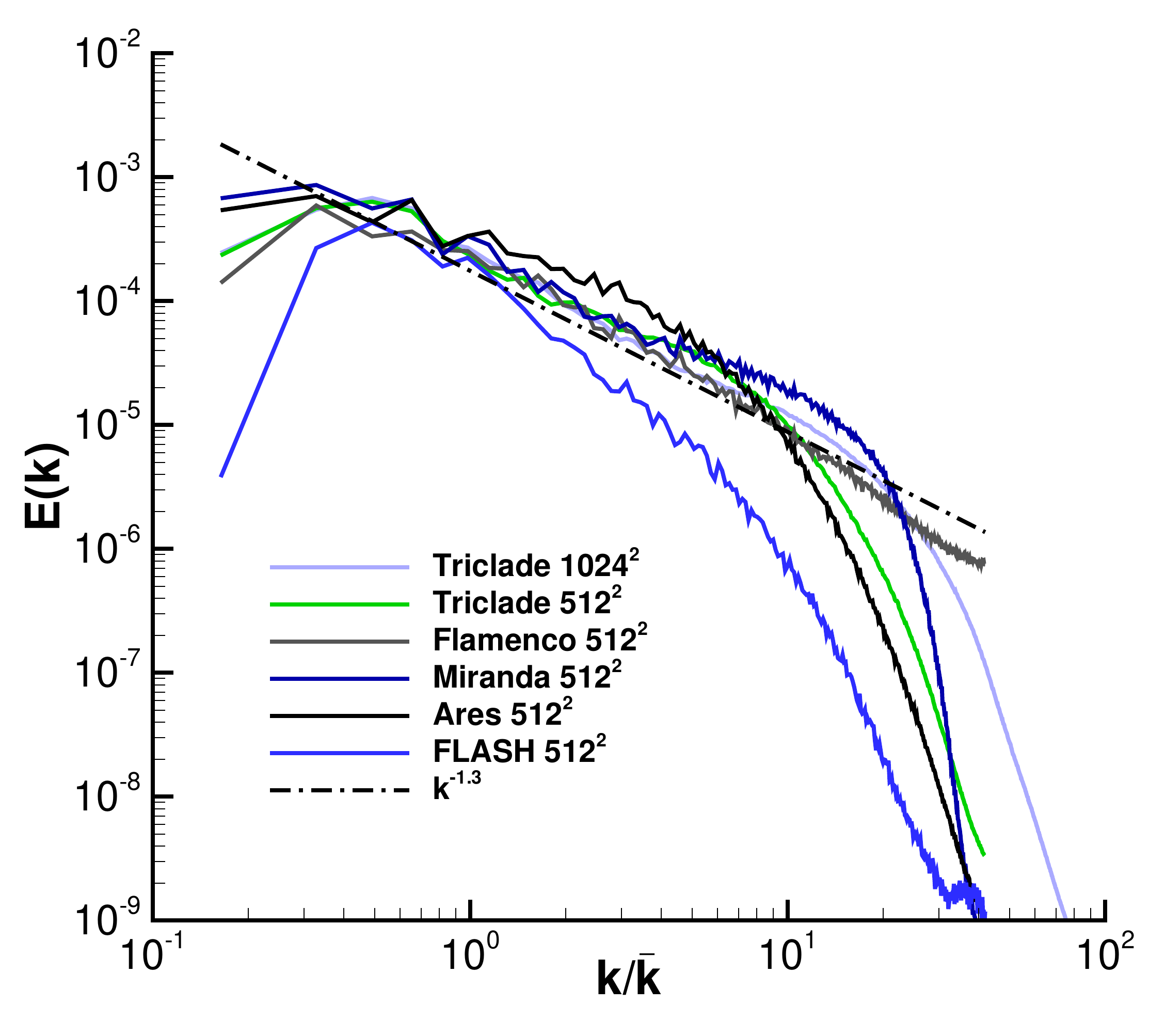}
\caption{Variable density kinetic energy spectra at $\tau=1.23$ ($t=0.1$ s) (left)  and $\tau=6.15$ ($t=0.5$ s) (right) for the standard problem \label{spectra}}
\end{centering}
\end{figure}

The variable density kinetic energy spectra taken at a single plane at the centre of the mixing layer, defined as $\avg{f_1}=0.5$ are shown in Figure \ref{spectra} at $\tau>1.23$ ($t>0.1$ s) and $\tau=6.15$ ($t=0.5$ s) for Ares, Flamenco, Flash, Miranda and Triclade at $720\times 512^2$ and Triclade at $1440\times 1024^2$. The wavenumbers have been normalised by $\bar k=6.11$. The spectra at $\tau=1.23$  peak at $k=3-5$, then for most algorithms show the early stages of a constant power--law region to $k/\bar k \approx 16$, dependent on algorithm. This behaviour is mirrored in the $\tau=6.15$ results, where the peak is now located at $k/\bar k=0.33-0.5$. The constant power-law region has a slope of $\approx k^{-1.3}$ for $k/\bar k=3-5$ which is shallower than a classical Kolmogorov spectrum, or that proposed by Zhou \cite{Zhou2001}.

Comparing the algorithms, the agreement at the large scales is remarkably good considering that this is a spectrum from a single plane located at the centre of the mixing layer. There are differences in the dissipative properties of the schemes, as expected from the different orders of accuracy and algorithmic constructions. The spectra from Flash depart from the bulk of the results at $k/\bar k\approx 1.3-1.6$, most likely due to numerical dissipation. Ares has a higher kinetic energy in the range $k=0.5-8.2$ which may be a numerical manifestation of the physically observed `bottleneck' phenomenon, and then damps the high wavenumber energy to zero. 

Triclade, Flamenco, and Miranda agree very well throughout the power-law region, with the key difference between the three in the approach to the grid cutoff. Relative to the Triclade simulation at  $1440\times 1024^2$, the Miranda and Triclade simulations have a small `bump' at the end of the power-law region, then uniformly damp fluctuations to zero as the cut-off is approached. Miranda has the sharpest damping at high wavenumbers. The spectrum in the Flamenco siulation does not damp to zero at the high wavenumbers. This high wavenumber behaviour is most likely responsible for the observed differences in the integral mix measures in the previous section, as the schemes with the highest energy at high wavenumbers have the highest values of $\Theta$ and $\Xi$ (see Figure \ref{StandardMix}).

\section{Quarter-scale Problem: Results and Discussion \label{quarter}}

\begin{table*}
\caption{$\theta$ for each algorithm in the quarter scale problem defined using non-linear regression [$W=A(t-t_0)^\theta$, tabulated data is dimensional], $\Theta$, $\Xi$, and anisotropy measure at the final time for the quarter scale problem. \label{thetasq}}
\begin{tabular}{ccccccccc}
Code & Exp. Err. \% & Final $\tau$ ($t$ (s)) &  $\theta (\tau>24.6)$& $A$ & $t_0$ & $\Theta$ & $\Xi$ & Aniso. \tabularnewline
\hline
\hline
Ares  &      5.8 & 123 (2.5) & 0.248 & 0.2972 & -0.01042 & 0.804 & 0.8138 & 1.53\tabularnewline
Flamenco  &  0.5 & 246 (5.0) & 0.296 & 0.2800 & -0.04327 & 0.807 & 0.8189 & 1.49\tabularnewline
Flash  &     5.1 & 64 (1.3) & 0.297 & 0.2787 & -0.08664 & 0.777 & 0.7882 & N/A\tabularnewline
Miranda  &   3.8 & 246 (5.0) & 0.331 & 0.2625 & -0.10366 & 0.7985 & 0.7963 & 1.55 \tabularnewline
NUT3D  &     1.8 & 60 (0.65) & 0.282 & 0.2819 & -0.07083 & 0.772 & 0.7794 & N/A\tabularnewline
Triclade  &  0.2 & 123(2.5) & 0.283 & 0.2852 & -0.04059 & 0.782 & 0.7946 & 1.55\tabularnewline
TURMOIL  &   1.6 & 108(2.2) & 0.301 & 0.2825 & -0.09909 & 0.800 & 0.8095 & 1.66\tabularnewline
Mean  &  &  & 0.291 & 0.2820 & -0.0649 & 0.792& 0.8000 & 1.55\tabularnewline
$\sigma$  &  &  & 0.025 & 0.0110 & -0.03465 & 0.0141 & 0.0144 & 0.06\tabularnewline
\end{tabular}
\end{table*}

Based on the observation that the key integral quantities were grid converged at $180\times 128^2$ resolution for several algorithms, a second simulation was undertaken at $720\times512^2$ resolution where all initial physical length scales were divided by four, with the computational domain size fixed. This permitted computations to more than eight times the dimensionless time ($t \dot{a}/\lambda_{min}\approx 380-760$). Due to the computational expense, each simulation was run for the longest time possible.  Most of the simulations terminated before $\tau=123$ ($t=2.5$ s), which is when parts of the leading spikes exit the domain. Some simulations were continued as far at $\tau=246$ ($t=5$ s). Table \ref{thetasq} details the expected maximum error in the integral width estimated from the standard problem, the final time each algorithm was run to, and  key quantities recorded or computed from the results. In this section all results are presented together.

\begin{figure*}
\begin{centering}
\subfigure[]{\includegraphics[width=0.4\textwidth]{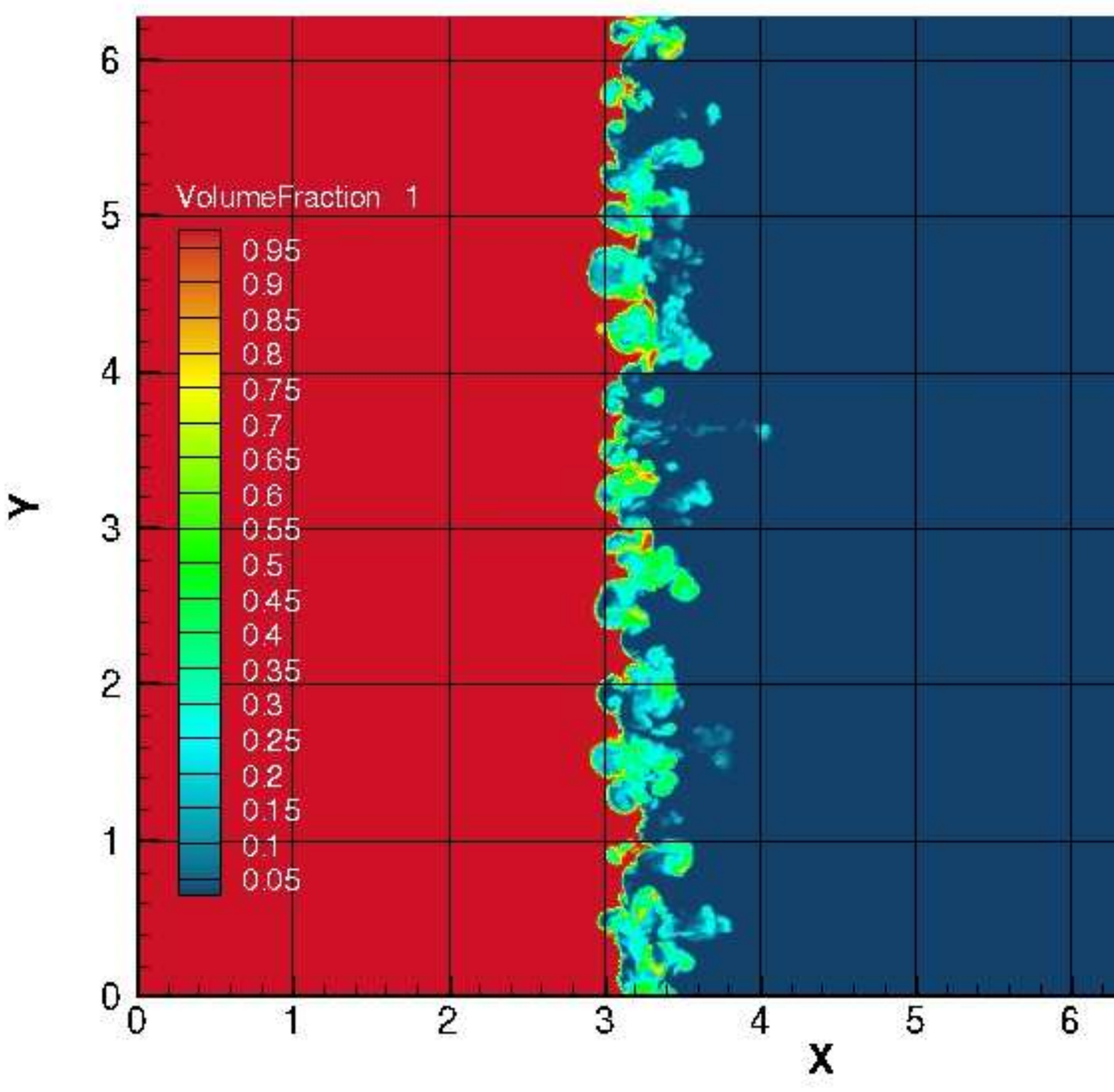}}
\subfigure[]{\includegraphics[width=0.4\textwidth]{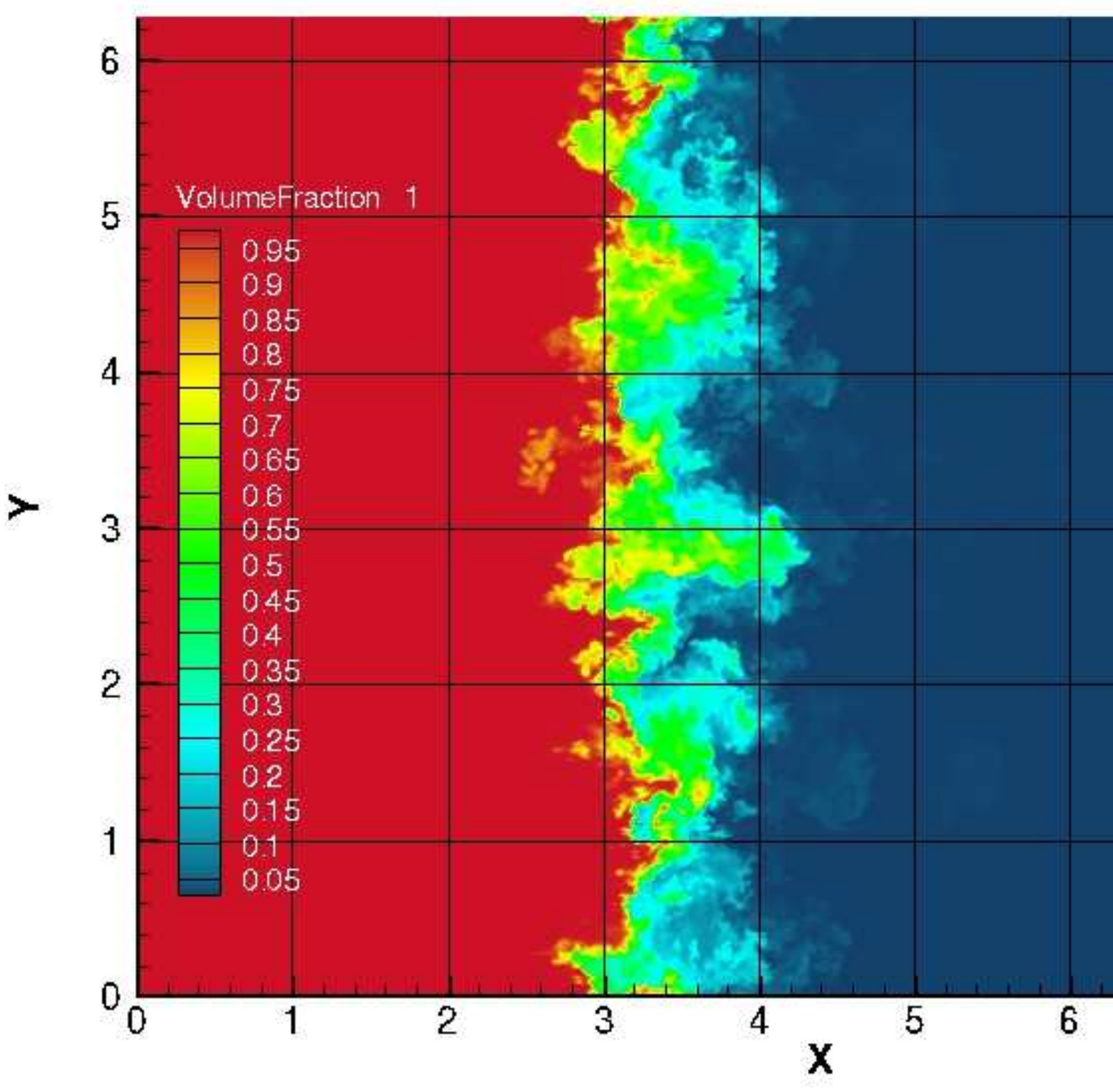}}
\subfigure[]{\includegraphics[width=0.4\textwidth]{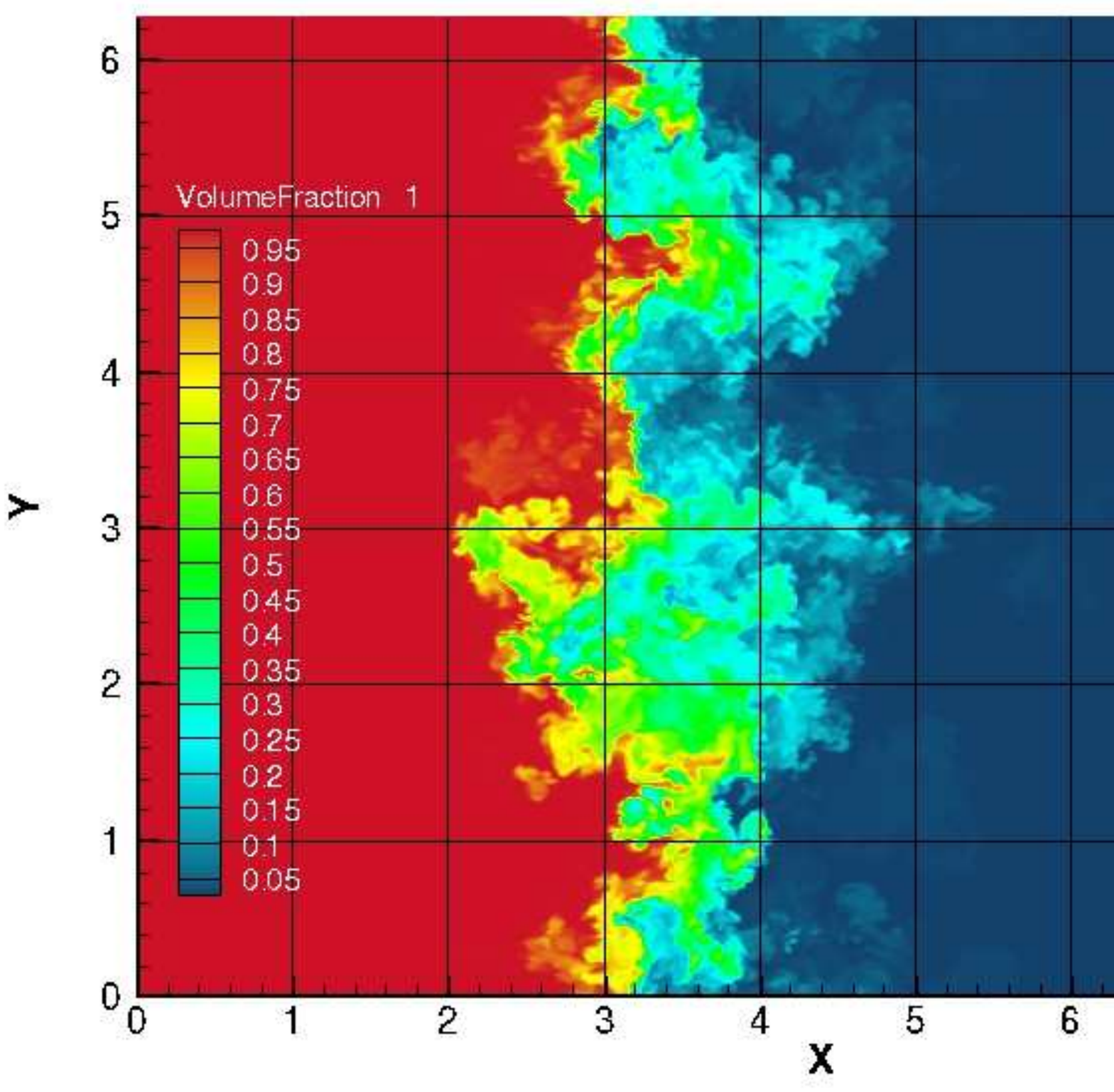}}
\subfigure[]{\includegraphics[width=0.4\textwidth]{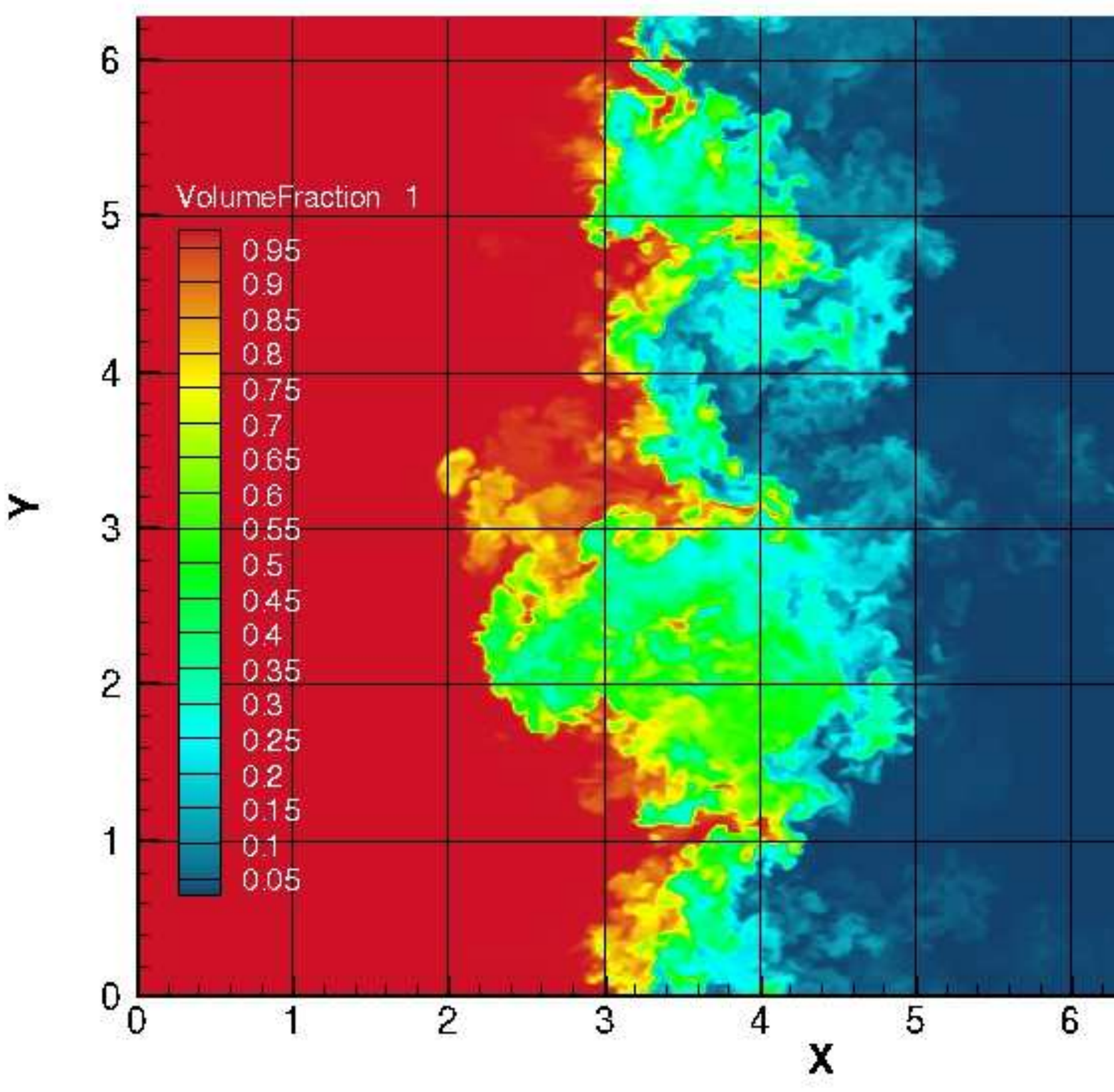}}
\caption{Visualisations of the quarter scale problem at $\tau=1.23, 25.6, 124$ and $246$ ($t=0.025, 0.52, 2.52$ and $5.00$ s) using Flamenco. \label{visq}}
\end{centering}
\end{figure*}

A visualisation of the resultant flow field is shown at four times from the Flamenco code in Fig. \ref{visq}. The earliest time is at the same dimensionless time as the first time instant of the standard problem shown in Figure \ref{vis01}. Here the contact surface between the two fluids is relatively well defined, although individual spikes and bubbles are breaking down. At the later times, the mixing layer develops through transition to the start of an established mixing layer.

\begin{figure*}
\begin{centering}
\includegraphics[width=0.49\textwidth]{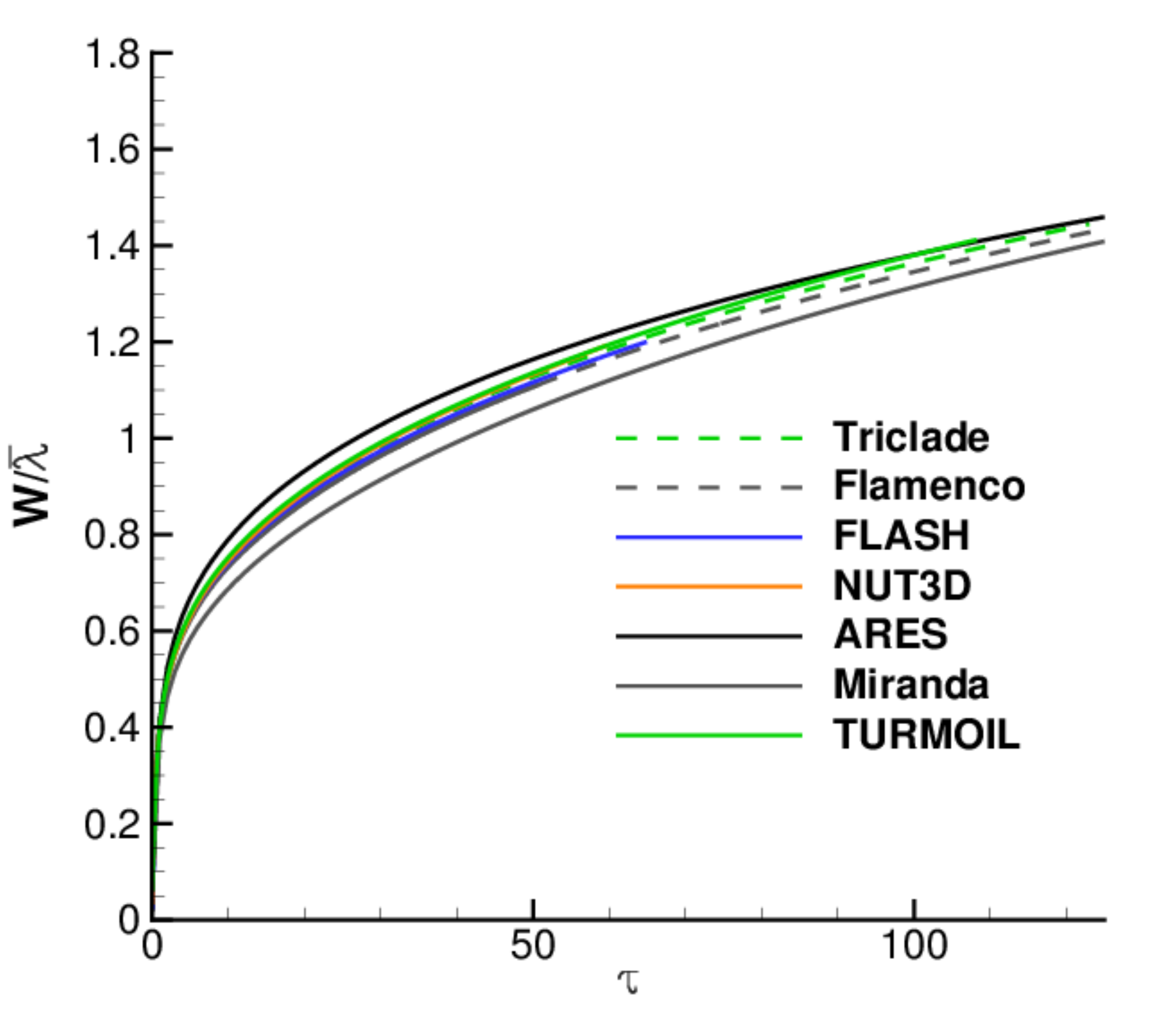}
\includegraphics[width=0.49\textwidth]{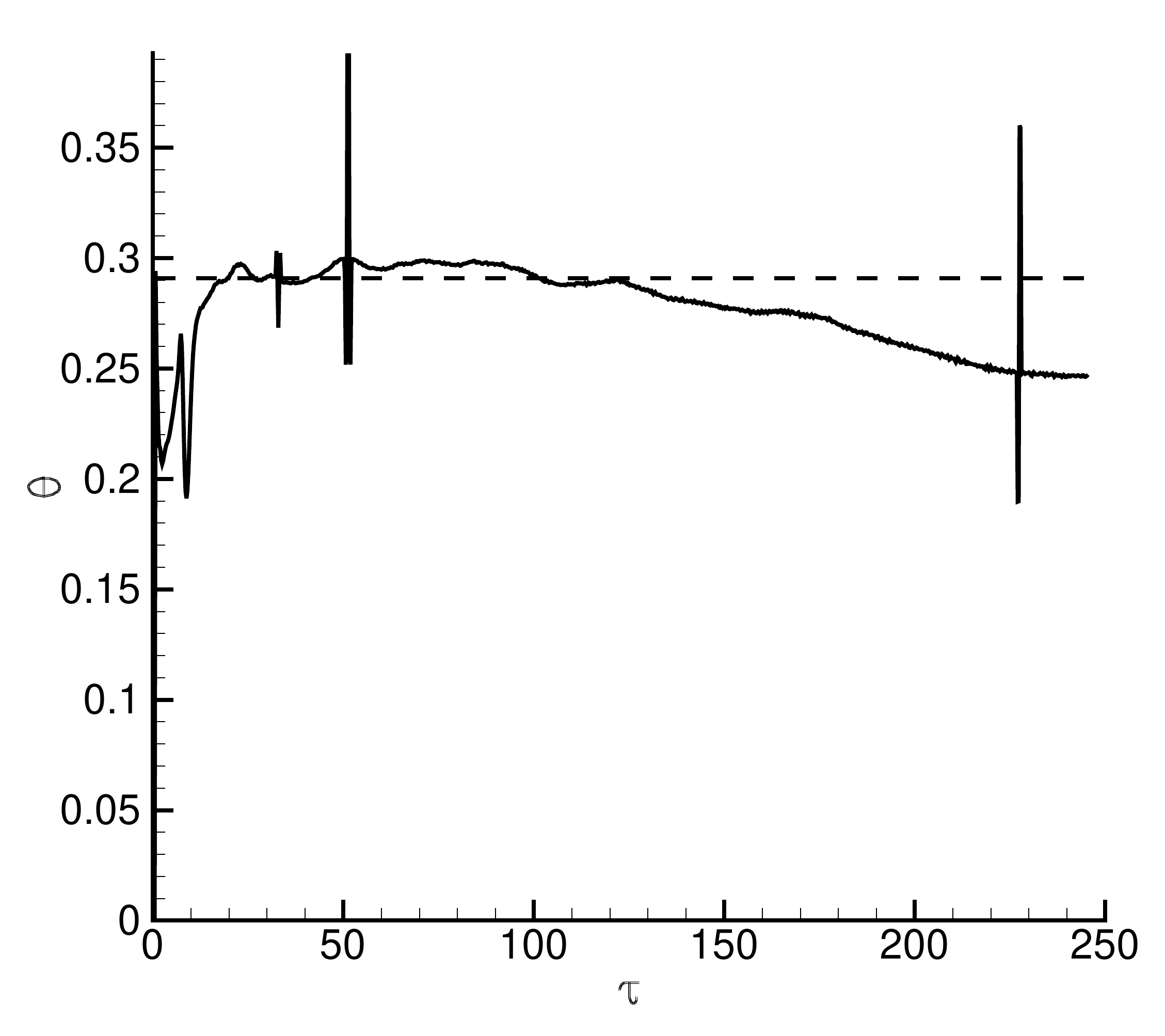}
\caption{Integral width, the time dependent value of $\theta$ computed from $W$ and it's derivatives (solid line) compared to non-linear regression (dashed line) from Flamenco for the quarter scale problem \label{widthq}}
\end{centering}
\end{figure*}

Figure \ref{widthq} (left) plots the integral width for all algorithms, and  Table \ref{thetasq} summarises the data fits for the integral width for each individual algorithm. 

\noindent The initial conditions were designed to be at the limit of expected convergence. As such, it is important to clarify the expected confidence in the solutions. Column one of Table \ref{thetasq} details the expected error based on the error between the $180\times 128^2$ and $720\times 512^2$ simulations on the standard problem (for NUT3D this was based on the error from $360\times 256^2$ to $720\times 512^2$). The results in Figure \ref{widthq} show that there is a grouping of algorithms whose time dependent integral width agree relatively closely, including Flamenco, Flash, NUT3D, Triclade and TURMOIL. This is consistent with the predicted convergence error. The standard case demonstrates that the Ares simulation should converge down towards the other algorithms, and that the Miranda simulation should converge upwards if a higher grid resolution was run. This behaviour has been confirmed by shorter simulations at higher resolution at early time, however not for the full simulation for which resources were not available.

The integral mixing measures vary by $<$5\% following $\tau=24.6$ ($t=0.5$ s), thus non-linear regression was employed again to determine $\theta$, $A$ and $t_0$ for each algorithm. The second column of Table \ref{thetasq} details the latest time to which the problem was simulated, which is the latest time of the data fit if the algorithm was not run past $\tau=123$ ($t=2.5$ s). If it was run to later time, the computed data fit used only data up to $\tau=123$ ($t=2.5$ s). The code-average was not employed due to the larger scatter relative to the standard problem. Table \ref{thetasq} plots these values and their means. The mean values were $\theta=0.291 \pm 0.025$ where all algorithms were considered, however if Miranda and Ares results are excluded then $\theta=0.292 \pm 0.009$. The computed values of $t_0$ and $A$ are equally consistent. Note that this value of $\theta$ is quite close to the value consistent with the self-similar solution of the multicomponent $k$--$\epsilon$ turbulence model, which gives $\theta = 0.3$ using standard coefficient values based on shear turbulence, and very good agreement with many experimentally measured pre-reshock mixing layer widths in numerically modeled Richtmyer--Meshkov instability  \cite{Moran2013,Moran2014}. 

\begin{figure*}
\begin{centering}
\includegraphics[width=0.49\textwidth]{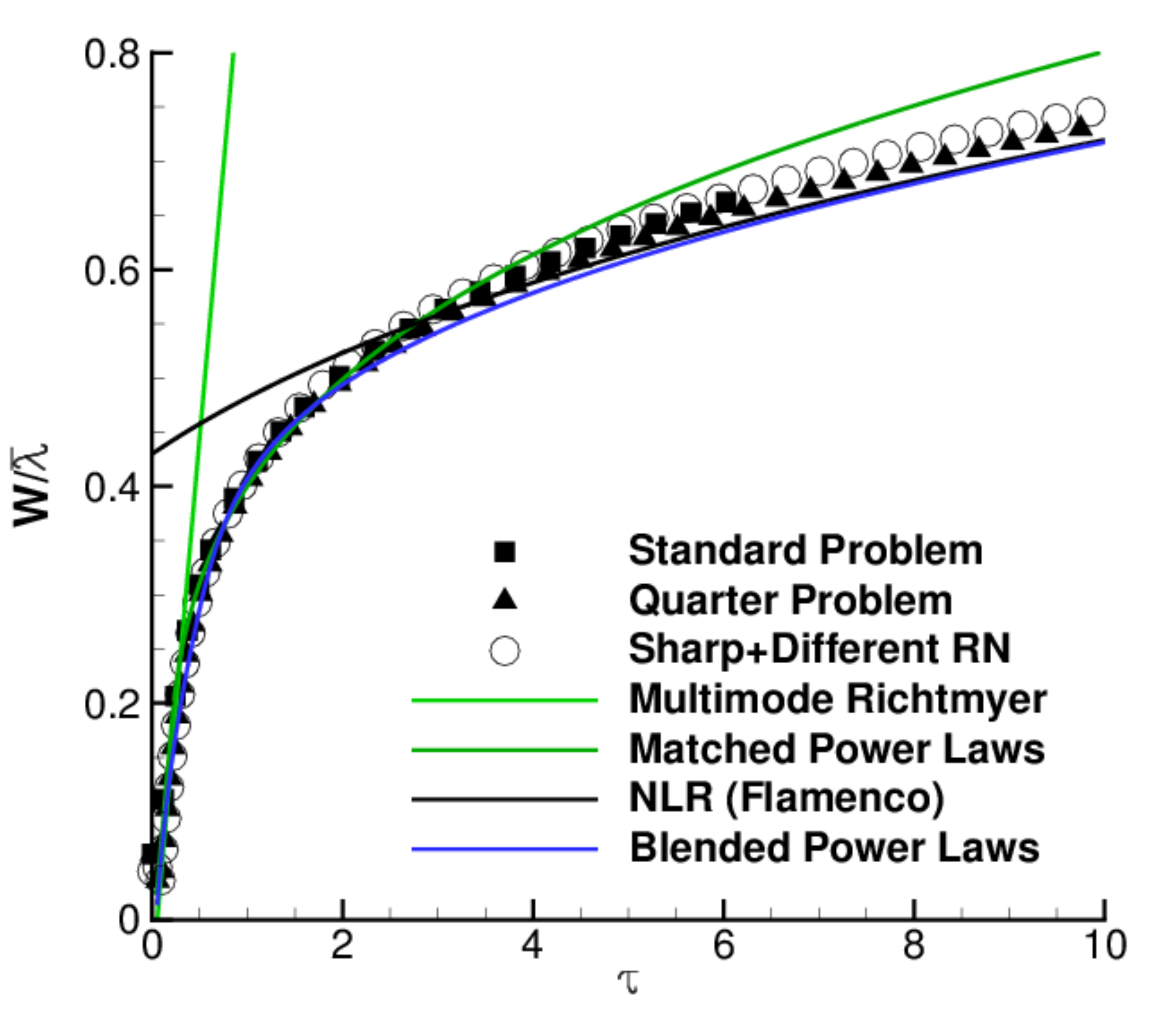}
\includegraphics[width=0.49\textwidth]{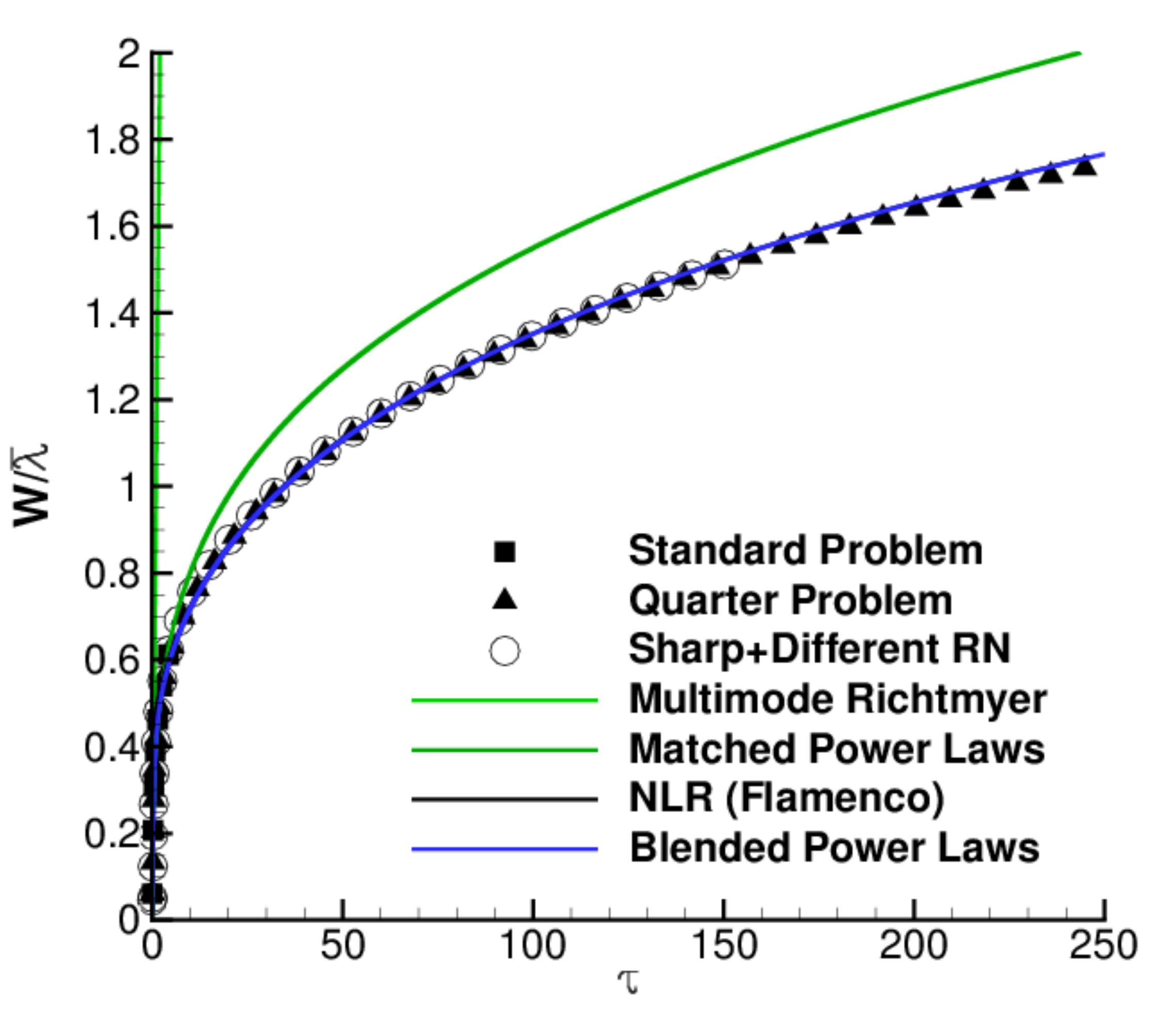}
\caption{Integral width from Flamenco for the standard and quarter scale problem and an additional simulation run with a sharp interface and different random numbers, plotted along with the linear multimode theory and two models [`matched' power law Eq. (\ref{mpl} and `blended' power law Eq. (\ref{bpl}] and the result of the non-linear regression. \label{widthqtheory}}
\end{centering}
\end{figure*}

Figure \ref{widthqtheory} plots the integral width for the quarter scale problem with the standard problem, and an additional simulation run with a sharp interface with different random numbers \cite{Thornber2016}. The integral width for the quarter scale problem varies by $<$0.1\% from the case with a different set of random numbers (and random number generator) and sharp interface, thus indicating that there is (i) little residual dependence on the phase relation of the initial conditions, (ii) little residual dependence on statistical variations in the individual modal amplitudes and (iii) a small impact of the diffuse layer on the vorticity deposited in the layer. The results from the standard case collapse on top of the quarter scale case under the chosen non-dimensionalisations. The `blended' and `matched' power-law model are plotted, and the agreement with the dataset is good, with the matched power laws showing a 13.7\% overestimation at the latest time and the blended within 1.2\% as expected, as it is based on the non-linear regression data fit. 

The Flamenco simulations were continued to $\tau=246$ to explore the late time behaviour of the mixing layer, however it must be noted that past $\tau=123$, spikes are observed exiting the domain. This extended data set permitted the plotting of $\theta(t)$, as shown in Figure \ref{widthq}. At late time there is a substantial reduction in $\theta$, however this is also mirrored in the non-linear regression, which as a function of starting time gives $\theta=0.285$ $(\tau>50)$, $\theta=0.274$ $(\tau >100)$, $\theta=0.263$ $(\tau >147)$, $\theta=0.251$ $(\tau >197)$ for example. Jumps in $\theta$ can be seen due to small oscillations in the mixing layer width (not visible by the naked eye) which the computed value of $\theta$ is very sensitive to. This provides additional justification for terminating simulations prior to  $\tau=123$.

Next, the values of $\theta$ have been computed using two alternative mixing measures, the  mixed mass $\mathcal{M}$ \cite{Zhou2016b} and the layer width $h$ \cite{Cook2001,Weber2013}:

\begin{equation}
\mathcal{M}(t)=\int \rho\, Y_1\, Y_2\, dV,\,\,\,\,\,\, h(2)=2 \int  \textrm{min}(\avg{f_1},\avg{1-f_1}) dx,
\end{equation}

\noindent  where $\avg{.}$ indicates a plane-averaged quantity and $Y_i$ denotes the mass fraction of gas $i$. These both give a power-law dependence similar in form to the integral width. Using the Flamenco data and non-linear regression, the $\theta (\tau >24.6)$ based on these quantities gives $0.292$ for $h$ and $0.287$ for $\mathcal{M}$ indicating a relative insensitivity of the computation of $\theta$ using these different mixing  measures ($\theta=0.291 \pm 0.005$).

\begin{figure*}
\begin{centering}
\includegraphics[width=0.49\textwidth]{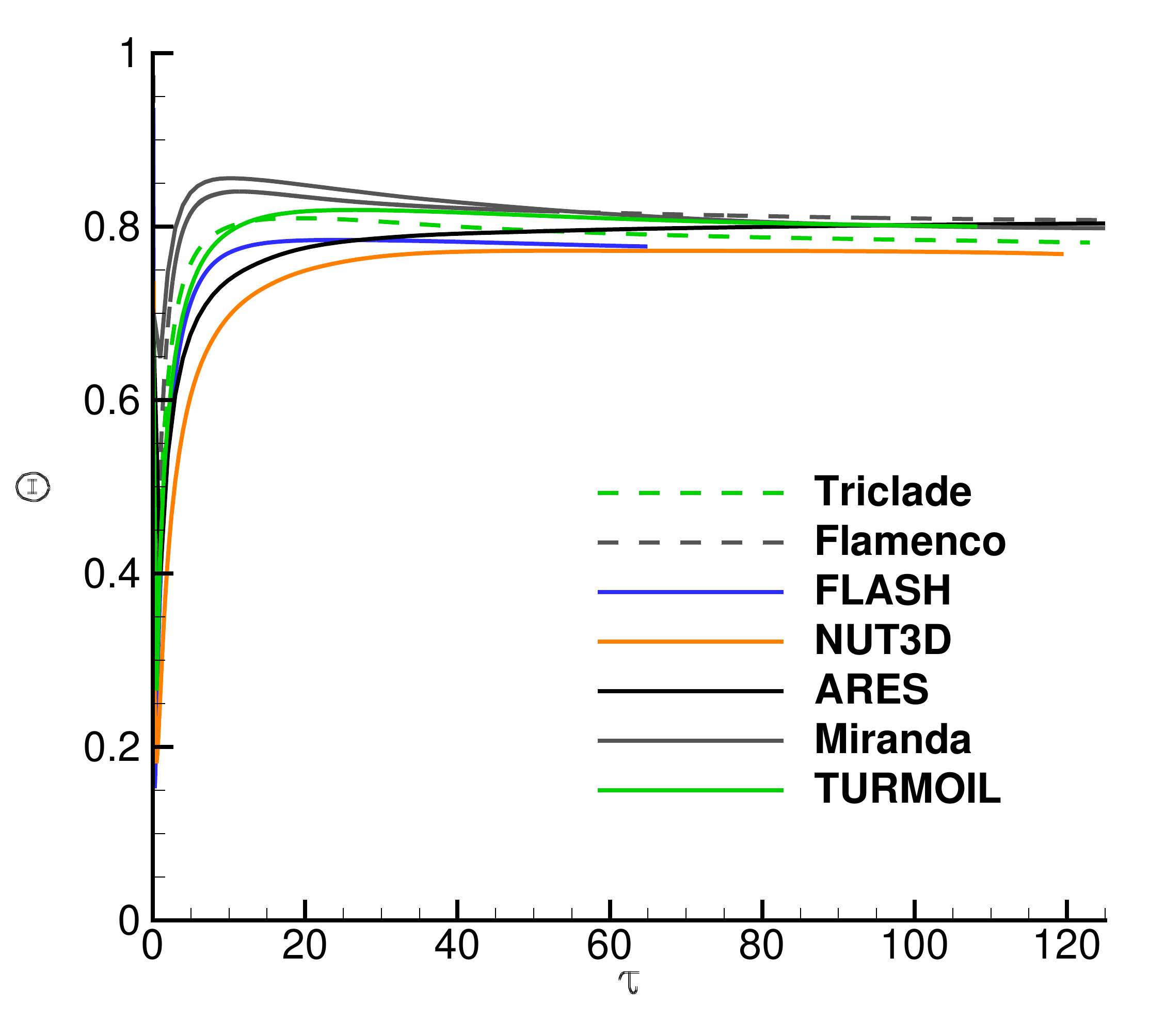}
\includegraphics[width=0.49\textwidth]{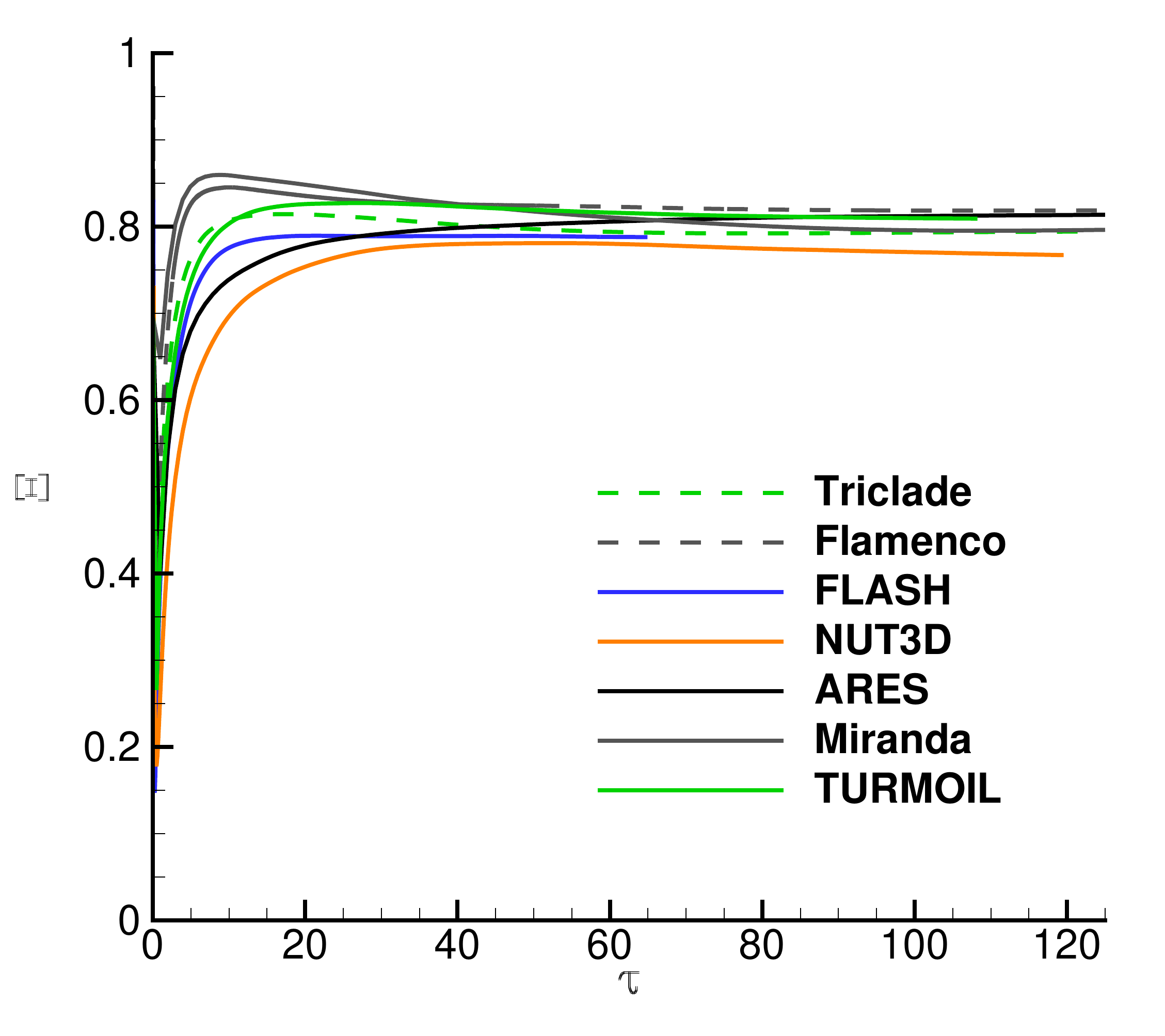}
\includegraphics[width=0.49\textwidth]{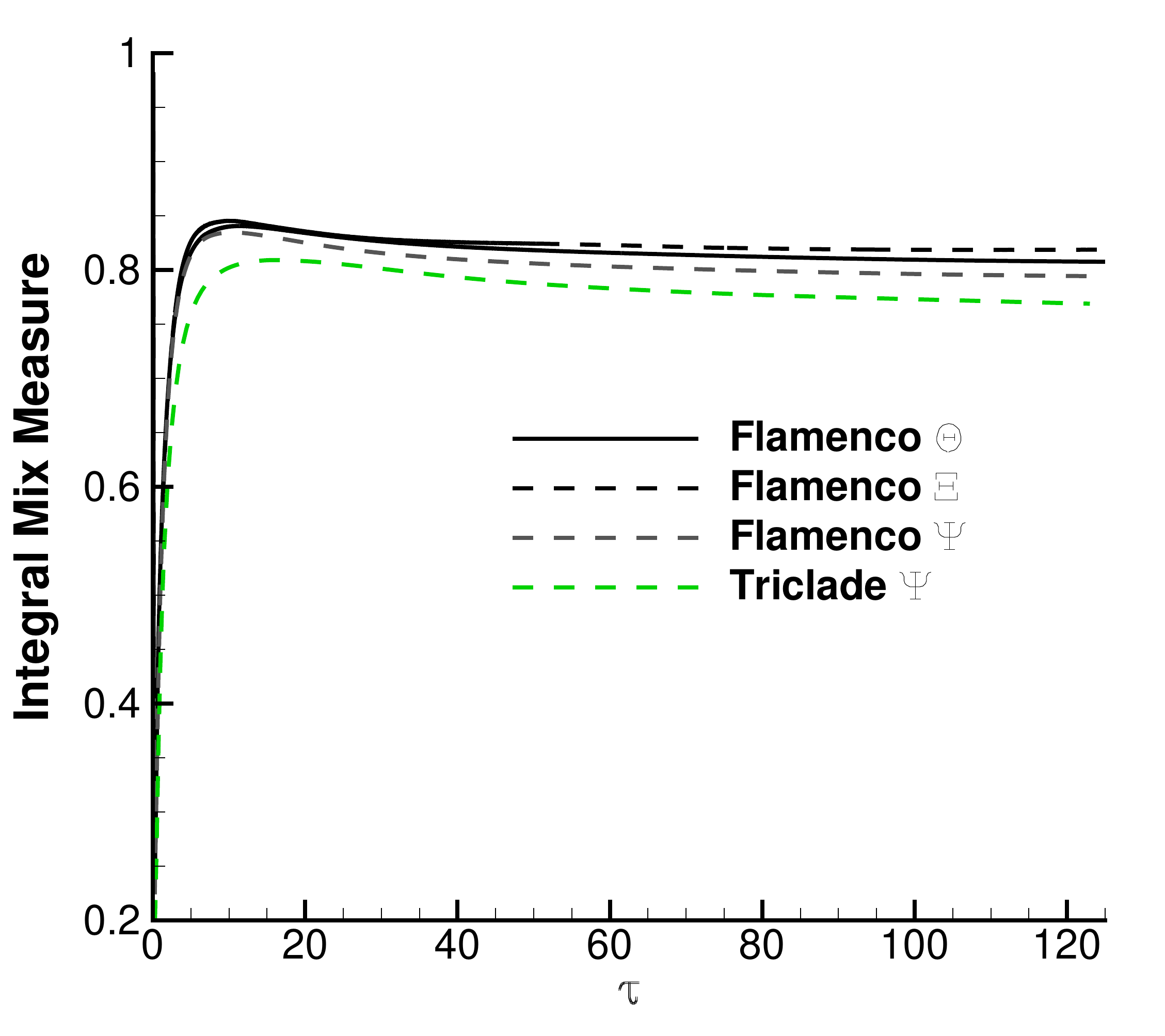}
\caption{Integral mixing measures for the quarter scale problem, and a comparison of three mix measures computed using the results from Flamenco.\label{mixq}}
\end{centering}
\end{figure*}

Figure \ref{mixq} plots the integral mixing measures and $\Theta$, $\Xi$ and shows a comparison of the results of $\Xi$ and $\Theta$ against a recently proposed normalised mixed mass measure \cite{Zhou2016b}:

\begin{equation}
\Psi(t)=\frac{\int\avg{\rho\, Y_1 \,Y_2}dx}{\int \avg{\rho}\avg{Y_1}\avg{Y_2}dx}.
\end{equation}

The integral mix measures $\Theta$ and $\Xi$ at the final time, or $\tau=123$, are detailed in Table \ref{thetasq} for each algorithm. The mean of both are 
\begin{equation}
\Theta=0.792\pm 0.01\,\,\,\Xi=0.800 \pm 0.014,
\end{equation}
\noindent showing excellent inter-algorithm agreement of these late time, quasi-self-similar statistics. As with the standard case, these are consistent with previous results \cite{Weber2012,Lombardini2012,Thornber2010,Oggian2015,Zhou2016,Thornber2016} and close to experimentally measured results of Krivets and Jacobs \cite{Krivets2017} but lower than those of Orlicz {\it et al.},\cite{Orlicz2013}. The maximum difference between the two at the final time for all algorithms is 1.6\%. All algorithms but Ares show decreasing mix measures as time progresses, however the maximum variation for $149\le \tau \le 123$ is $<$1\%, and for $374 \le t \le 748$ Miranda and Flamenco show variations of $<$0.4\%. Normalised mixed mass $\Psi$ mirrors the behaviour of $\Theta$, lying between 0.987$\Theta$ and 0.982$\Theta$ for the entire simulation, and asymptoting to $0.795$ and $0.769$ for Flamenco and Triclade, respectively, at the latest time. 

\begin{figure*}
\begin{centering}
\includegraphics[width=0.49\textwidth]{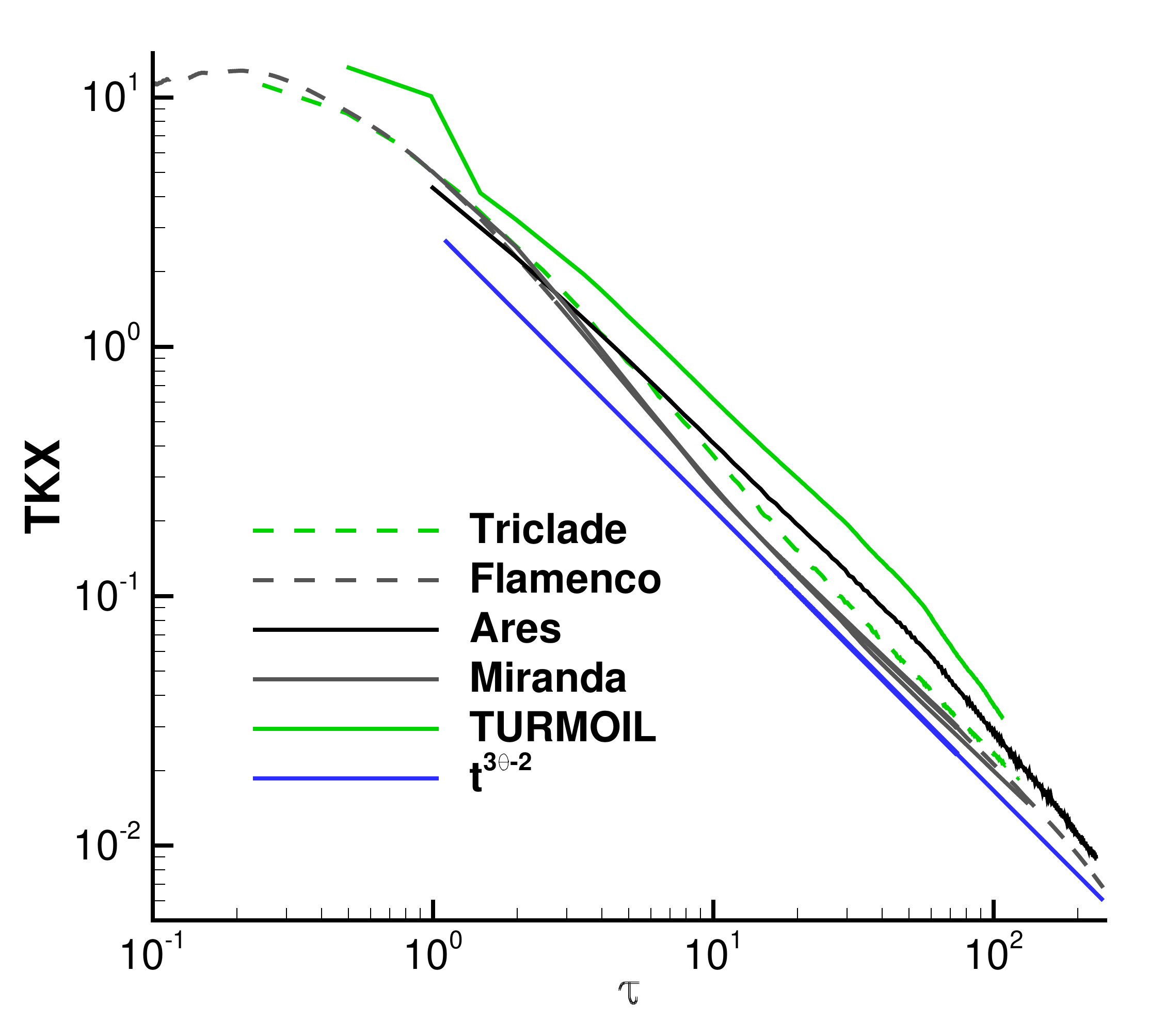}
\includegraphics[width=0.49\textwidth]{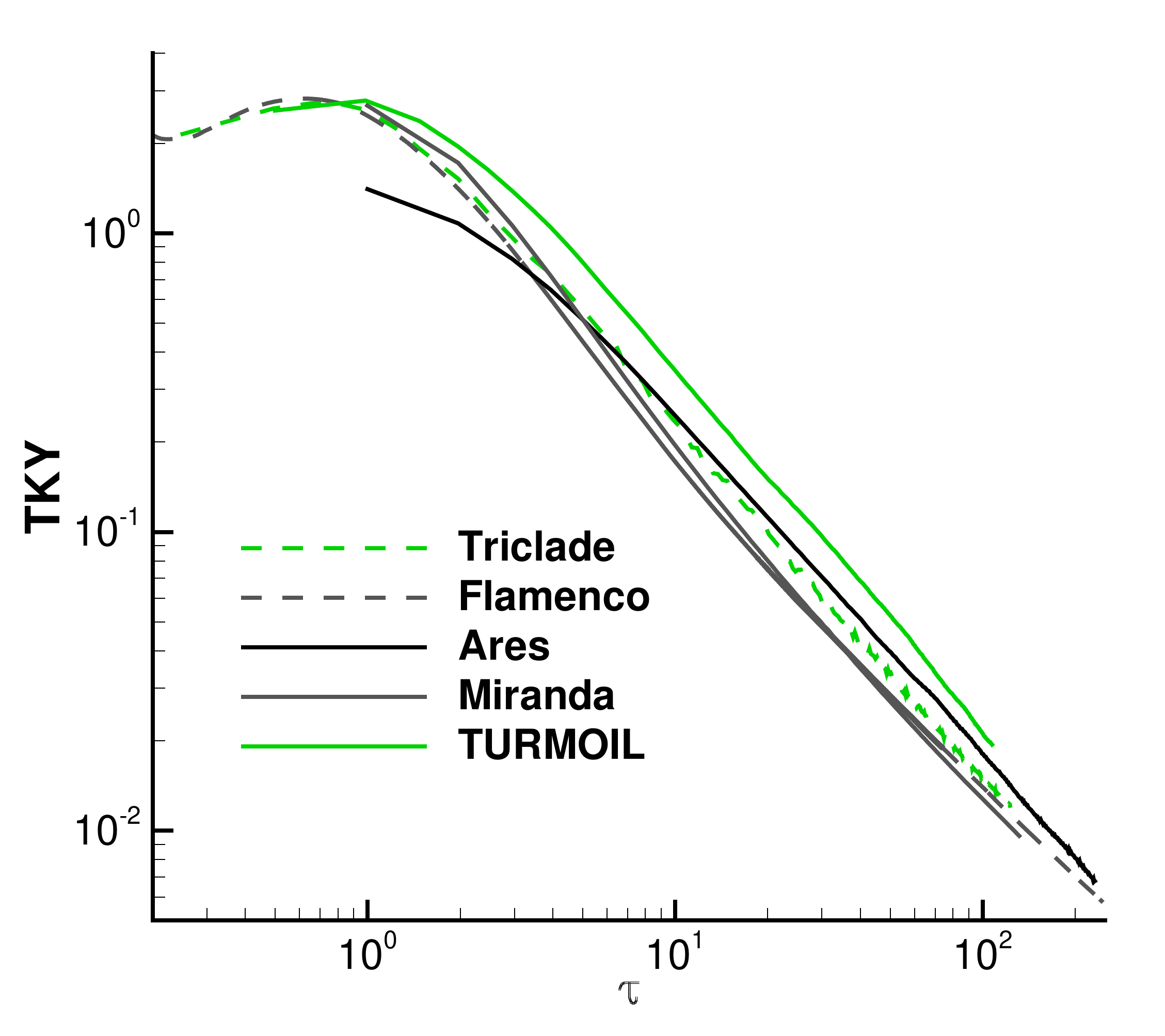}
\includegraphics[width=0.49\textwidth]{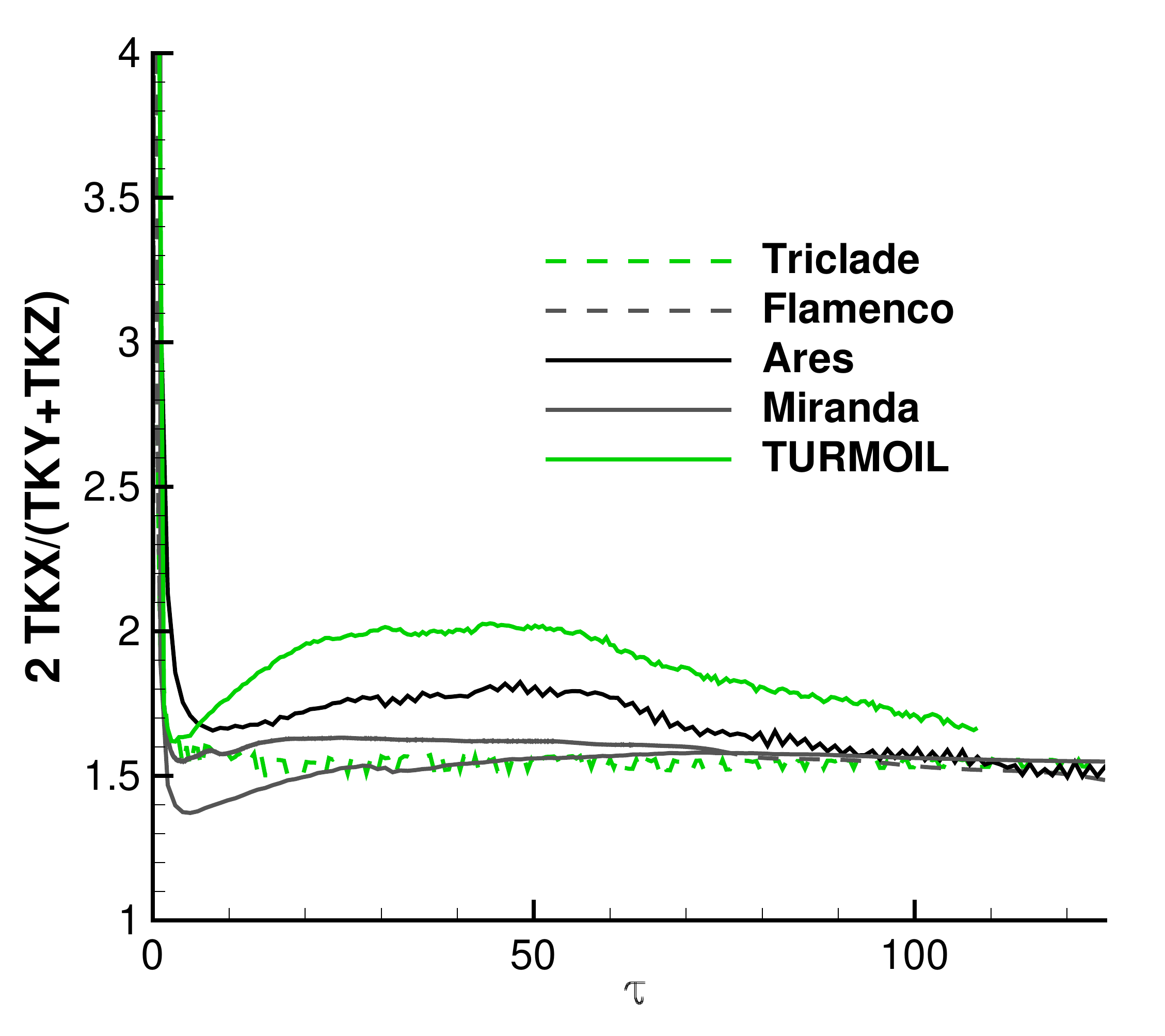}
\caption{Turbulent kinetic energy components and ratio $2{\rm TKX}/({\rm TKY}+{\rm TKZ})$ for the quarter scale problem.\label{TKEq}}
\end{centering}
\end{figure*}

The turbulent kinetic energy components are plotted in Fig. \ref{TKEq}, along with the previously defined anisotropy measure (also detailed at the final time in Table \ref{thetasq}). There is a greater scatter between codes as a result of the marginal grid convergence in this test case. All algorithms show power-law behaviour in time for the kinetic energy components (TKZ is omitted since it is negligibly different from TKY), however the exponents differ by small amounts. Plotted on Figure \ref{TKEq} is the power-law $t^{3\theta-2}$  with the code-averaged $\theta=0.291$ which shows reasonable agreement with Flamenco, Miranda and Triclade at late time $\tau >24.6$. 

The observed anisotropy in this case increases following the minimum observed for the standard case at the latest time (Fig. \ref{TKE}), then ranges from $1.49$ to $1.66$ over all algorithms. From this result, it can be concluded that the decrease in anisotropy at `late' times in the standard case was a transient phenomena, as a minimum is reached in the quarter scale problem at $\tau \approx 6$ which is close to the final time of the standard problem. This recovery can also be seen in the results of Oggian {\it et al.} albeit at a shorter dimensionless time \cite{Oggian2015}. There is no return to isotropy and the anisotropy is maintained at a nearly constant value after the initial transients within the simulation time scale.

\begin{figure*}
\begin{centering}
\subfigure[]{\includegraphics[width=0.3\textwidth]{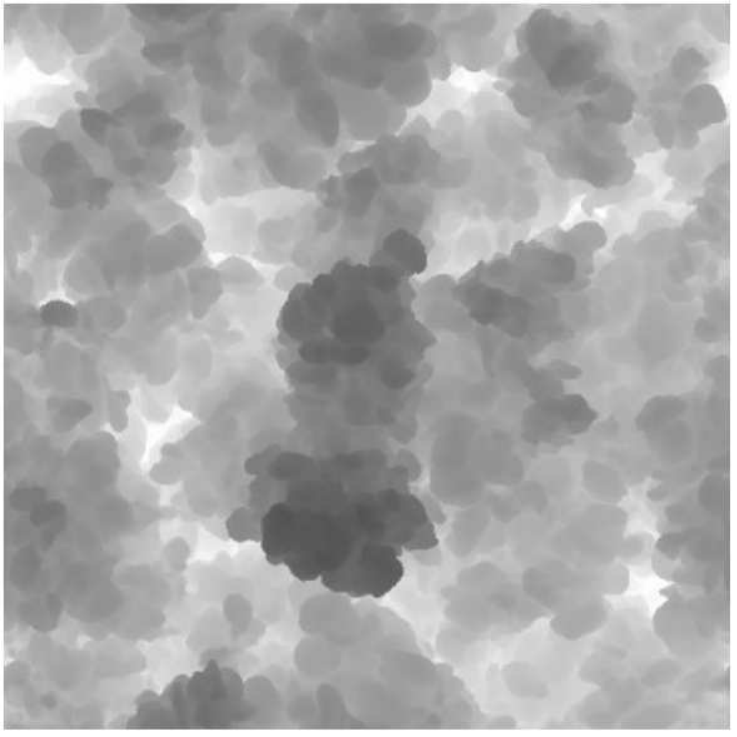}}
\subfigure[]{\includegraphics[width=0.3\textwidth]{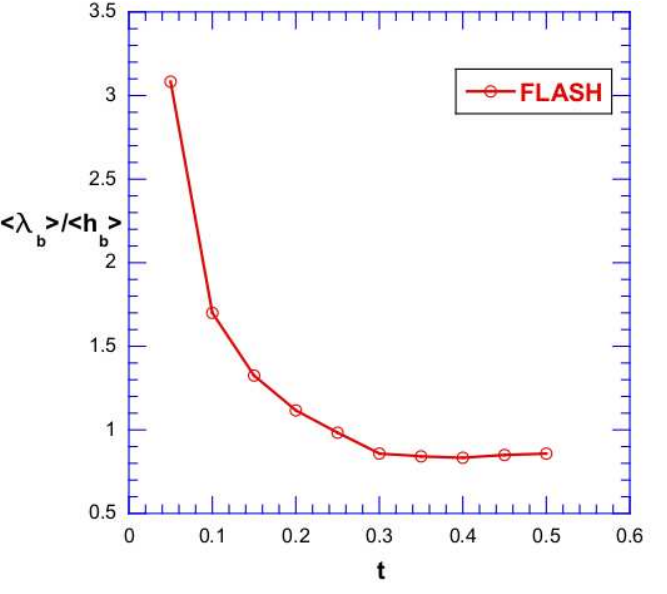}}
\subfigure[]{\includegraphics[width=0.3\textwidth]{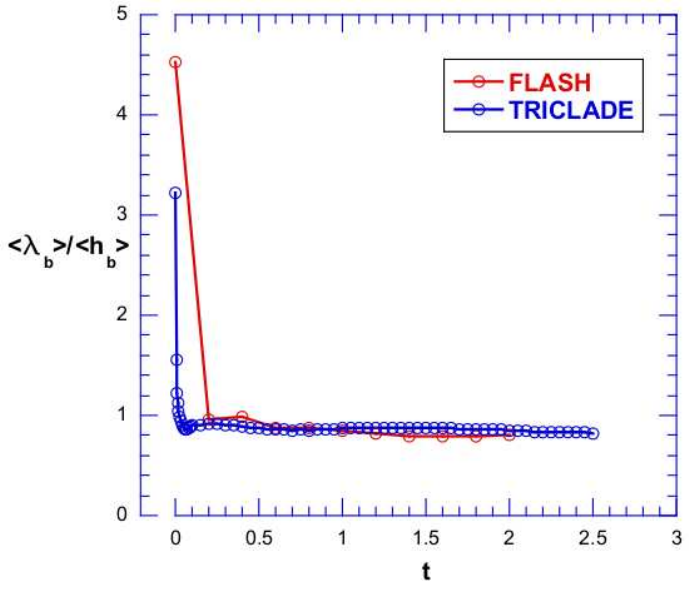}}
\caption{(a) Visualization of the bubble front at late time $\tau=6.2$ ($t = 0.5 $ s) from the standard scale problem using FLASH. Time evolution of the self-similarity ratio  $\avg{\lambda_b}/\avg{h_b}$ 􀵘 from (b) the standard problem (FLASH) and (c) the quarter scale problem (FLASH and Triclade).\label{bub}}
\end{centering}
\end{figure*}

The ratio $\avg{\lambda_b}/\avg{h_b}$  of the dominant bubble wavelength to the bubble front amplitude is a measure of self-similarity in the flow. For self-similar flows, $ \avg{\lambda_b}/\avg{h_b}$ saturates to a constant value, but may still depend weakly on other factors such as the initial conditions or the Atwood number of the flow. The self-similarity ratio is also an input parameter in buoyancy-drag models \cite{Alon1995,Srebro2003}, where it is used to evolve the lateral length scales beyond the point of nonlinear transition in the flow. In Fig. \ref{bub} (a), the late-time  $\tau=6.2$ ($t = 0.5$ s) bubble front from FLASH simulations of the standard problem is plotted, visualized here as the x-heights of the iso-surfaces of the 1\% volume-fraction. The autocorrelation procedure developed in \cite{Dimonte2000,Ramaprabhu2005} is adopted to extract the dominant bubble wavelengths from such images. Briefly, the autocorrelation function $\eta(x,y)$ of the bubble front $Z_b(x,y)$ is computed according to 

\begin{equation}
\eta(x,y)=\frac{\sum_{x',y'} \left(Z_b(x',y')-\avg{Z_b}\right)\left(Z_b(x'+x,y'+y)-\avg{Z_b}\right)}{\sum_{x',y'} \left(Z_b(x',y')-\avg{Z_b}\right)^2}.
\end{equation}

\noindent The dominant bubble diameter is then obtained as the radial length, where the azimuthally-averaged function $\avg{\eta}_b$ drops below a threshold value of $0.3$ (calibrated from several test images). The corresponding bubble wavelength is then obtained from the relation $\avg{\lambda_b}\approx \avg{D_b}(\rho_h+\rho_L)/\rho_h$, where the post-shock densities are used \cite{Daly1967}.

The results of this analysis are shown in  Fig. \ref{bub} (b) and (c) for the standard problem and the quarter-scale problem, respectively. For the standard problem, the analysis was performed for FLASH simulation data, and shows a late-time  $\tau=3.69$ ($t > 0.3$ s) saturation of  to a nearly constant value of $0.85\pm 0.01$. This time-to-saturation may reasonably be taken as the time required by the dominant wavelength in the initial wavepacket to reach nonlinearity, thus triggering the onset of self-similarity for the flow. For the quarter-scale problem in Fig. \ref{bub} (c), the analysis was performed on FLASH and Triclade data, but they both obtain very similar saturated values of $0.85\pm 0.07$ and $0.86\pm 0.02$ respectively. Furthermore, since the dominant wavelength from the initial condition is smaller for the quarter-scale problem, the time required by that mode to reach nonlinearity is also shorter as expected from the $W\propto 1/\lambda_{min}$ scaling.


\begin{figure*}
\begin{centering}
\includegraphics[width=0.49\textwidth]{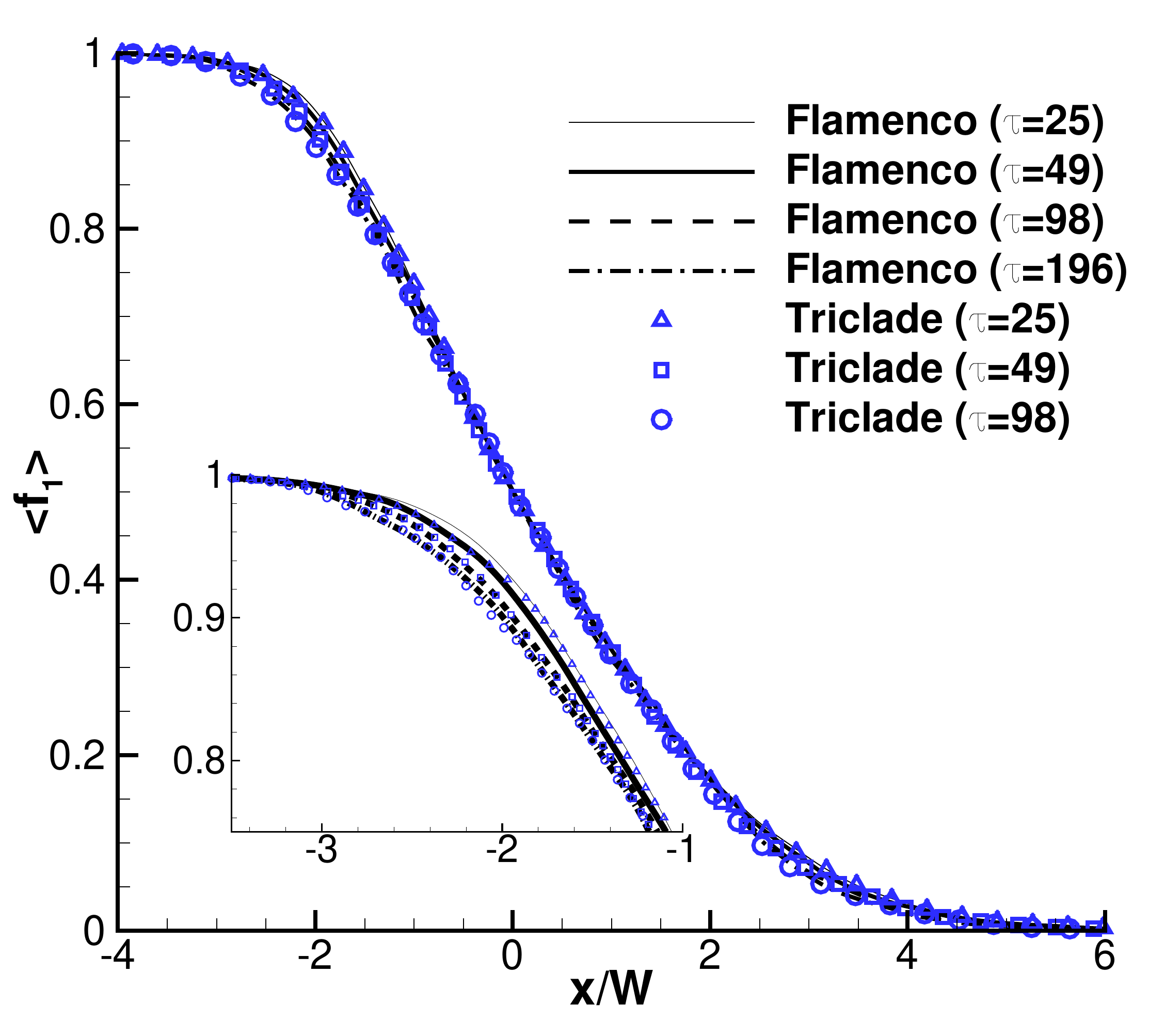}
\includegraphics[width=0.49\textwidth]{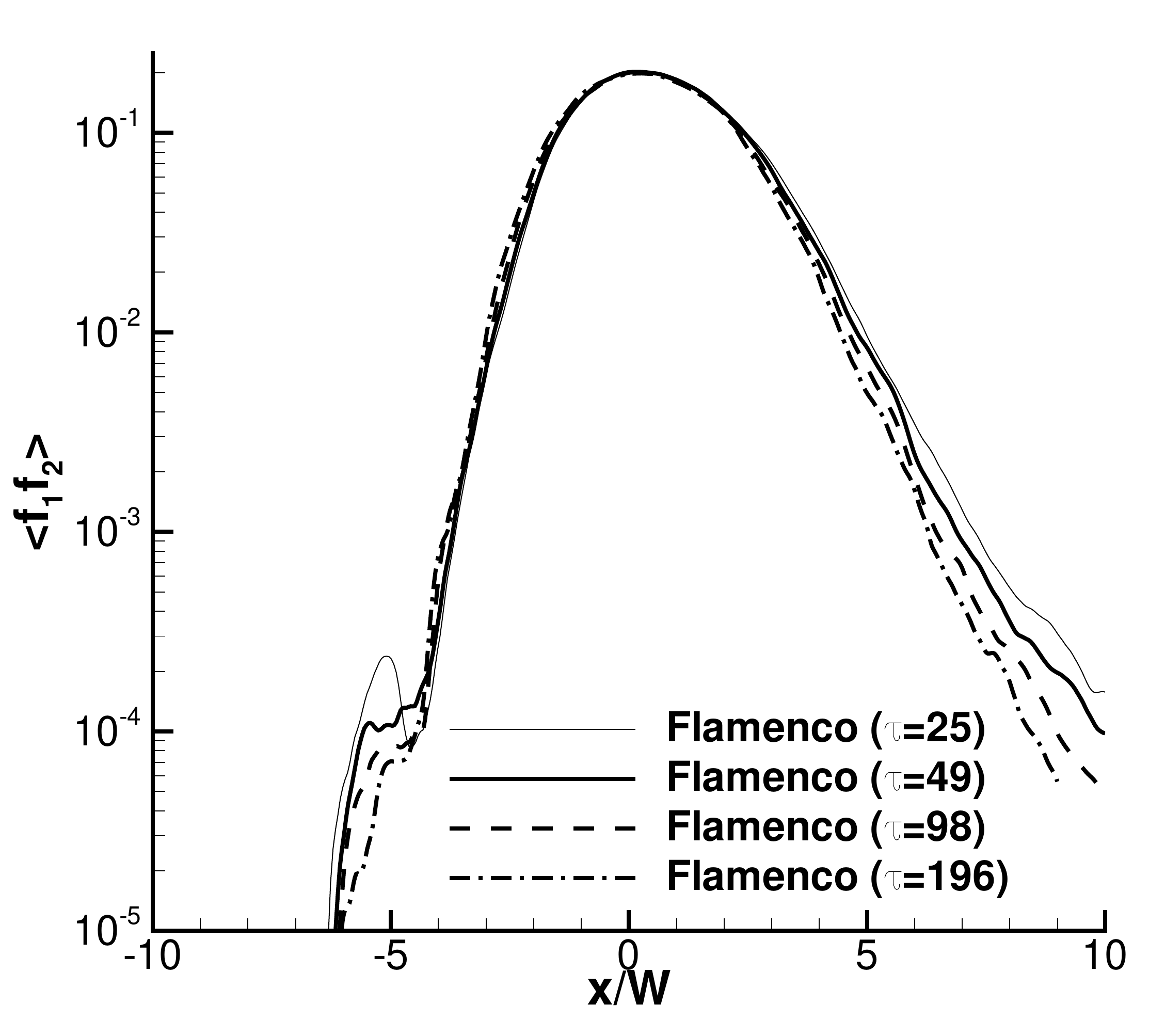}
\caption{Plane-averaged volume fraction $\langle f_1 \rangle$ plotted against scaled distance $x/W$ for several times for the Flamenco and Triclade simulations (left), and the equivalent for $\langle f_1 f_2 \rangle$ for Flamenco only (right) \label{scaledpaf1}}
\end{centering}
\end{figure*}

Finally,  $\langle f_1 \rangle$ and $\langle f_1 f_2 \rangle$ are plotted against scaled distance $x/W$ for several times for Flamenco and Triclade in Figure \ref{scaledpaf1}. By scaling the spatial dimension by the layer width, this figure illustrates the validity of the assumption that the mixing layer profiles can be collapsed via a single length scale. Should the bubbles and spikes evolve with a differing power law (different $\theta^b$ and $\theta^s$), there should be a noticeable lack of collapse under a single scaling. Examining the profiles of $\langle f_1 \rangle$ show that the agreement between Flamenco and Triclade is very good, and that both are showing a slight lack of self-similarity, most notably a smoothing of the bubble side (close to $\langle f_1 \rangle=1$) at later $\tau$, as shown in the inset of Figure \ref{scaledpaf1} (left). Examining the $\langle f_1 f_2 \rangle$ profiles show that the the extremes of the spike side ($x>0$), and to a lesser extent, the extremes of the bubble side ($x<0$) have not reached self-similarity. Despite this, the core of the mixing layer  $-4 \le x/W \le 4$ is reasonably collapsed at all times indicating that a single length scale and, equivalently a single $\theta$, is a reasonable approximation.

\section{Conclusions \label{concl}}

A carefully designed narrowband multimode Richtmyer--Meshkov diffuse interface test case was proposed and results were obtained for eight independent algorithms representing a broad range of numerical approaches. A new analytical estimation of the initial linear growth of a narrowband multimode perturbation was also presented. 

A detailed study of the integral measures for a standard case for which grid convergence can be demonstrated has been presented, showing excellent agreement for all algorithms at the highest grid resolution. Code-averages and standard deviations were then computed, enabling an uncertainty estimate not possible without multiple independent simulations. This standard case demonstrated that the transitional RM problem has width $W=0.807(t-0.0309)^{0.219}$ ($0.210<\theta<0.228$) for diffuse interface algorithms, mix measures $\Theta=0.769\pm 0.0225$, $\Xi= 0.767\pm 0.0229$, and a power-law decay of fluctuating kinetic energy approximately proportional to $t^{3\theta-2}$. It is highly anisotropic, where the shock direction kinetic energy is 43\% greater than each in-plane component. Spectra demonstrated a short power-law range of the form $\approx k^{-1/.3}$, shallower than expected from previous results, although the mixing layer had not achieved self-similarity. 

Based on these results, a second problem was run by a subset of the codes with one quarter the physical length scales, but the same computational domain size, enabling the simulations to reach more than eight times the dimensionless time. From the converged algorithms, the growth exponent was $\theta=0.292 \pm 0.009$ computed using non-linear regression, however a time dependent $\theta$ indicated that the exponent was decreasing with time. This value of $\theta$ was insensitive to the choice of width measure when comparing against $h$ or normalised mixed mass $\mathcal{M}$. The integral mix measures were $\Theta=0.792\pm 0.014$ and $\Xi=0.800 \pm 0.014$, and whilst not constant during the simulation, vary by $<$2.5\% in the period during which the growth exponent is fit. The fluctuating kinetic energy showed greater uncertainty in this case, however all algorithms clearly showed a minimum anisotropy at early times, followed by a marginal increase in anisotropy at later times to $1.49\le 2{\rm TKX}/({\rm TKY}+{\rm TKZ})\le 1.66$.

The analytical estimation of the linear growth rate of a multimode narrowband perturbation was found to be in excellent agreement with numerical data, and two models of the growth rate are presented and validated, which merge linear and non-linear growth relations into a single expression. 

Thus this study has quantified the value of $\theta$ for a narrow band mixing layer, including clear uncertainty bands based on an ensemble of simulations with state-of-the-art algorithms. The RMI layer is predicted to have late time anisotropy for a wide range of algorithms for the time scales simulated. The value of $\theta$ reflects a likely lower-bound on the growth rate due to a multimode pure RM instability in the planar case, whereas the mixing parameters reflect an upper bound. Plotting the plane averaged volume fraction profiles against $x/W$ indicate that the bulk of the layer collapses self-similarly through a single length-scale, implying that there is a single $\theta$, or that $\theta^b\approx \theta^s$ for this problem. Furthermore, the proposed case is now sufficiently documented and analysed to benchmark future algorithms for shock-induced turbulent mixing. 

An important conclusion of this study is that with current computational power, it is very difficult to converge the integral properties over a number of algorithms. Thus when examining higher order moments such as probability distribution functions and spectra, great care must be taken to ensure that they are well converged such that the underlying physics can be correctly elucidated. 

\section*{Acknowledgments}

The authors would like to strongly acknowledge Dr. Karnig Mikaelian who suggested this collaboration at the International Workshop on the Physics of Compressible Turbulent Mixing in 2014. This research was supported under Australian Research Council's Discovery Projects funding scheme (project number DP150101059). The author would like to acknowledge the computational resources at the National Computational Infrastructure through the National Computational Merit Allocation Scheme which were employed for all cases presented here. FLASH was developed by the DOE-sponsored ASC/Alliance Center for Astrophysical Thermonuclear Flashes at the University of Chicago. P. Ramaprabhu was partially supported by the (U.S.) Department of Energy (DOE) under Contract No. DE-AC52-06NA2-5396. Results from TURMOIL are \copyright  British Crown Owned Copyright 2017/AWE. This work was performed under the auspices of the U.S. Department of Energy by Lawrence Livermore National Laboratory under Contract DE-AC52-07NA2734.

\bibliographystyle{unsrt}
\bibliography{bibliography}

\begin{appendix}

\section{Derivation of Initial Conditions \label{deriv}}

The perturbed interface $A(y,z)$ is determined according to a specified power spectrum and standard deviation following previous studies \cite{Youngs2004,Thornber2010} . The power spectrum is 

a power spectrum of the form

\begin{equation}
P(k)=
\left\{ \begin{array}{cc}
C & k_{\rm min}<k<k_{\rm max}\\
0 & \text{otherwise}
\end{array} \right.,
\end{equation}

\noindent where $k=\sqrt{k_y^2+k_z^2}$ is the wave vector of the perturbation. First, the power spectrum is rewritten as a two-dimensional
power spectrum in wave space

\begin{equation}
\sigma^2  =\int_{0}^{\infty} P(k)dk=  \int_{-\infty}^{\infty}\int_{-\infty}^{\infty}\frac{1}{2\pi}\frac {P(k)}{k} dk_{y}dk_{z}.
\label{eqn: SurfacePowerSpectrum}  
\end{equation}

 The specified
 power spectrum gives an equivalent amplitude $a({\mathbf k})$ in wave space:

\begin{equation}
a({\mathbf k}) \propto \sqrt {\frac {P(k)}{k}}.
\label{eqn: A(k)}
\end{equation}

\noindent To initialise modes within a certain band, the inverse Fourier transform
of this relation can be taken, noting that the amplitude is a real function:

\begin{equation}
a(y,z)=\sum_{m,n=-N}^{N}\text{Re}\left\{ c_{m,n}\exp\left[ik_{0}\left(my+nz\right)\right]\right\}, \label{eqn: InvFourSurface}\end{equation}

\noindent where $k_{\rm max}=2N \pi/L $, the cross-section is $L\times
L$ and $c_{m,n}$ are the amplitudes the
mode with wavenumber $m$ in the $y$ direction and $n$ in the $z$ direction. To satisfy the given power spectrum, Eq. (\ref{eqn: InvFourSurface})
can be simplified considerably by expanding using the Euler formula,
and considering only the real part:

\begin{equation}
A(y,z)=\sum_{m,n=-k_{\rm max}}^{k_{\rm max}}r_{m,n}\cos\left[k_{0}\left(my+nz\right)+\phi_{m,n}\right],
\end{equation}

\noindent where $c_{m,n}=c_1+ic_2$, thus $r_{m,n}=\sqrt{c_1^{2}+c_2^{2}}$ and
$\phi_{m,n}=\tan\left(c_2/c_1\right)$. Next, the cosine term is expanded using
trigonometric relations giving

\begin{eqnarray}
A(y,z)  = 
\sum_{m,n=0}^{N}&a_{m,n}\cos\left(k_{0}my\right)\cos\left(k_{0}nz\right)+ b_{m,n}\cos\left(k_{0}my\right)\sin\left(k_{0}nz\right)+\nonumber\\
&c_{m,n}\sin\left(k_{0}my\right)\cos\left(k_{0}nz\right)+d_{m,n}\sin\left(k_{0}my\right)\sin\left(k_{0}nz\right).
\label{eqn: FourierExpansion}
\end{eqnarray}

To initialise a random field, $a_{m,n}$, $b_{m,n}$.... must be
chosen from a distribution randomly so that the standard deviation is
proportional to the Fourier coefficients in Equation (\ref{eqn: A(k)}). The random
variables are selected from a Gaussian distribution to give the
desired mean amplitude at a given wave number $k_{m,n}$, in this case

\begin{equation}
\frac{1}{4}(\bar a^2_{m,n}+\bar b^2_{m,n}+\bar c^2_{m,n}+\bar d^2_{m,n})=
  \frac{1}{2\pi}\,\frac {P(k_{m,n})}{k_{m,n}}\, \Delta k_{y}\Delta k_{z},
\label{discretemn}
\end{equation}

\noindent where $k_{m,n}=\sqrt{k^2_{ym}+k^2_{zn}}$, the wavenumber in
the $y$ direction is $k_{ym}=2\pi m/L$, the wavenumber in the
$z$-direction $k_{zn}=2\pi n/L$ and $L$ is the width of domain, which
in this case is assumed to have a square cross-section. 

The random seeds are generated using a Mersenne Twister routine which is deterministic. The initial coefficients $a_{m,n}$, $b_{m,n}$.... are rescaled such that the initial perturbation has an rms amplitude of $0.1 \lambda_{min}$ when perfectly resolved.

\section{Grid Convergence \label{gc}}

This section shows the grid  convergence for each of the algorithms on the standard problem where results for more than one grid level were provided. Grid convergence is shown for two properties which are functions of the large scales and thus are expected to converge ($W$ and ${\rm TKX}$) and one which is a function of the small scales so will be more sensitive to grid spacing ($\Theta$).

\begin{figure*}
\begin{centering}
\includegraphics[width=0.325\textwidth]{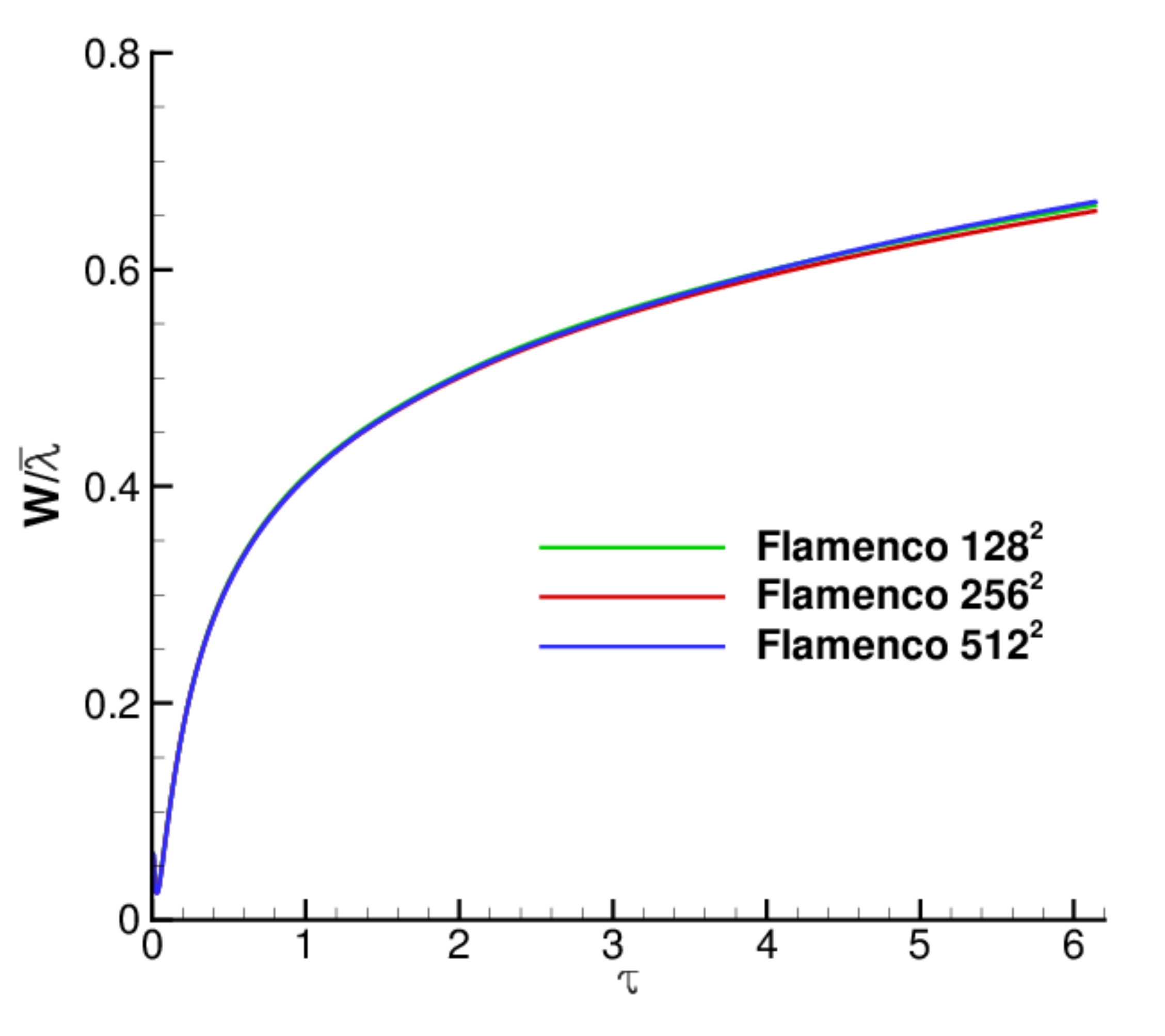}
\includegraphics[width=0.325\textwidth]{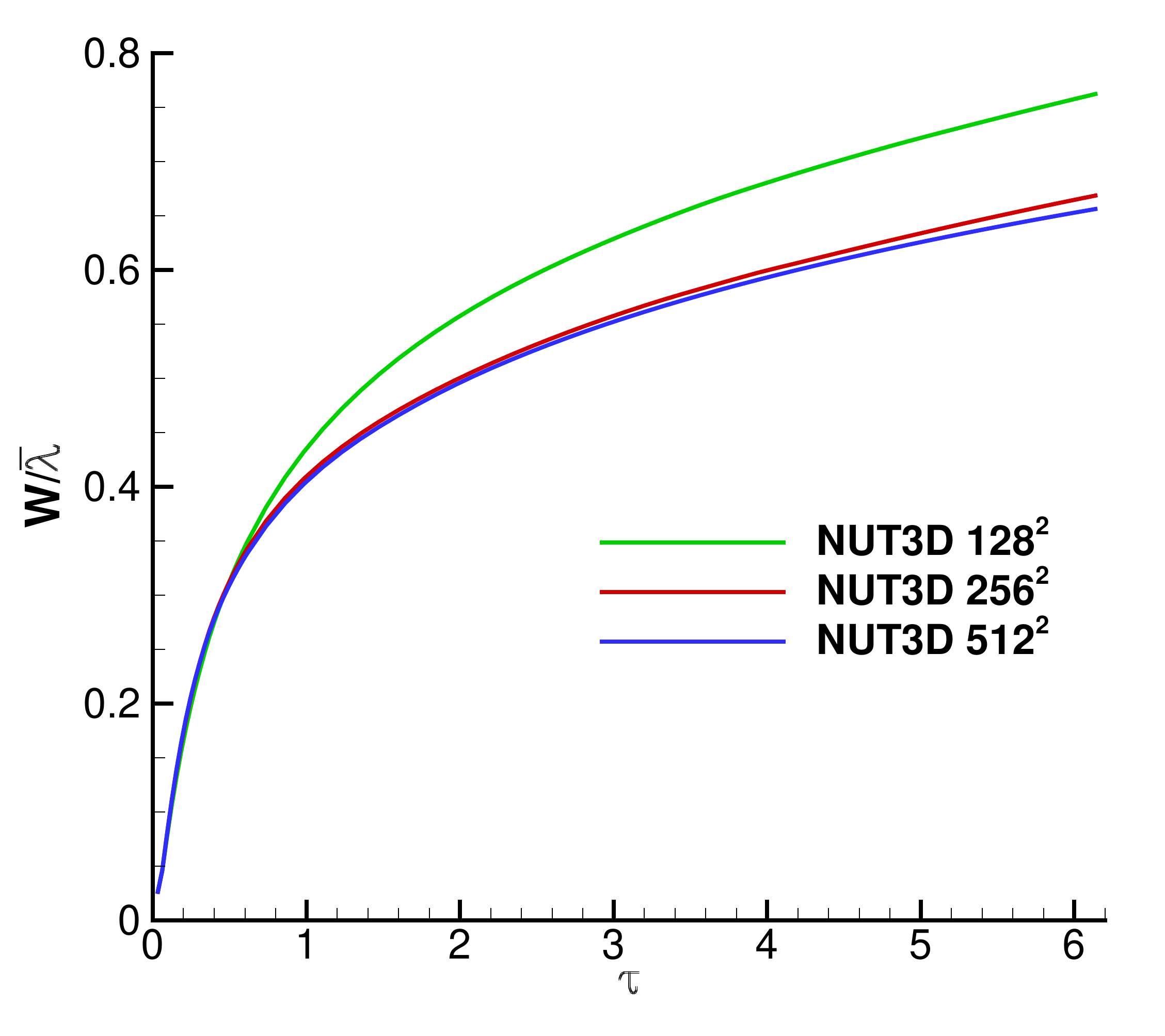}
\includegraphics[width=0.325\textwidth]{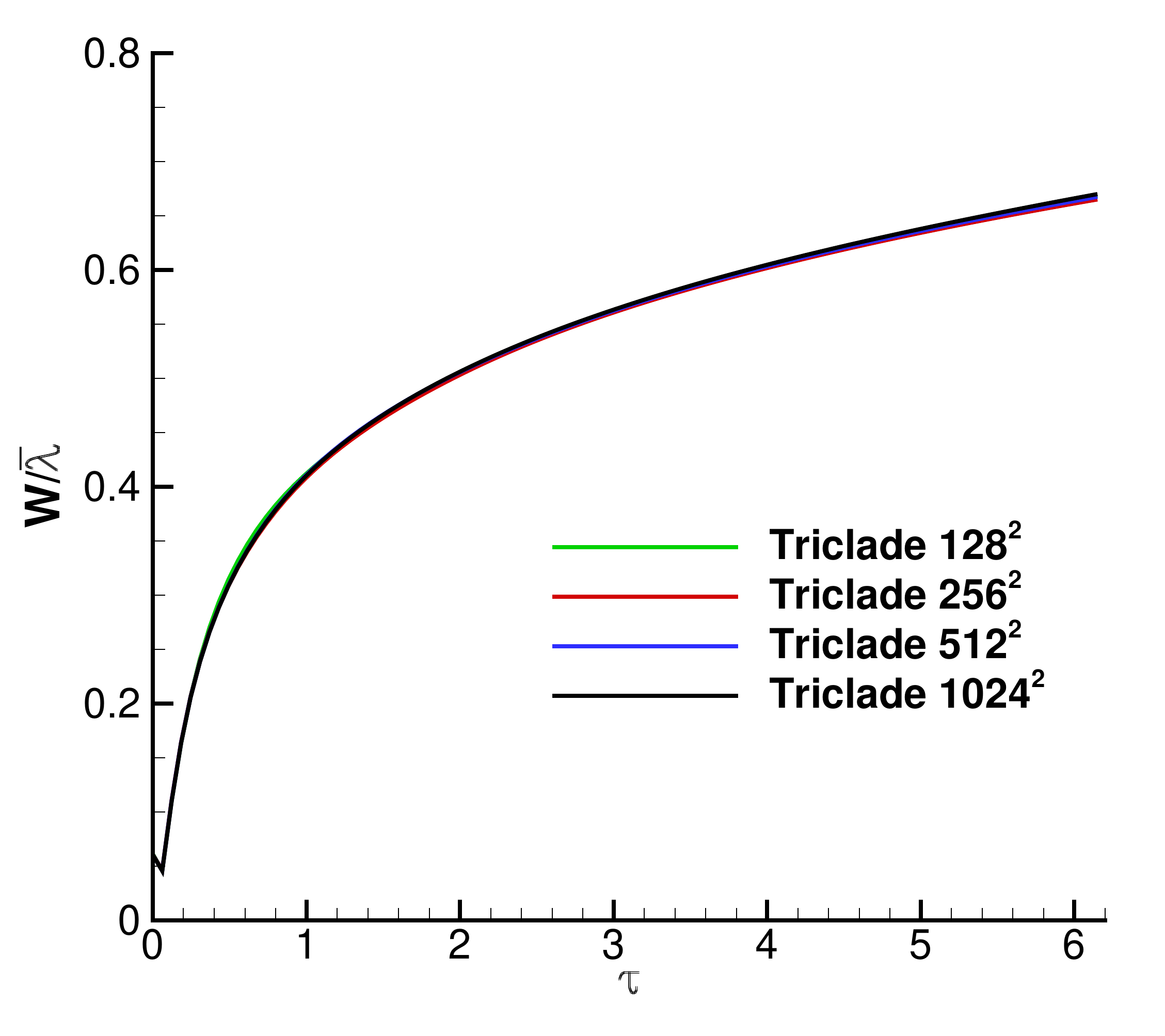}
\includegraphics[width=0.325\textwidth]{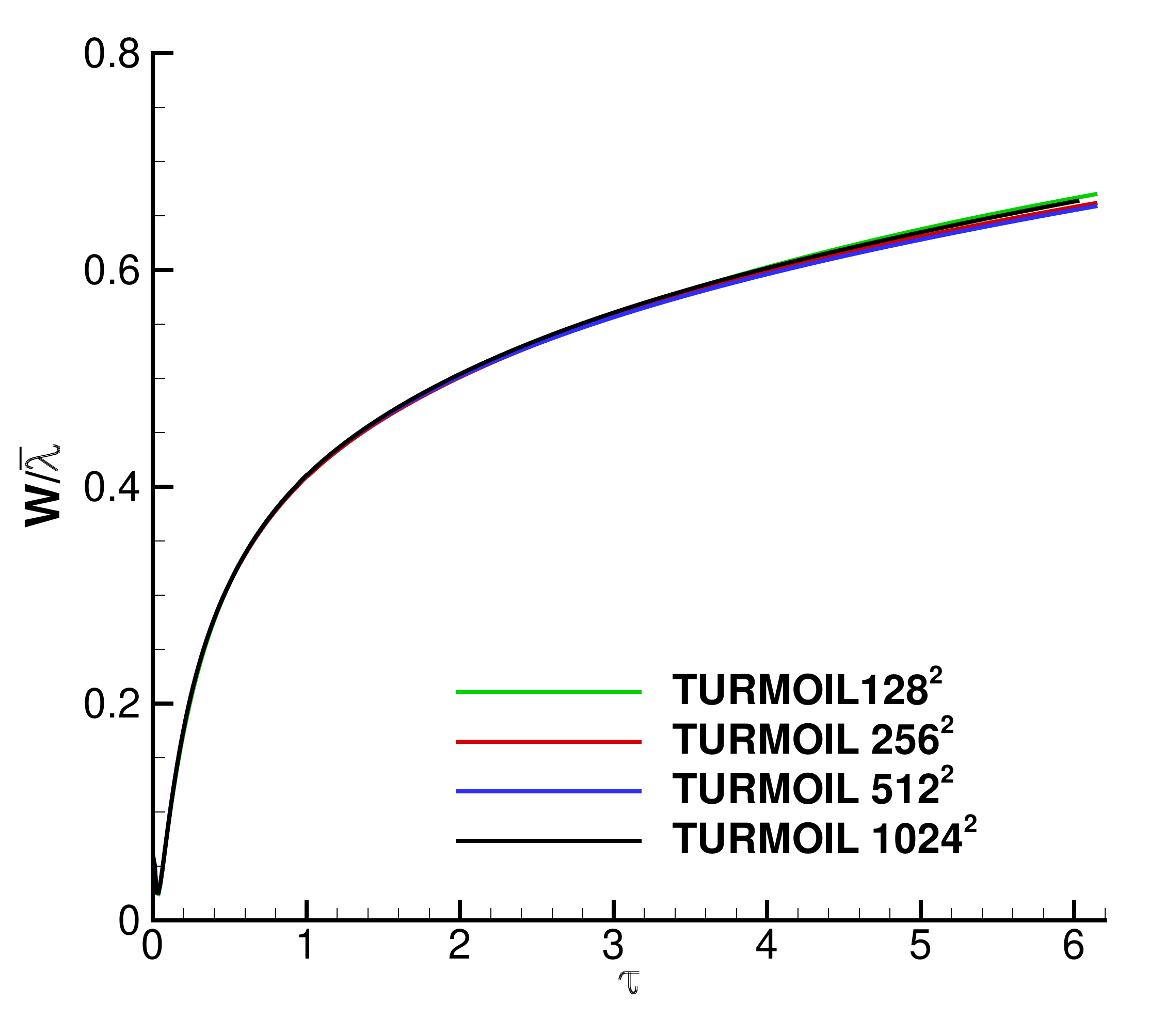}
\includegraphics[width=0.325\textwidth]{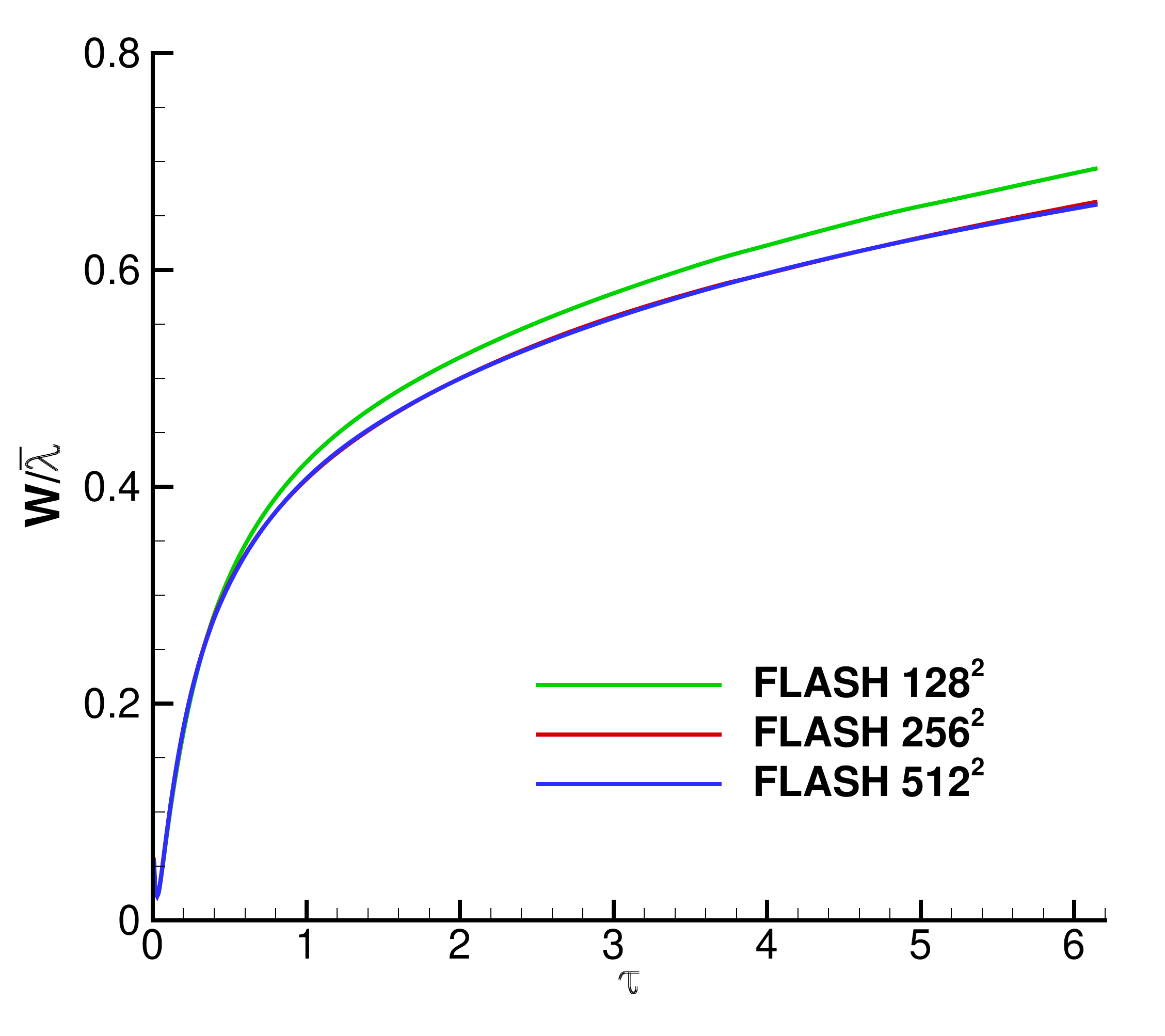}
\includegraphics[width=0.325\textwidth]{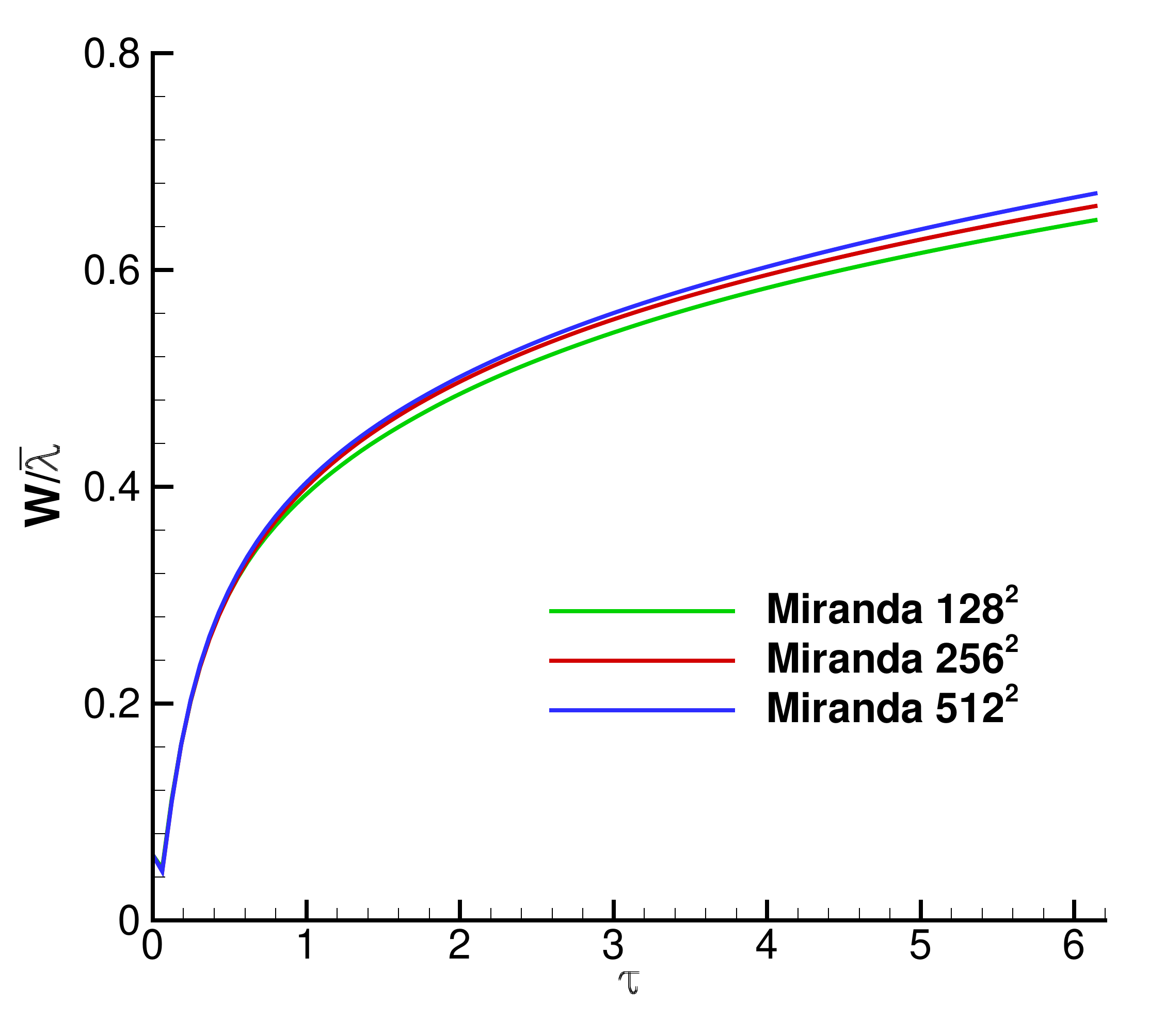}
\includegraphics[width=0.325\textwidth]{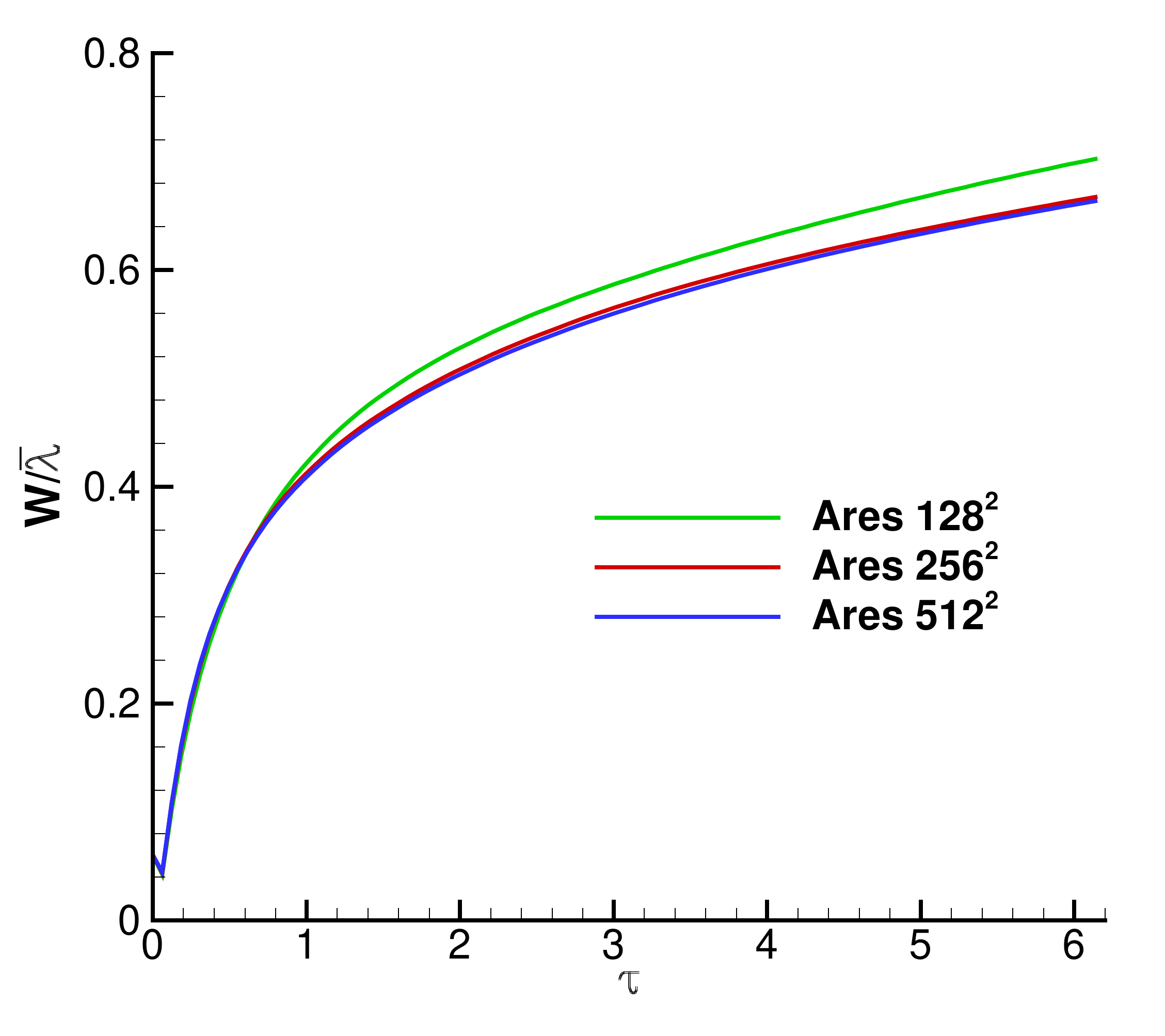}
\caption{Convergence of W. \label{convW}}
\end{centering}
\end{figure*}

\begin{figure*}
\begin{centering}
\includegraphics[width=0.325\textwidth]{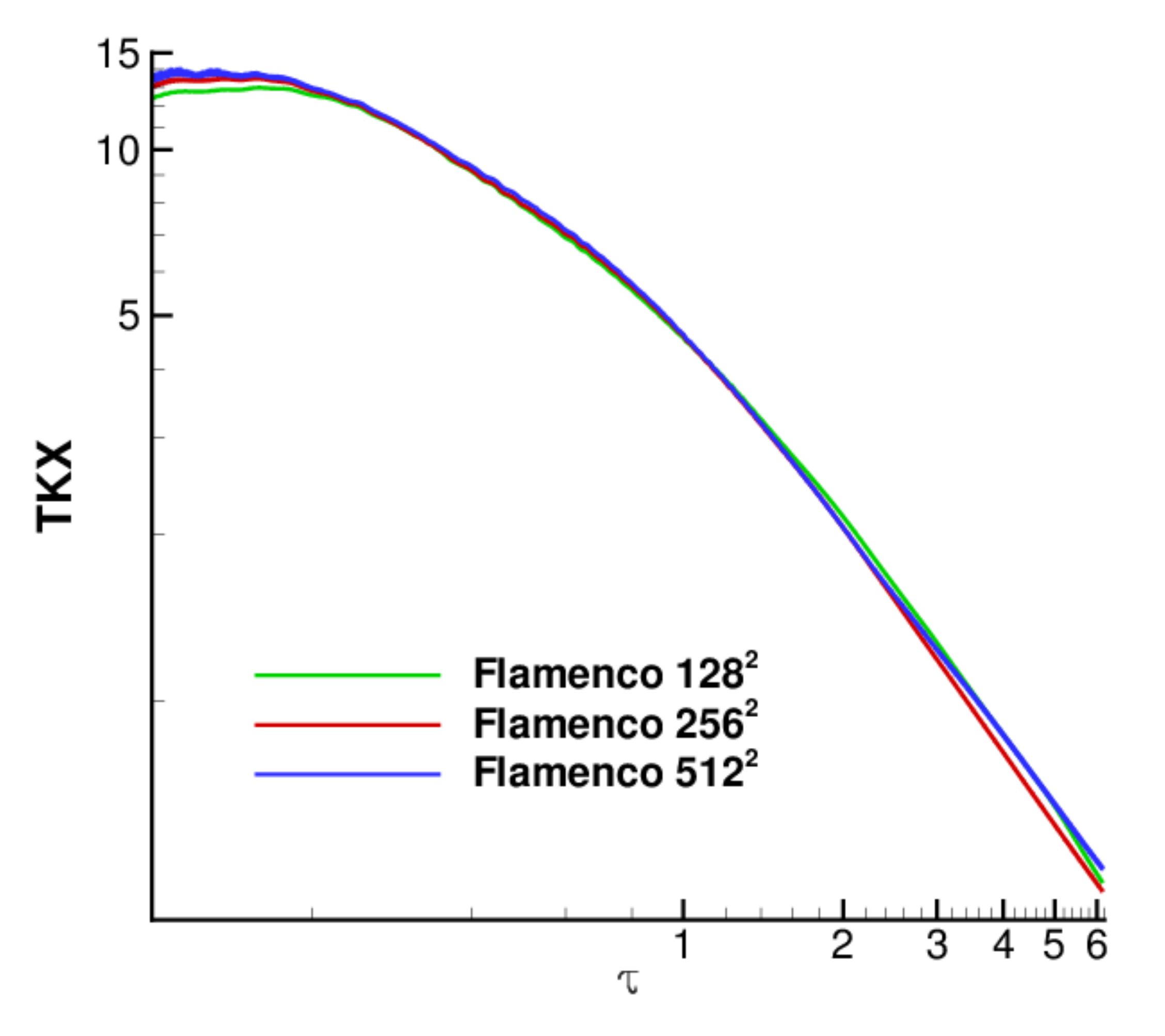}
\includegraphics[width=0.325\textwidth]{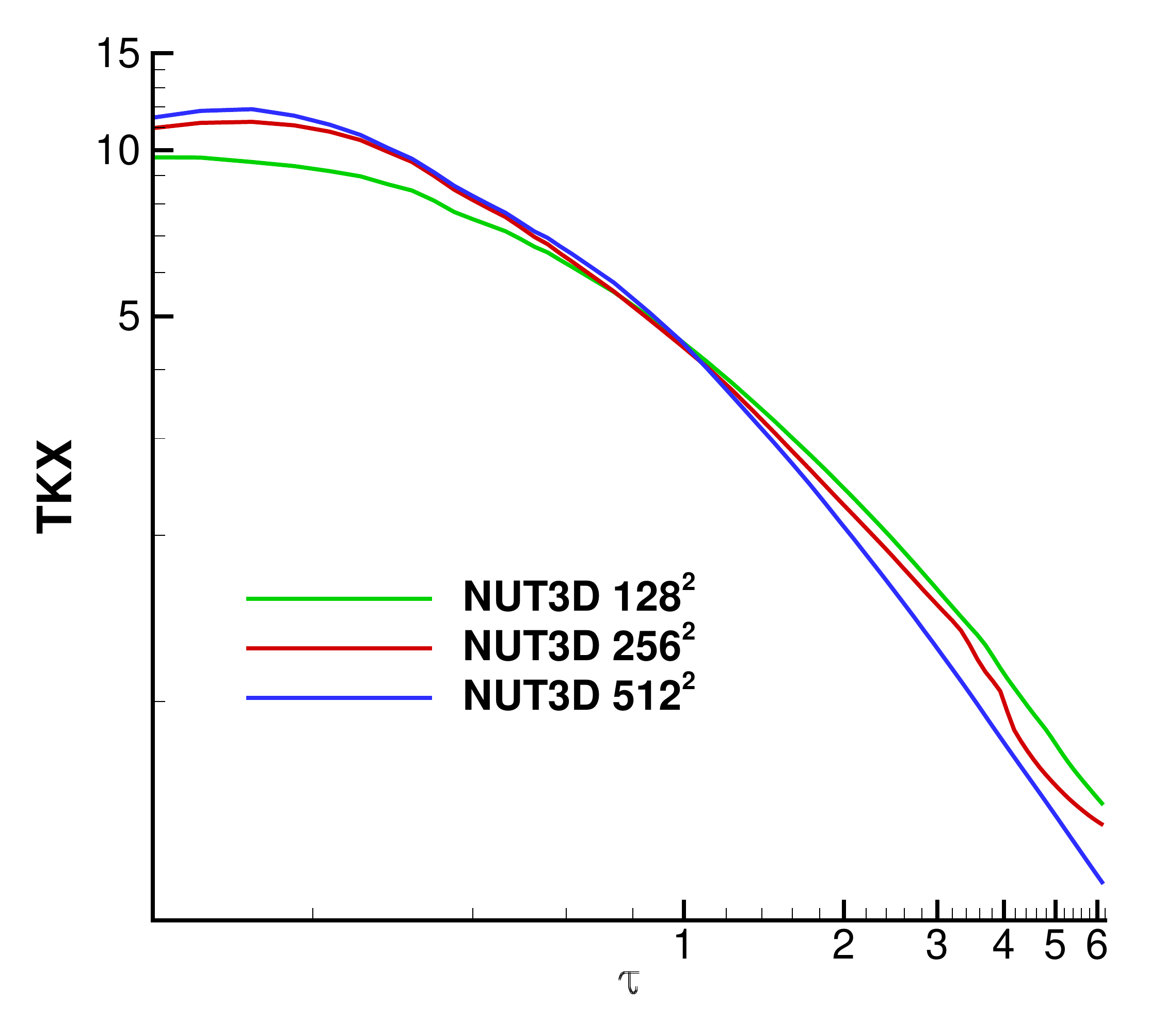}
\includegraphics[width=0.325\textwidth]{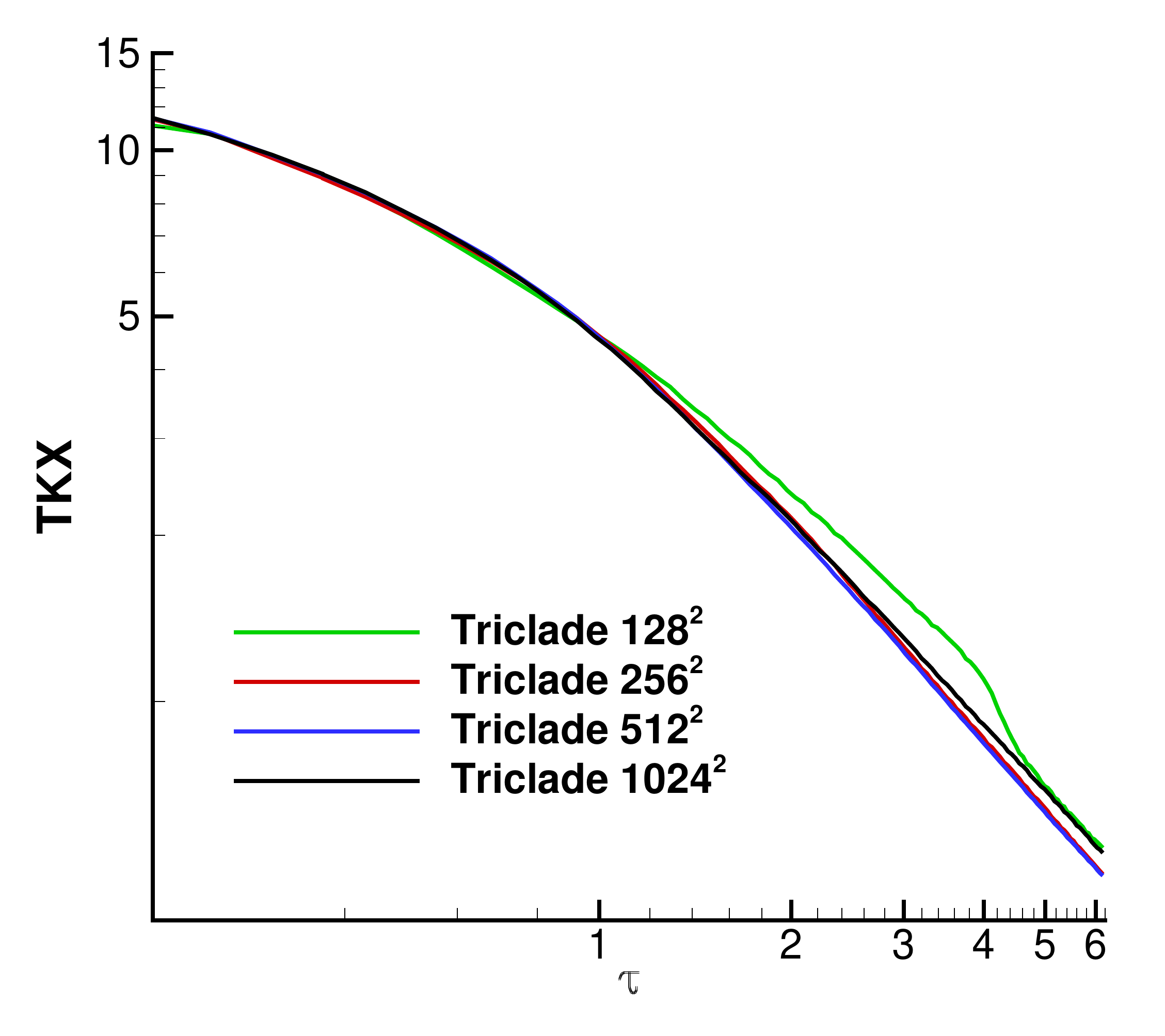}
\includegraphics[width=0.325\textwidth]{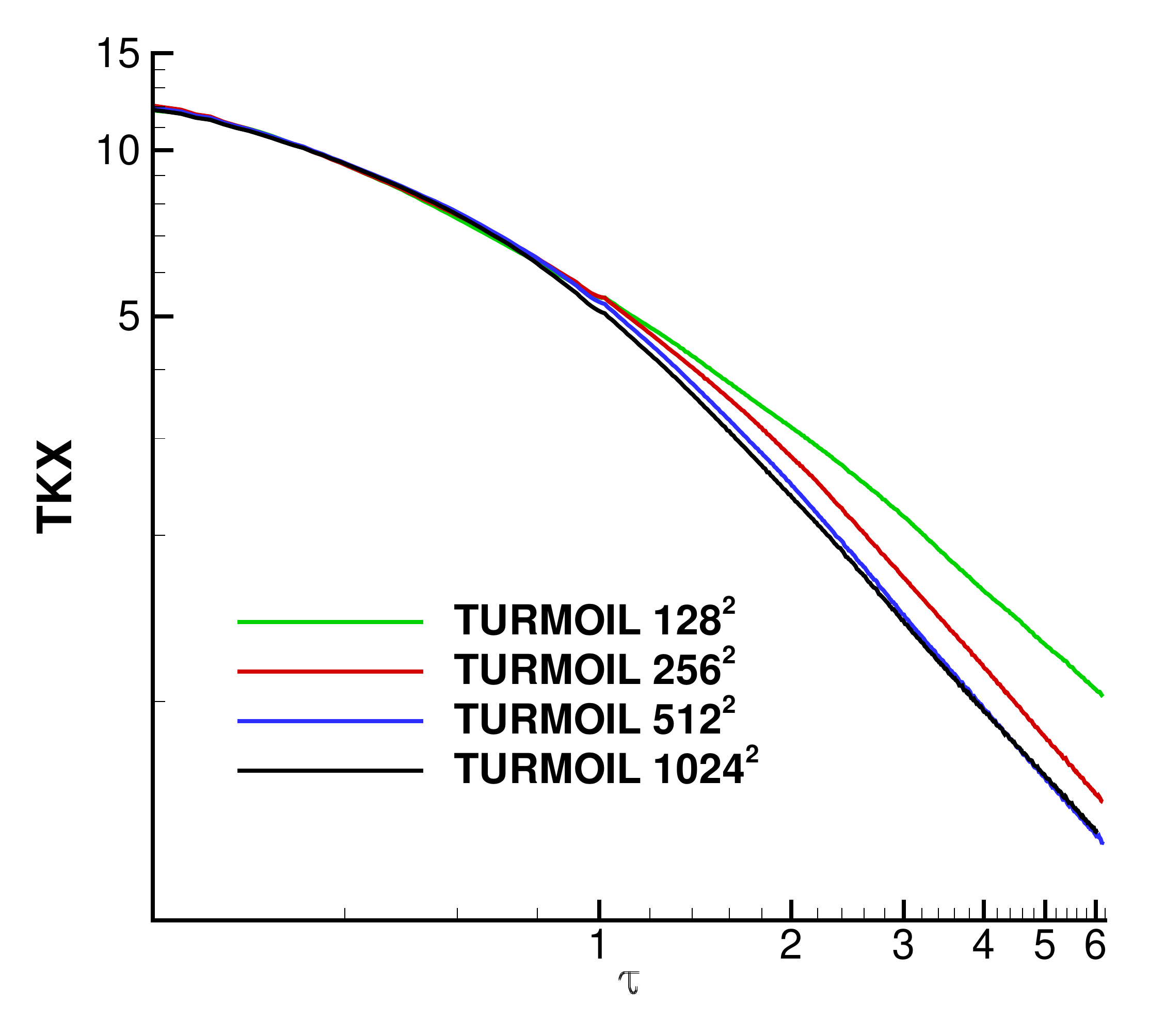}
\includegraphics[width=0.325\textwidth]{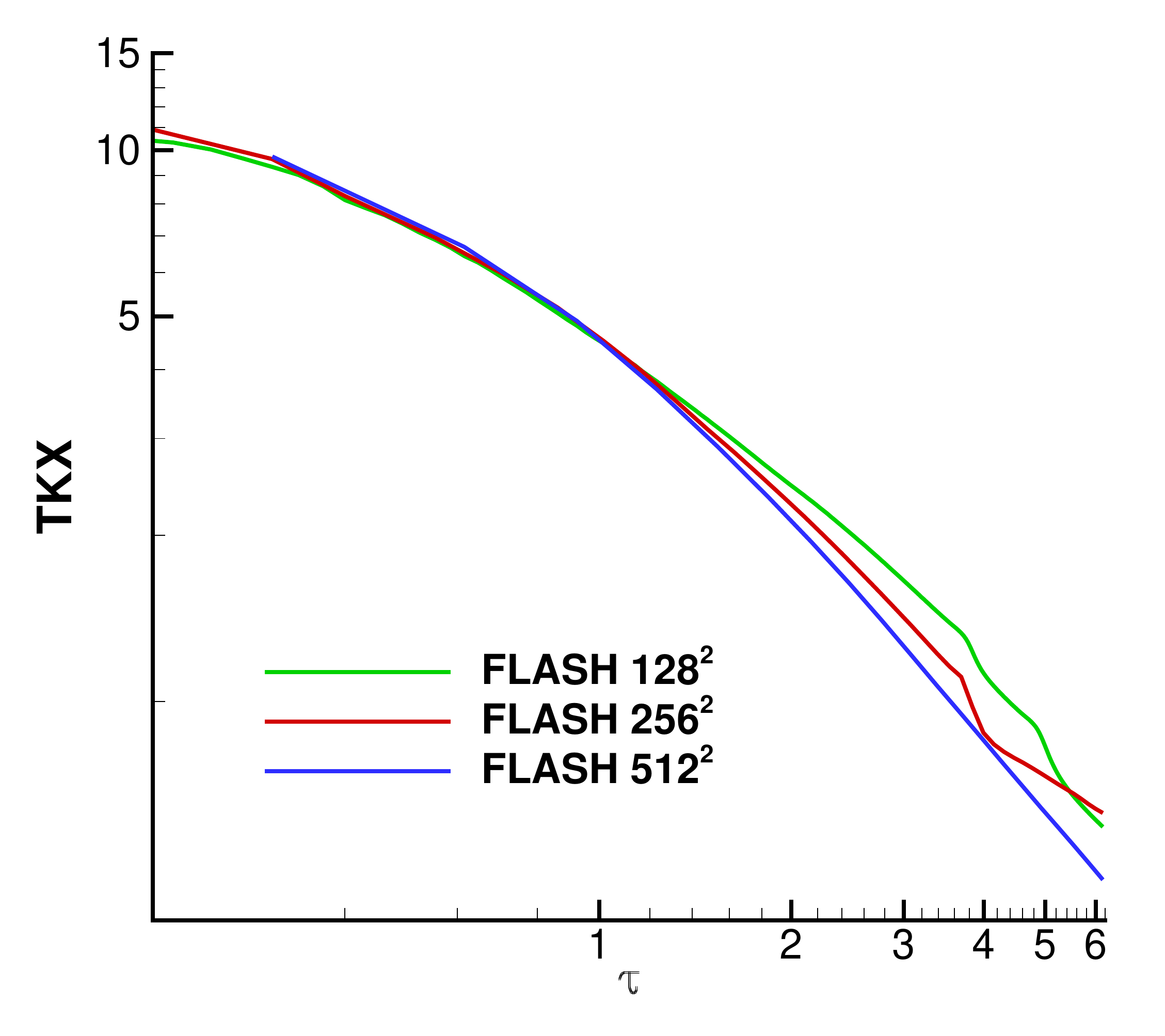}
\includegraphics[width=0.325\textwidth]{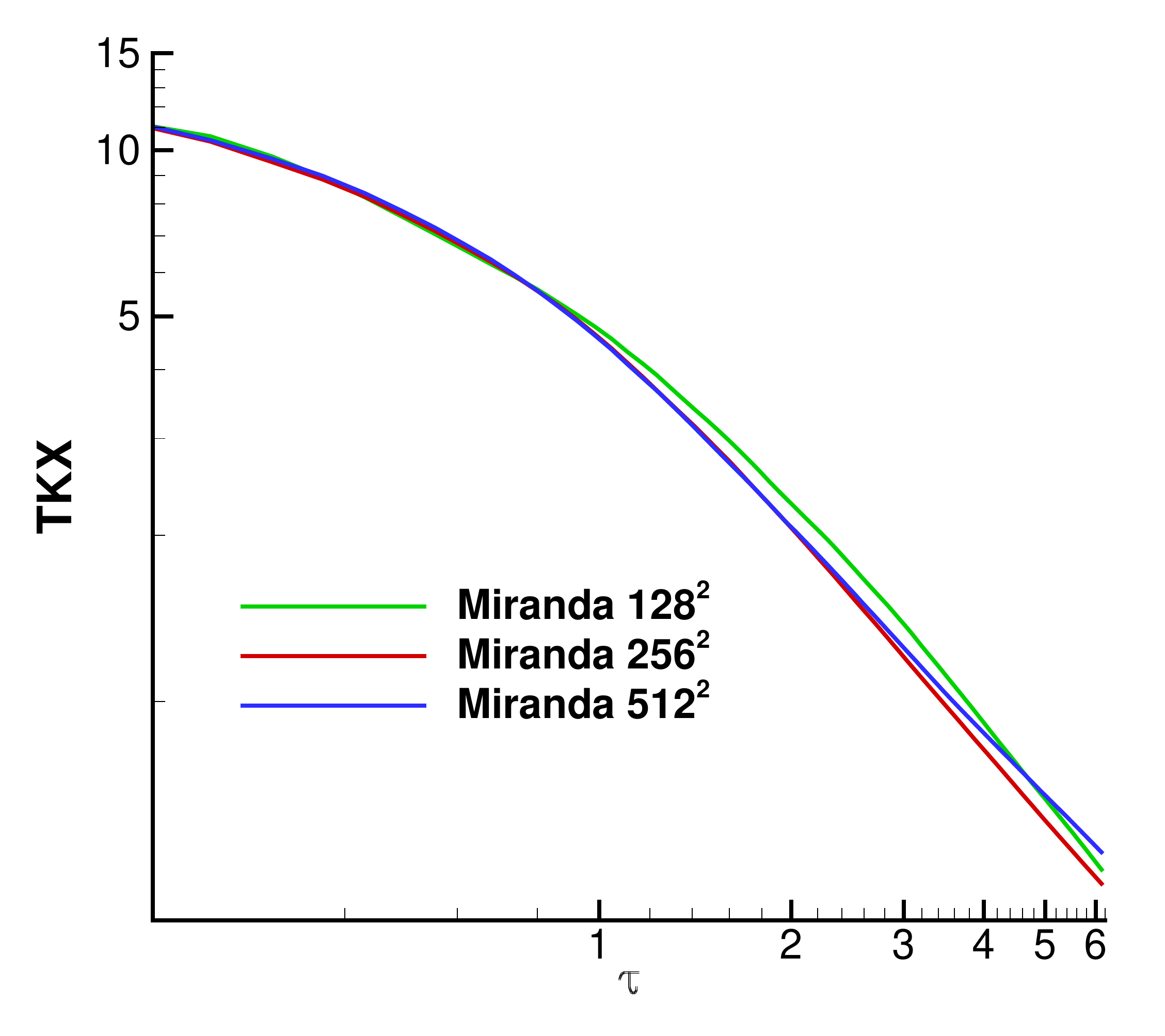}
\includegraphics[width=0.325\textwidth]{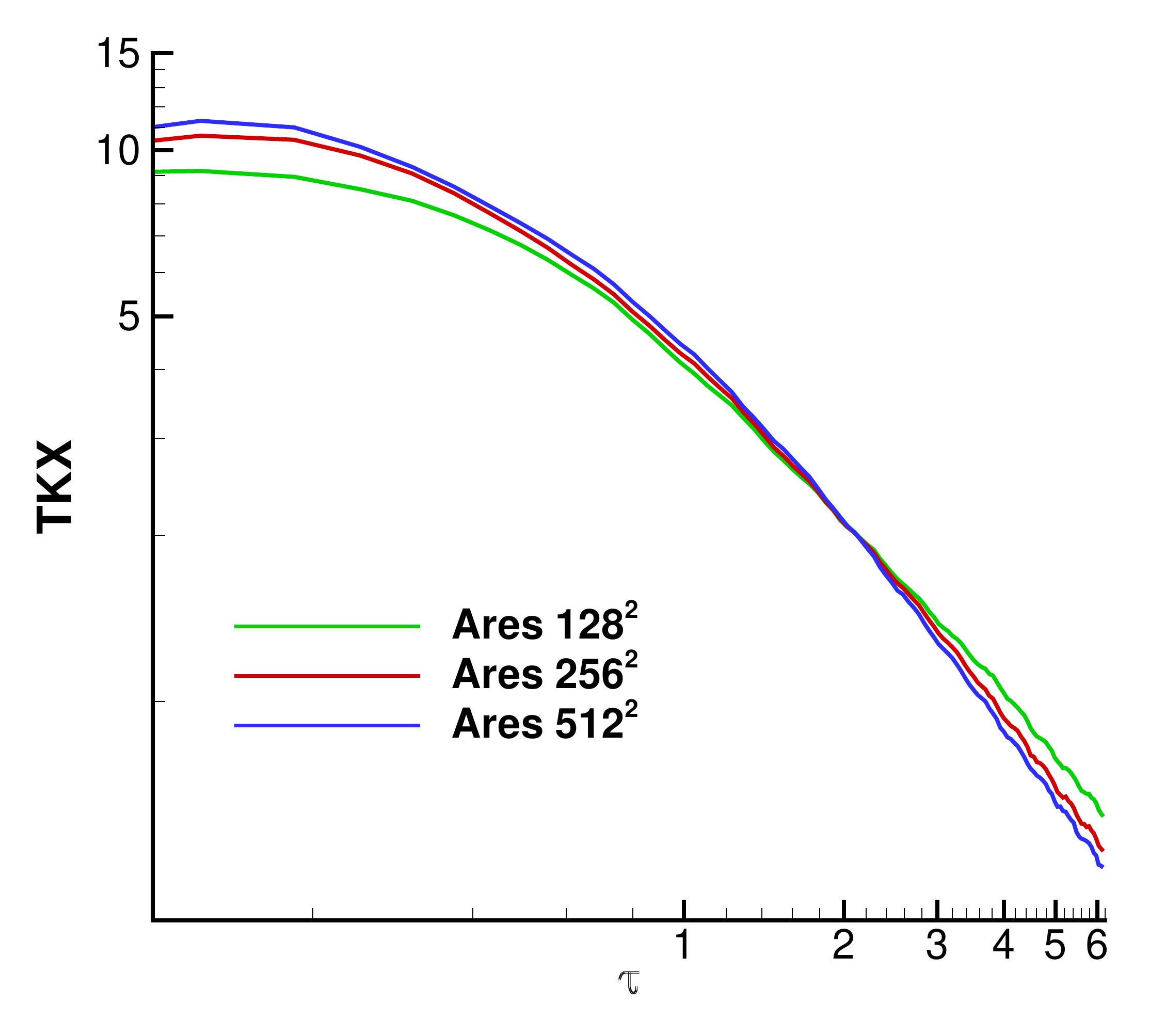}
\caption{Convergence of TKX. \label{convTKX}}
\end{centering}
\end{figure*}
\begin{figure*}
\begin{centering}
\includegraphics[width=0.325\textwidth]{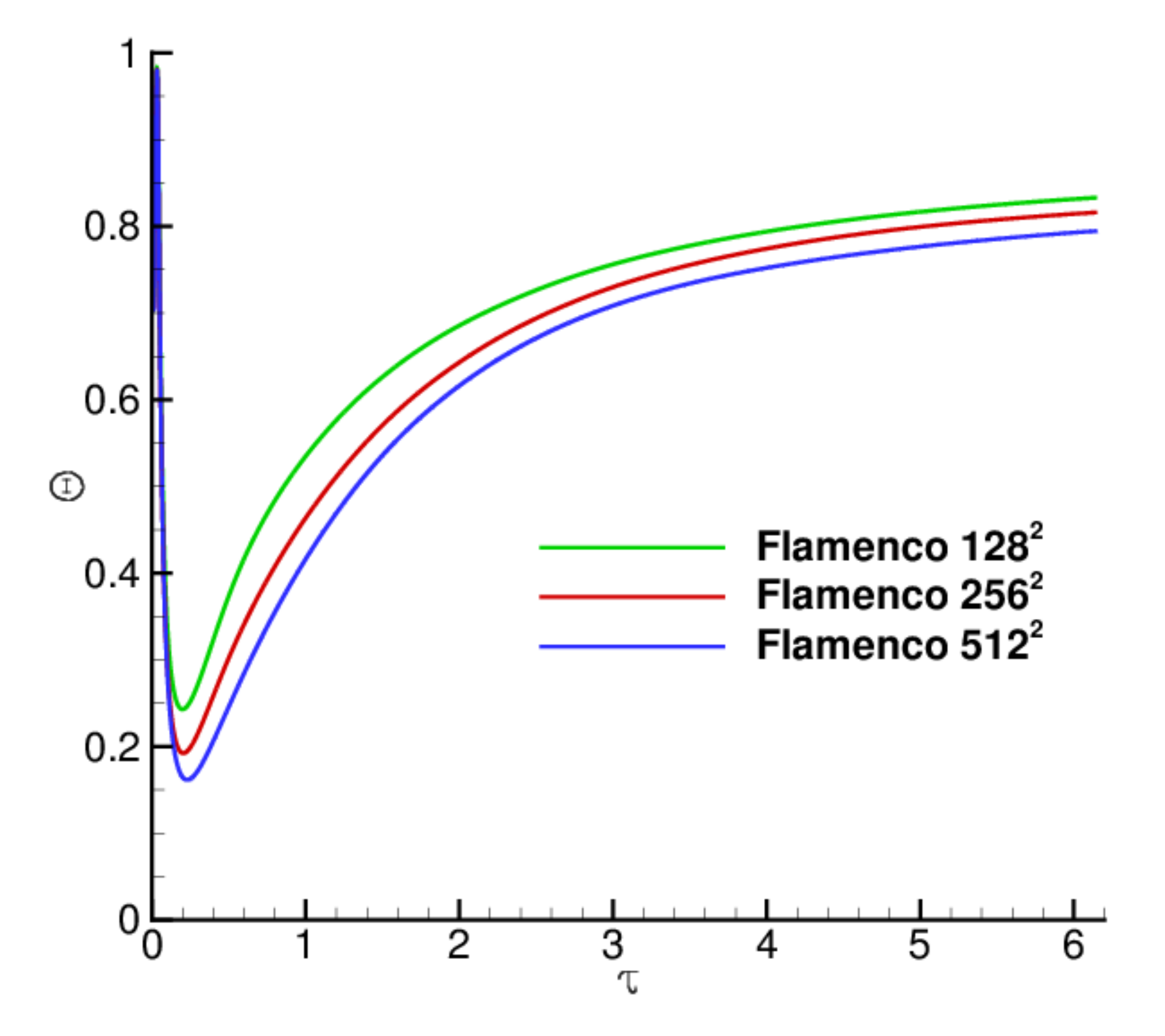}
\includegraphics[width=0.325\textwidth]{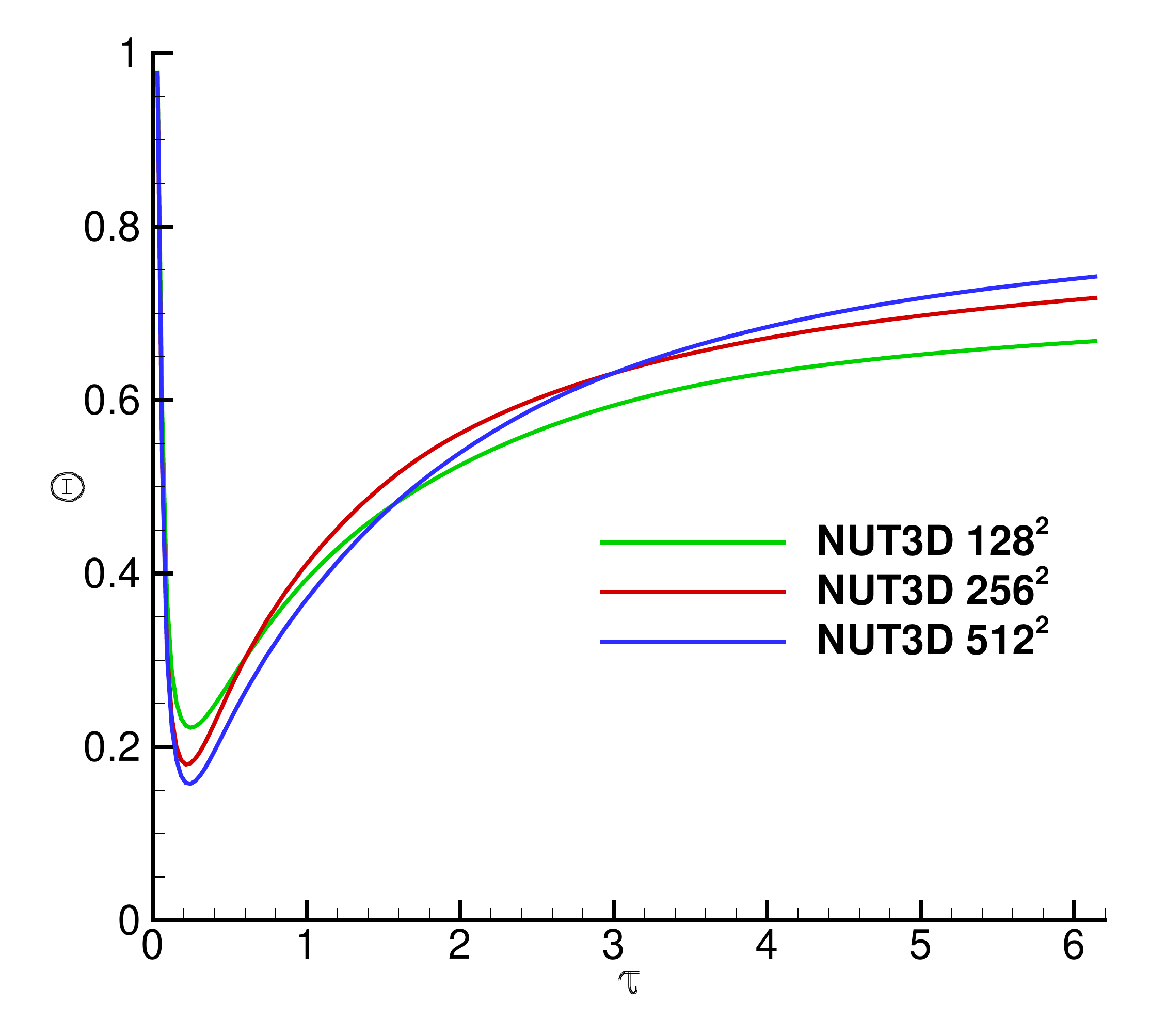}
\includegraphics[width=0.325\textwidth]{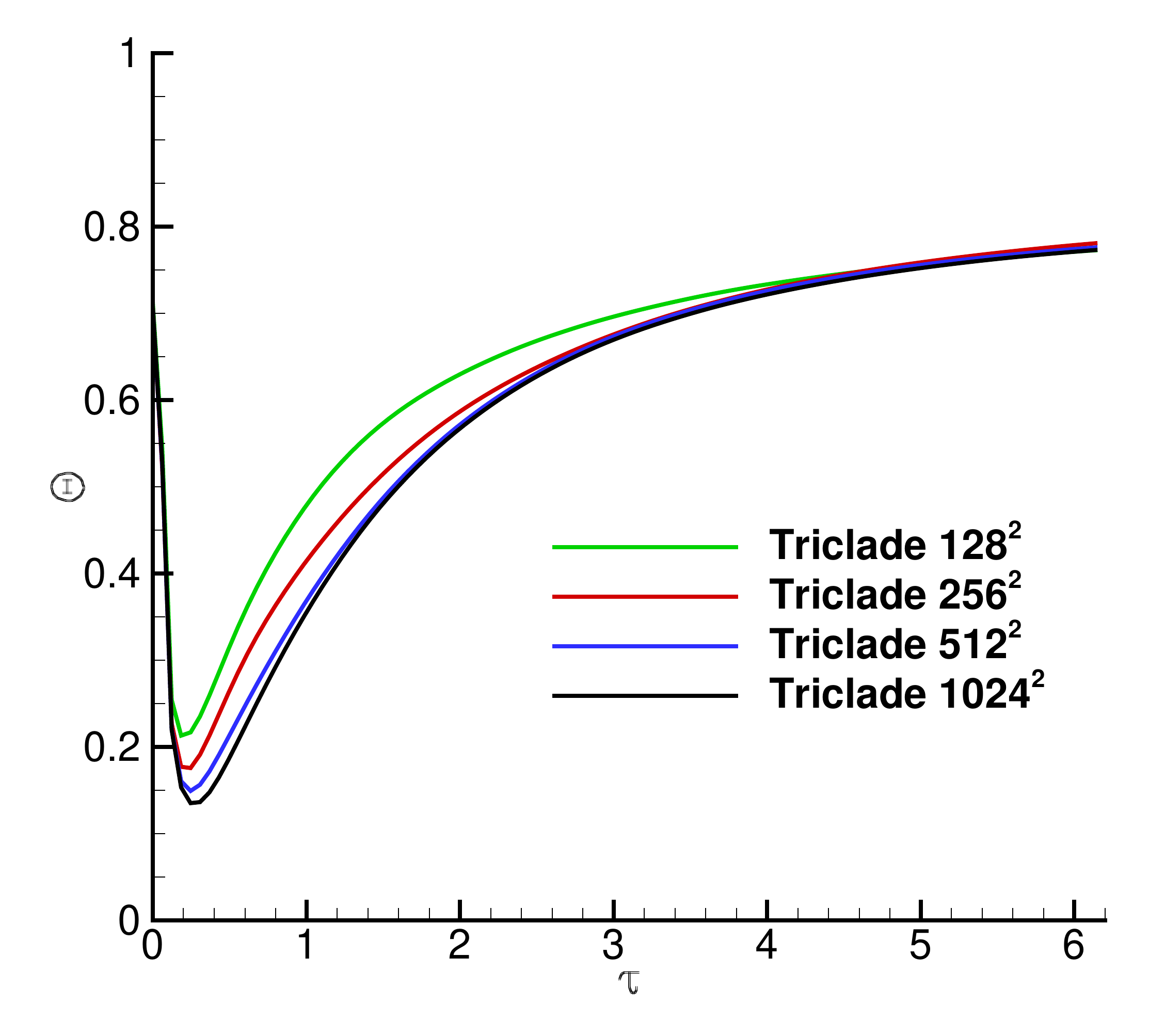}
\includegraphics[width=0.325\textwidth]{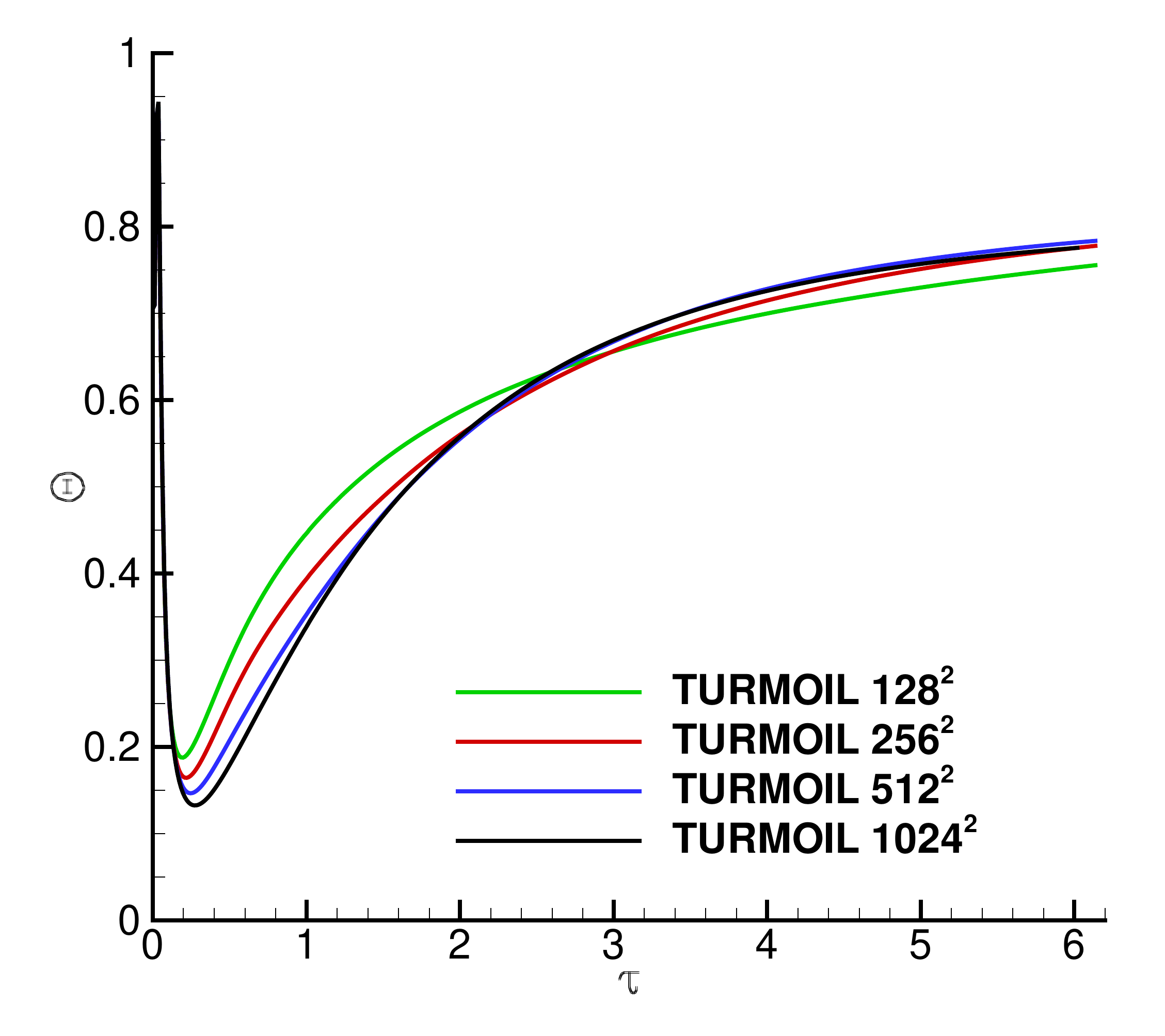}
\includegraphics[width=0.325\textwidth]{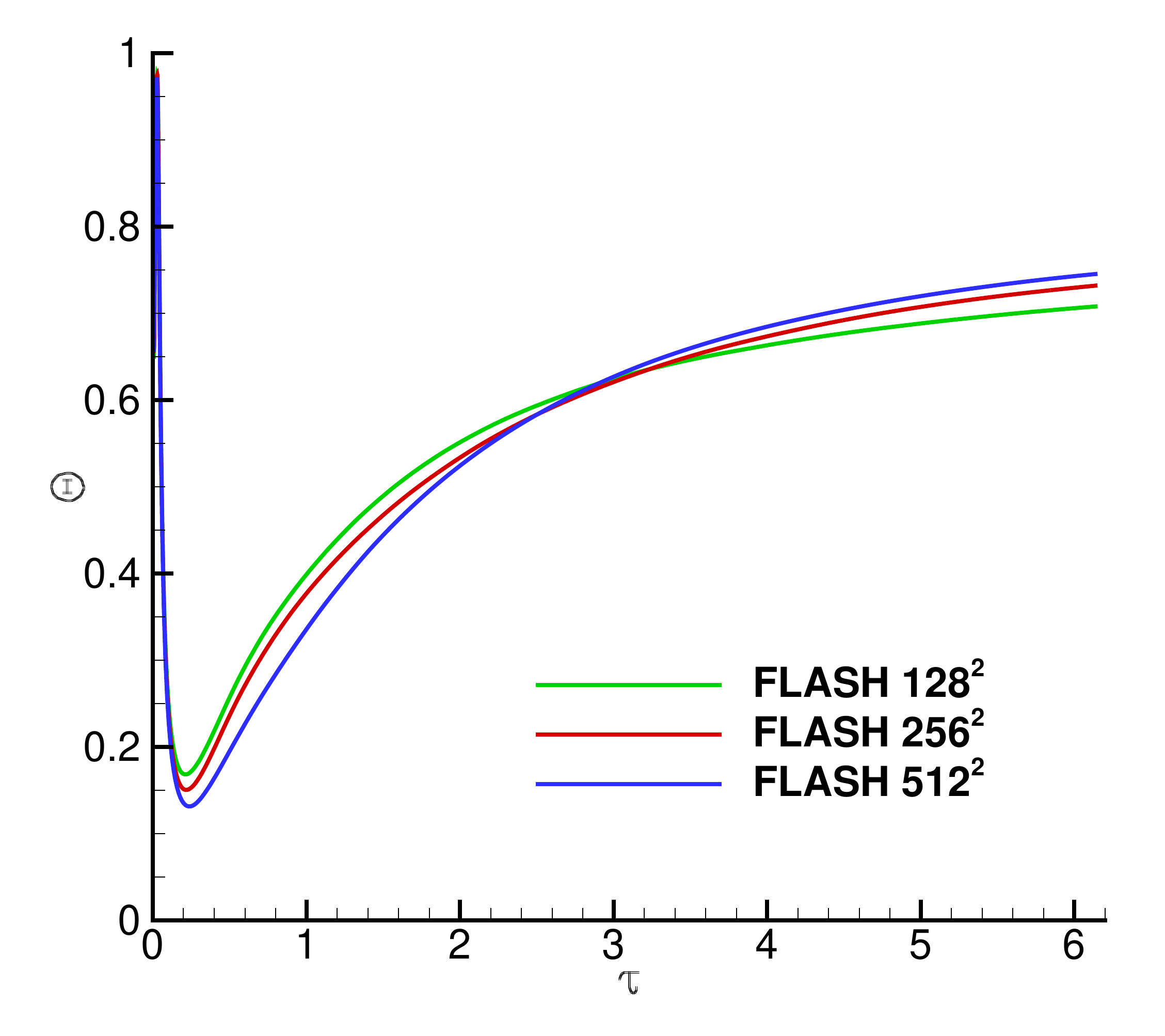}
\includegraphics[width=0.325\textwidth]{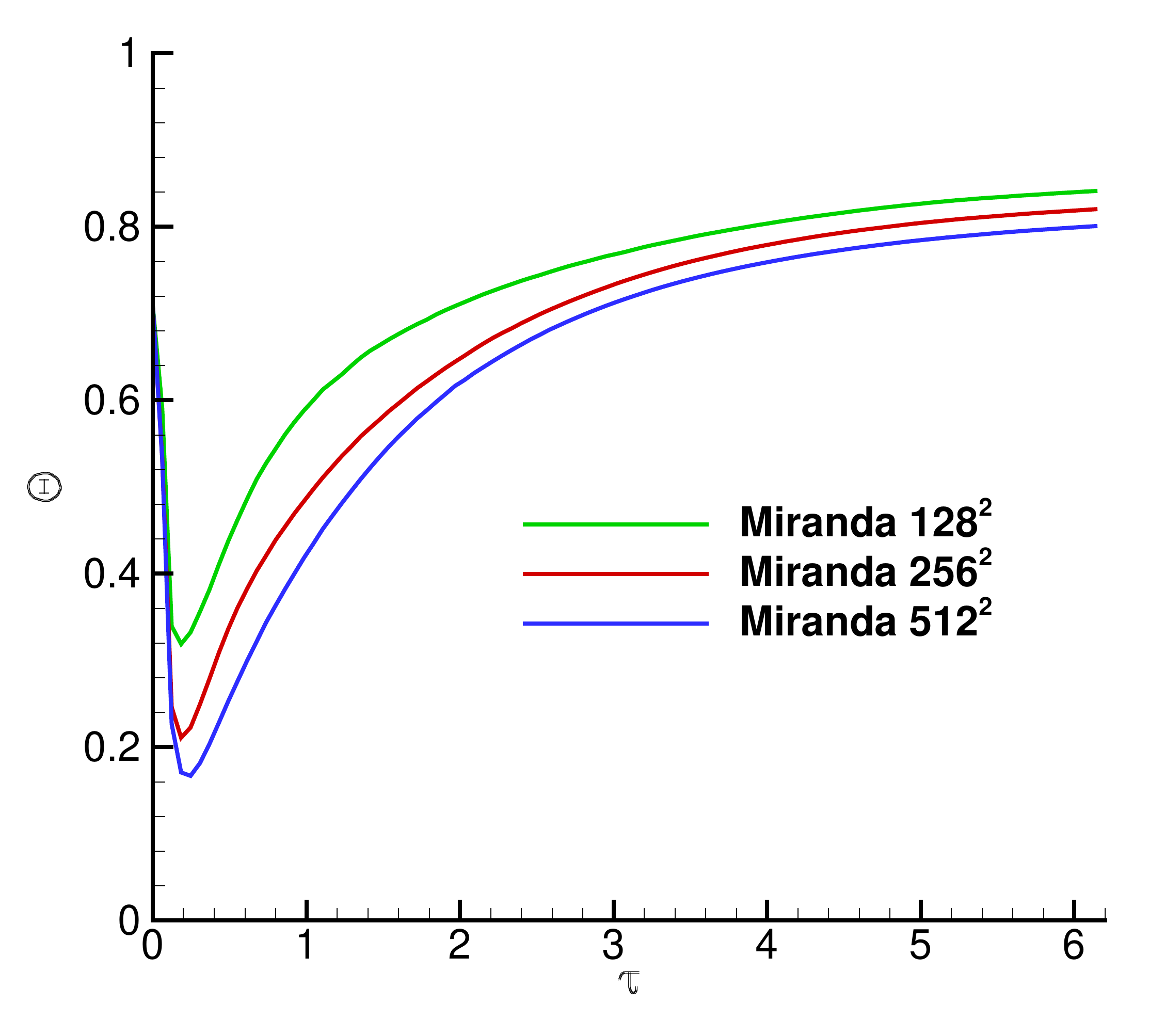}
\includegraphics[width=0.325\textwidth]{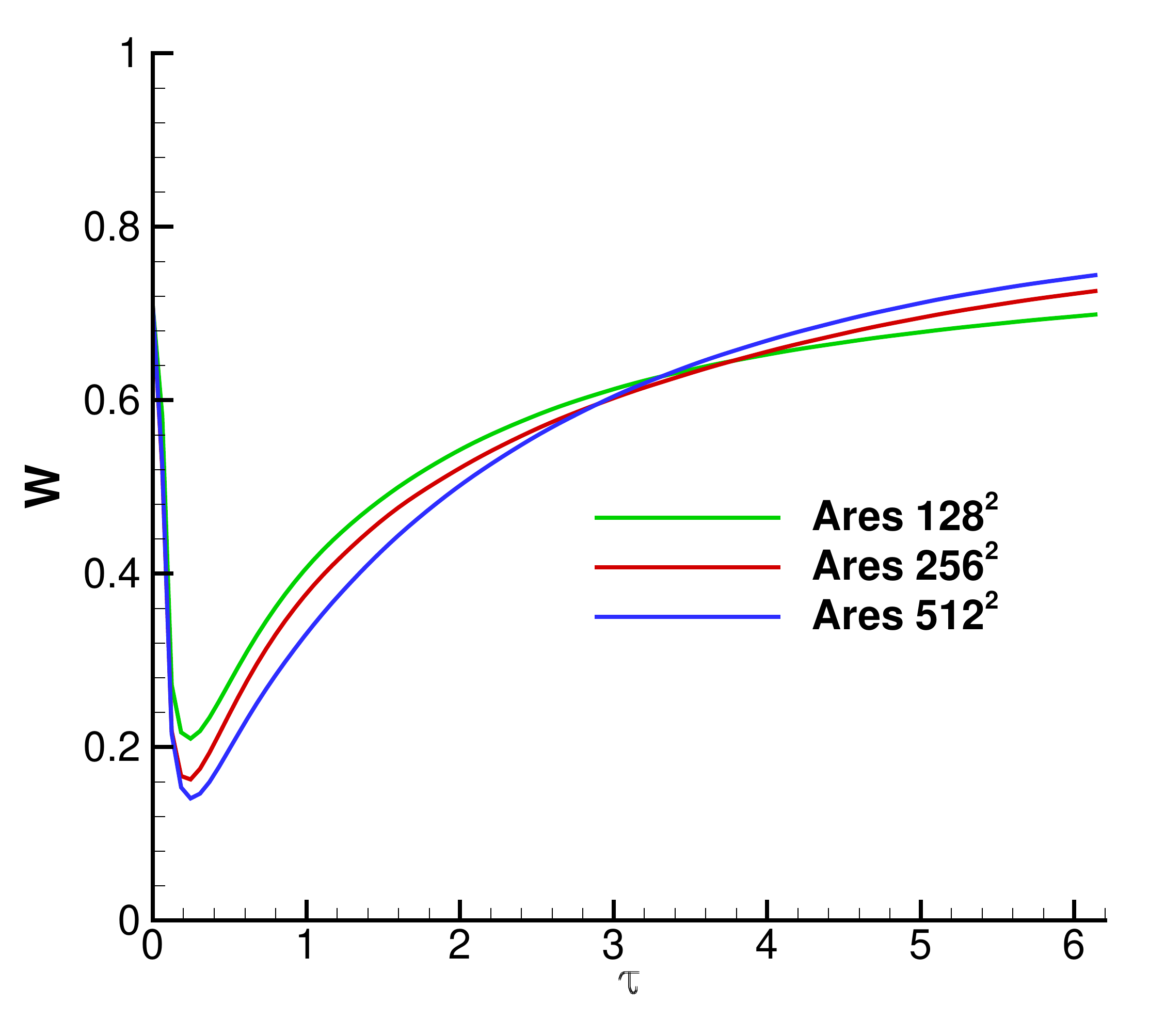}
\caption{Convergence of mix measure $\Theta$. \label{convTheta}}
\end{centering}
\end{figure*}

\end{appendix}

\end{document}